\documentclass[useAMS,usenatbib]{mn2e}
\usepackage[centering,totalwidth=480pt,totalheight=680pt]{geometry}
\usepackage{psfig}
\usepackage[dvips]{graphicx}
\usepackage{natbib}
\usepackage[english]{babel}
\usepackage{amssymb}
\usepackage{xspace}
\usepackage{amsmath}

\newcommand{\LCDM}{$\Lambda$CDM}
\newcommand{\msun}{\mbox{${\rm M}_{\odot}$}}
\newcommand{\msunpcsq}{\mbox{${\rm M}_{\odot} {\rm pc}^{-2}$}}

\newcommand{\HII}{\mbox{${\rm H}_{\rm II}$}}
\newcommand{\HI}{\mbox{${\rm H}_{\rm I}$}}
\newcommand{\Htwo}{\mbox{${\rm H}_{2}$}}

\def\HI{\ifmmode {\mbox H{\scshape i}}\else H{\scshape i}\fi\xspace}
\def\HII{\ifmmode {\mbox H{\scshape ii}}\else H{\scshape ii}\fi\xspace}
\def\Htwo{\ifmmode {\mbox H$_2$}\else H$_2$\fi\xspace}

\def\lesssim{\lower.5ex\hbox{$\; \buildrel < \over \sim \;$}}
\def\gtrsim{\lower.5ex\hbox{$\; \buildrel > \over \sim \;$}}

\begin{document}

\title[Star Formation in Models with Multiphase Gas]{Star Formation in Semi-Analytic Galaxy Formation Models with Multiphase Gas}

\author[Somerville, Popping, \& Trager] {
Rachel S. Somerville$^{1}$\thanks{e-mail: somerville@physics.rutgers.edu},
Gerg\"o Popping$^{2,3}$, Scott C. Trager$^{2}$\\
$^1$Department of Physics and Astronomy, Rutgers University, 136
Frelinghuysen Road, Piscataway, NJ 08854, USA\\
$^2$ Kapteyn Astronomical Institute, University of Groningen, Postbus 800, NL-9700 AV Groningen, the Netherlands\\
$^3$European Southern Observatory, Karl-Schwarzschild-Strasse 2,
85748, Garching, Germany\\
}

\maketitle

\begin{abstract}
We implement physically motivated recipes for partitioning cold gas
into different phases (atomic, molecular, and ionized) in galaxies
within semi-analytic models of galaxy formation based on cosmological
merger trees. We then model the conversion of molecular gas into stars
using empirical recipes motivated by recent observations. We explore
the impact of these new recipes on the evolution of fundamental galaxy
properties such as stellar mass, star formation rate (SFR), and gas
and stellar phase metallicity. We present predictions for stellar mass
functions, stellar mass vs. SFR relations, and cold gas phase and
stellar mass-metallicity relations for our fiducial models, from
redshift $z\sim 6$ to the present day. In addition we present
predictions for the global SFR, mass assembly history, and cosmic
enrichment history.  We find that the predicted stellar properties of
galaxies (stellar mass, SFR, metallicity) are remarkably insensitive
to the details of the recipes used for partitioning gas into \HI\ and
\Htwo. We see significant sensitivity to the recipes for
\Htwo\ formation only in very low mass halos, which host galaxies that
are not detectable with current observational facilities except very
nearby. The properties of low-mass galaxies are also quite insensitive
to the details of the recipe used for converting \Htwo\ into stars,
while the formation epoch of massive galaxies does depend on this
significantly. We argue that this behavior can be interpreted within
the framework of a simple equilibrium model for galaxy evolution, in
which the conversion of cold gas into stars is balanced on average by
inflows and outflows. Star formation in low mass galaxies is strongly
self-regulated by powerful stellar driven outflows, so the overall
galaxy-scale star formation efficiency is nearly independent of the
\Htwo\ depletion time. Massive galaxies at high redshift have not yet
had time to come into equilibrium, so the star formation efficiency is
strongly affected by the \Htwo\ depletion time.

\end{abstract}


\section{Introduction}
\label{sec:intro}

While the \LCDM\ (Cold Dark Matter plus cosmological constant
$\Lambda$) model \citep{blumenthal:84} now provides us with a
well-motivated framework for predicting the abundances and properties
of dark matter halos and the large scale structures in which they are
embedded, all galactic or larger scale simulations must rely on
``sub-grid'' recipes in order to treat processes such as star
formation and stellar feedback. Cosmological simulations are unable to
directly resolve individual stars or, usually, even Giant Molecular
Clouds (GMC).  In order to model the conversion of cold gas into
stars, up until recently, both numerical and semi-analytic
cosmological simulations typically utilized a very simple empirical
sub-grid recipe based on observations most famously by
\citet{Schmidt:1959,Schmidt:1963} and
\citet{Kennicutt:1989,kennicutt:98} (often referred to as the
``Kennicutt-Schmidt'' (KS) relation). These observations showed that
the surface density of star formation $\Sigma_{\rm SFR}$ was
proportional to the surface density of cold gas to a power $N_{KS}$.
Observations also showed that the efficiency of star formation dropped
rapidly below a critical gas surface density
\citep{Martin:2001}. There has been debate about whether this critical
surface density is best described in terms of a Toomre stability
criterion \citep{Toomre:1964} or a constant critical density, and
indeed about the physical origin of this critical density
\citep{Schaye:2004,Leroy:2008}.

From the pioneering work of \citet{Katz:1992} up until recently,
cosmological simulations of galaxy formation, both numerical and
semi-analytic, have implemented a star formation recipe in which
``cold'' gas (typically with $T \lesssim 10^4$ K) with volume density
$\rho_{\rm gas}$ is assumed to form stars at a rate per unit volume:
\begin{equation}
\dot{\rho}_* = \epsilon_* \rho_{\rm gas}^{N}
\label{eqn:sflaw}
\end{equation}
with $N \simeq 1.5$ and $\epsilon_*$ usually treated as a free
parameter, tuned to match the observed Kennicutt relation. A common
variant assumes $\dot{\rho}_* \propto \rho_{\rm gas}/t_{\rm ff}$,
which is approximately equivalent because the local free-fall time
$t_{\rm ff}\propto \rho^{-0.5}$. Motivated by the observational
evidence described above, many modelers incorporated either a critical
surface density or volume density into their star formation recipe,
which proved to be important in order to reproduce the observed high
gas fractions in low-mass galaxies.

Beginning about a decade ago, our understanding of how star formation
on $\sim 100$ pc--kpc scales depends on local conditions began to
undergo a revolution. \citet{Wong:2002} showed that the correlation
between $\Sigma_{\rm SFR}$ and the surface density of \emph{molecular}
gas $\Sigma_{H2}$ was stronger than that between $\Sigma_{\rm SFR}$
and the \emph{total} gas density $\Sigma_{\rm gas}$ in molecule rich
galaxies.  In the past five years, this field has advanced rapidly
with the availability of galaxy-wide, high resolution maps of the star
formation and multi-phase (\HI\ and \Htwo) gas in reasonably large
samples of nearby galaxies, e.g. from the THINGS (The HI nearby galaxy
survey; \citealt{Walter:2008}) combined with BIMA SONG (BIMA survey of
Nearby Galaxies; \citealt{Helfer:2003}) and HERACLES (HERA CO-Line
Extragalactic Survey; \citealt{Leroy:2009}). Based on these
observations, it has been shown
\citep{Bigiel:2008,Leroy:2008,Bigiel:2011,Schruba:2011} that, when
averaged over scales of $\sim 700$ pc, the star formation density is
tightly correlated with the \emph{molecular} gas density to a nearly
linear power, and that there is almost \emph{no} correlation between
$\Sigma_{\rm SFR}$ and the density of atomic gas, so that the
correlation between $\Sigma_{\rm SFR}$ and $\Sigma_{\rm gas}$ (the
traditional KS relation) breaks down badly in the \HI-dominated parts
of galaxies (typically in galaxy outskirts). These results highlight
the importance of modeling the partition of gas into different
phases, i.e. atomic vs. molecular, which has not been attempted in
most cosmological simulations of galaxy formation to date.

At the same time, there has been significant progress in understanding
and modeling the formation of molecular hydrogen and star formation on
galactic scales.  \citet{BR:04,BR:06} showed that the fraction of
atomic to molecular gas in a sample of nearby disk galaxies was
tightly correlated with the midplane pressure (determined by the
density of both stars and gas), and this result has been confirmed in
larger samples such as THINGS
\citep{Leroy:2008}. \citet{Robertson:2008} implemented low-temperature
($T <10^4$ K) cooling, photo-dissociation of \Htwo, and an \Htwo-based
SF recipe in hydrodynamic simulations of isolated disk galaxies of
various masses.  \citet{KMT:09} presented analytic models for the
formation of \Htwo\ as a function of total gas density and
metallicity, supported by numerical simulations with simplified
geometries \citep{Krumholz:2008,Krumholz:2009}, emphasizing the
importance of metallicity as a controlling parameter in
\Htwo\ formation. \citet{Gnedin:2010,GK:11} included detailed
chemistry and low temperature cooling as well as a simplified
treatment of radiative transfer and an \Htwo-based SF recipe in
cosmological ``zoom-in'' Adaptive Mesh Refinement (AMR) simulations of
small regions, and presented analytic fitting functions to their
results as a function of total gas density, metallicity, and the
strength of the local UV background. \citet{Christensen:2012} used a
similar approach to implement chemistry and simplified radiative
transfer in Smoothed Particle Hydrodynamics (SPH) zoom-in simulations
of galaxy-sized regions, which include a blast-wave treatment of
supernova feedback.

A somewhat different view has been presented by \citet{Ostriker:2010},
who propose that heating of the Interstellar Medium (ISM) by the
stellar UV background plays a key role in regulating star
formation. In their model, the thermal pressure in the diffuse ISM,
which is proportional to the UV heating rate, adjusts until it
balances the midplane pressure set by the vertical gravitational
potential. This could provide an explanation for the strong empirical
correlation between \Htwo\ fraction and disk midplane pressure found by
\citet{BR:06}.

Although detailed simulations are crucial in order to understand the
complex physical processes involved, extremely high resolution is
required in order to obtain reliable results \citep[see
  e.g.][]{Christensen:2012}, implying that it will be feasible to
simulate only small numbers of galaxies with these techniques for the
next few years. Meanwhile, large surveys of cold gas in nearby and
distant galaxies with new and upcoming facilities such as the Atacama
Large Millimeter/submillimeter Array (ALMA) and the SKA (Square
Kilometer Array) and its pathfinders are already being planned and
pilot projects are underway. As a result, it is important to develop
computationally efficient techniques that can incorporate physically
motivated treatments of gas partioning into its atomic, molecular, and
possibly ionized phases and \Htwo-based star formation recipes into
simulations of cosmological volumes.

Semi-analytic models (SAMs) provide an alternative approach to this
problem. In semi-analytic merger tree models, a merger tree represents
the formation and growth of a dark matter halo that is identified at
some redshift of interest; these merger trees may be extracted from
dissipationless N-body simulations or created using analytic
techniques \citep[e.g.][]{sk:99,Parkinson:2008}. Simplified but
physically motivated recipes are used to track the rate of gas cooling
into galaxies, and these recipes have been tested against fully
numerical hydrodynamic simulations
\citep[e.g.][]{Hirschmann:2012a}. These models use angular momentum
based arguments to track the radial sizes of forming disks
\citep{mo:98,somerville:08a}, and can then implement recipes for how
cold gas is converted into stars, and how energy and momentum from
massive stars and supernovae is returned to the Interstellar Medium
(ISM). This ``feedback'' from stars and SNae is assumed to drive
large-scale winds that can remove gas from the galaxy. The production,
ejection, and recycling of metals is also tracked. Thus our existing
semi-analytic modeling framework provides the main quantities (total
gas density in disks, gas metallicity) needed to implement physically
motivated recipes for partitioning gas into an atomic and molecular
component and then implementing an \Htwo-based star formation recipe.

Several efforts along these lines have already been presented in the
literature. \citet{obreschkow:09} implemented a prescription to
estimate the \Htwo\ fraction based on the empirical pressure-based
recipe of \citet{BR:06} applied in post-processing to the Millennium
simulations of \citet{Delucia:2007}. However, in this approach, the
star formation in the simulations was still based on a traditional KS
recipe using the total gas density, not self-consistently on the
estimated \Htwo\ gas density. \citet{fu:10,fu:12} modeled the
partitioning of gas into \HI\ and \Htwo\ in radial bins in each
galaxy, using both the metallicity-dependent recipes of \citet{KMT:09}
and the pressure-based recipe of \citet{BR:06}, and self-consistently
implemented an \Htwo-based star formation recipe, within the
semi-analytic modeling framework of
\citet{Guo:2011}. \citet{lagos_sflaw:11,lagos_cosmic:11} also
estimated gas partitioning into an atomic and molecular component, and
implemented an \Htwo-based star formation recipe, within the GALFORM
semi-analytic models \citep{Baugh:2005,Bower:2006}.  Similar modeling
efforts utilizing a somewhat more simplified framework (i.e., only the
mass accretion history of the largest progenitor is tracked, rather
than the full merger tree) have been presented by \citet{Dutton:2010}
using the pressure-based \citet{BR:06} approach, and
\citet{Krumholz:2012} using the \citet{KMT:09} metallicity-based
approach.

It is already clear that the results of this kind of exercise may
depend on the other ingredients of the modeling, in particular on the
treatment of stellar feedback, chemical evolution, and potentially on
feedback from Active Galactic Nuclei (AGN). In this work, we present
new models that incorporate a metallicity or pressure oriented
treatment of atomic-molecular gas partitioning and an \Htwo-based star
formation recipe within the semi-analytic modeling framework developed
by the Santa Cruz group \citep{sp:99,spf:01,s08,Somerville:2012}.

The current generation of semi-analytic models (incorporating some
form of ``quenching'' in massive halos, e.g. from AGN feedback) has
been fairly successful at reproducing a variety of galaxy
observations, but suffer from generic problems. Both the successes and
problems seem to be common to the semi-analytic models developed by
many different groups as well as to large-volume cosmological
hydrodynamic simulations \citep[see ][for a
  discussion]{Somerville_Dave:2014}. Significant successes include the
ability to match the observed stellar mass function or luminosity
functions from the UV to the NIR at $z=0$, while simultaneously
matching the gas fraction as a function of stellar mass for nearby
disk galaxies
\citep[e.g.][]{s08,Somerville:2012,Lu:2014}. Observations show that
massive galaxies form their stars early, and that the star formation
in many of these massive objects is \emph{quenched} early, so that
their stars evolve largely passively. There is some tension in the
ability of models to produce enough massive galaxies at early times
($z\gtrsim 2$), and a dearth of very rapidly star forming objects
observed in the sub-mm and FIR
\citep{Somerville:2012,Niemi:2012}. However, the evolution of the
number of massive ``quenched'' galaxies in models with AGN feedback
seems to match observations reasonably well
\citep{Kimm:2009,Brennan:2015}.

Low mass galaxies seem to present a more thorny set of problems, which
we refer to collectively as the ``dwarf galaxy conundrum''. Models
that reproduce the low-mass end of the observed stellar mass function
locally, generically \emph{overproduce} low-mass ($m_* \lesssim
10^{10}\, \msun$) galaxies at redshifts $0.5 \lesssim z \lesssim
2$. Moreover, low-mass galaxies apparently have (specific) star
formation rates that are too low over the same redshift range. The
stellar ages predicted by our models for these galaxies are too old
compared to those derived for nearby galaxies based on
`archaeological' evidence. A summary of these problems, demonstrated
for several independently developed semi-analytic models, was
presented in \citet{fontanot:09}. \citet{weinmann:12} presented a
similar study that showed that the same problems also occur in
numerical hydrodynamic simulations, and recently
\citet{Somerville_Dave:2014} showed that the problem persists to
varying degrees in most state-of-the-art SAMs and cosmological
hydrodynamical simulations. It has been suggested that these problems
might be due to inaccurate recipes for star formation, and that they
might be cured by implementing metallicity dependent recipes for
\Htwo\ formation and \Htwo-based star formation
\citep{Krumholz:2012,kuhlen:12}. This was one of the original
motivations for the work we present here.

The purpose of this paper is to present the details of how we
incorporate partitioning of gas into an atomic, molecular and
(optionally) ionized component in our existing semi-analytic models,
how we self-consistently implement an \Htwo-based star formation
recipe, and how sensitive our results are to details of the
implementation. We explore three different recipes for the
partitioning of gas into different phases: the pressure-based recipe
of \citet[][BR]{BR:06} and two metallicity-based recipes, that of
Krumholz et al. (KMT) and that of \citet[][GK]{GK:11}. We compare the
predictions of these three new models with those using the ``classic''
Kennicutt-Schmidt (KS) star formation recipe with no gas
partitioning. In addition, we explore several different empirically
motivated \Htwo-based star formation recipes.

This paper is part of a series of related works. In
\citet[][PST14]{Popping:2014}, we presented predictions for the atomic
and molecular gas content of galaxies, and its evolution with redshift
from $z=6$--0, using the same models presented
here. \citet{Popping:2014c} extended these models by carrying out
radiative transfer calculations to predict sub-mm line emission
luminosities from several atomic and molecular species, including CO,
HCN, C$^+$, and [OI]. In \citet{Berry:2014}, we presented predictions
for the properties of objects that would be selected as Damped
Lyman-$\alpha$ systems (DLAS) in absorption against background
quasars, again using the same model framework described here. In this
paper, we focus on quantities pertaining to the stellar content, SFR,
and metal content of galaxies and their evolution since $z\sim 6$. In
addition, we explore a wider variety of model variants than presented
in the earlier works.

The outline of the paper is as follows. In \S\ref{sec:oldsam} we
outline the basic framework of the semi-analytic models and the
treatment of structure formation, gas cooling and infall, chemical
evolution, and starbursts and morphological transformation via galaxy
mergers. In \S\ref{sec:gaspart} we describe our approaches for
partitioning cold gas into an atomic, molecular, and (optionally)
ionized component, in \S\ref{sec:sam:sfrecipes} we describe the new
\Htwo-based star formation recipes, and in \S\ref{sec:metalwind} we
describe our implementation of metal enhanced winds. In
\S\ref{sec:results:norm} we describe how we choose the values of the
free parameters in our models, and summarize their values. In
\S\ref{sec:trace}, we show how the star formation histories and
build-up of stars, gas, and metals as a function of halo mass are
impacted by the different recipes for gas partioning and star
formation, and other details of our model implementation. In
\S\ref{sec:sfrel}, we show predictions for the relationship between
total gas density and SFR density in our models. In
\S\ref{sec:results:pops}, we present predictions for the stellar mass
functions and stellar fractions, specific star formation rates, gas
depletion timescales, and gas and stellar phase metallicities over
cosmic time from $z\sim 6$ to the present. We discuss our results in
\S\ref{sec:discussion} and summarize and conclude in
\S\ref{sec:conclusions}.


\section{Model Description}
\label{sec:model}

The semi-analytic models used here have been described in detail in
\citet{sp:99}, \citet{spf:01} and most recently in \citet[hereafter
  S08]{s08} and \citet[S12]{Somerville:2012}. The Santa Cruz modeling
framework has also recently been described in \citet{Porter:2014}. We
refer the reader to those papers for details.

\subsection{The Semi-Analytic Model Framework}
\label{sec:oldsam}

This section describes the aspects of the semi-analytic models that
have been documented in previous papers. Therefore we give a
relatively brief description of these ingredients here. 

In this work, the merging histories (or merger trees) of dark matter
haloes are constructed based on the Extended Press-Schechter (EPS)
formalism using the method described in \citet{sk:99}, with
improvements described in S08. These merger trees record the growth of
dark matter haloes via merging and accretion, with each ``branch''
representing a merger of two or more haloes. We follow each branch
back in time to a minimum progenitor mass $M_{\rm res}$, which we
refer to as the mass resolution of our simulation. Our SAMs give
nearly identical results when run on the EPS merger trees or on merger
trees extracted from dissipationless N-body simulations
\citep{Lu:2014,Porter:2014}. We use EPS merger trees here because they
allow us to attain extremely high resolution, which is important for
this study. We resolve halos down to $M_{\rm res}=10^{10}\, \msun$ for
all root halos, and below root halo masses of $M_{\rm res}=10^{10}\,
\msun$, we set $M_{\rm res} = 0.01\, M_{\rm root}$, where $M_{\rm root}$
is the mass of the root halo. Our root halos cover a range from $M_h =
5\times 10^8\, M_\odot$ to $5\times 10^{14} M_\odot$.

When dark matter haloes merge, the central galaxy of the largest
progenitor becomes the new central galaxy, and all others become
`satellites'. Satellite galaxies lose angular momentum due to
dynamical friction as they orbit and may eventually merge with the
central galaxy. To estimate this merger timescale we use a variant of
the Chandrasekhar formula from \citet{boylan-kolchin:08}. Tidal
stripping and destruction of satellites are also modeled as described
in S08.

Before the Universe is reionised, each halo contains a mass of hot gas
equal to the universal baryon fraction times the virial mass of the
halo. After reionisation, the photo-ionising background suppresses the
collapse of gas into low-mass haloes.  We use the fitting functions
provided by \citet{gnedin:00} and \citet{kravtsov:04}, based on their
hydrodynamic simulations, to model the fraction of baryons that can
collapse into haloes of a given mass after reionisation, and assume
that the universe was fully reionized by $z=11$.

When a dark matter halo collapses, or experiences a merger that at
least doubles the mass of the largest progenitor, the hot gas is
assumed to be shock-heated to the virial temperature of the new
halo. This radiating gas then gradually cools and collapses. The
cooling rate is estimated using a simple spherically symmetric model
similar to the one originally suggested by \citet{White:1991}. Details
are provided in S08.

We assume here that the cold gas is accreted only by the central
galaxy of the halo, although in reality satellite galaxies probably
also continue to accrete some cold gas after they cross the virial
radius of their host. In addition, we assume that all newly cooling
gas initially collapses to form a rotationally supported disc. The
scale radius of the disc is computed based on the initial angular
momentum of the gas and the halo profile, assuming that angular
momentum is conserved and that the self-gravity of the collapsing
baryons causes contraction of the matter in the inner part of the halo
\citep{blumenthal:86,flores:93,mo:98}. This approach has been shown to
reproduce the observed size versus stellar mass relation for
disc-dominated galaxies from $z\sim 0$--2 \citep{somerville:08a}. In
PST14 we also showed that our models reproduce the sizes of \HI\ disks
in nearby galaxies, and sizes of CO disks out to $z\sim 2$.

Star formation occurs in two modes, a normal ``disc'' mode in isolated
discs, and a merger-driven ``starburst'' mode. Star formation in the
disc mode is modelled as described in Section~\ref{sec:sam:sfrecipes}
below. The efficiency and timescale of the merger driven burst mode is
modeled as described in S08 and is a function of the merger mass ratio
and the gas fractions of the progenitors. The treatment of
merger-driven bursts is based on the results of hydrodynamic
simulations of binary galaxy mergers
\citep{robertson:06a,hopkins:09a}.

Some of the energy from supernovae and massive stars is assumed to be
deposited in the ISM, resulting in the driving of a large-scale
outflow of cold gas from the galaxy. The mass outflow rate is
\begin{equation}
\dot{m}_{\rm out} = \epsilon_{\rm SN} \left(\frac{V_0}{V_c} \right)^{\alpha_{\rm rh}} \dot{m}_*
\end{equation}
where $V_c$ is the maximum circular velocity of the galaxy (here
approximated by $V_{\rm max}$ of the dark matter halo), $\dot{m}_*$ is
the star formation rate, $\epsilon_{\rm SN}$ and $\alpha_{\rm SN}$ are
free parameters, and $V_0=200$ km/s is an arbitrary normalization
constant.  Some fraction of this ejected gas escapes from the
potential of the dark matter halo, while some is deposited in the hot
gas reservoir within the halo, where it becomes eligible to cool
again. The fraction of gas that is ejected from the disc but retained
in the halo, versus ejected from the disc and halo, is a function of the
halo circular velocity (see S08 for details), such that low-mass
haloes lose a larger fraction of their gas.

The gas that is ejected from the halo is kept in a larger reservoir,
along with the gas that has been prevented from falling in due to the
photoionizing background. This gas is assumed to accrete onto the halo
on a timescale that is proportional to the halo dynamical time (see
S08 for details).

Each generation of stars produces heavy elements, and chemical
enrichment is modelled in a simplified manner using the instantaneous
recycling approximation. For each parcel of new stars ${\rm d}m_*$, we
also create a mass of metals ${\rm d}M_Z = y \, {\rm d}m_*$, which we
assume to be instantaneously mixed with the cold gas in the disc. The
yield $y$ is assumed to be constant, and is treated as a free
parameter. When gas is removed from the disc by supernova driven winds
as described above, a corresponding proportion of metals is also
removed and deposited either in the hot gas or outside the halo,
following the same proportions as the ejected gas. Ejected metals also
``re-accrete'' into the halo along with the ejected gas, as described
above.

Mergers are assumed to remove angular momentum from the disc stars and
to build up a spheriod. The efficiency of disc destruction and
spheroid growth is a function of progenitor gas fraction and merger
mass ratio, and is parameterized based on hydrodynamic simulations of
disc-disc mergers \citep{hopkins:09a}. These simulations indicate that
more ``major'' (closer to equal mass ratio) and more gas-poor mergers
are more efficient at removing angular momentum, destroying discs, and
building spheroids. Note that the treatment of spheroid formation in
mergers used here has been updated relative to S08 as described in
\citet{hopkins:09b} and \citet{Porter:2014}. We do not include a disk
instability driven mode for spheroid growth in the models presented
here.

In addition, mergers drive gas into galactic nuclei, fueling black
hole growth. Every galaxy is born with a small ``seed'' black hole
(BH; here we adopt $M_{\rm seed} \sim 1.0 \times 10^{4}\,
\msun$). Following a merger, any pre-existing black holes are assumed
to merge immediately, and the resulting hole grows at its Eddington
rate until the energy being deposited into the ISM in the central
region of the galaxy is sufficient to significantly offset and
eventually halt accretion via a pressure-driven outflow. This results
in self-regulated accretion that leaves behind black holes that
naturally obey the observed correlation between BH mass and spheroid
mass or velocity dispersion. Our models produce good agreement with
the observed luminosity function of X-ray/optical/IR detected quasars
and AGN \citep{Hirschmann:2012b}.

A second mode of black hole growth, termed ``radio mode'', is
associated with powerful jets observed at radio frequencies. In
contrast to the merger-triggered, radiatively efficient mode of BH
growth described above (sometimes called ``bright mode'' or ``quasar
mode''), in which the BH accretion is fueled by cold gas in the
nucleus, here, hot halo gas is assumed to be accreted according to the
Bondi-Hoyle model \citep{bondi:52}. This leads to accretion rates that
are typically $\la 10^{-3}$ times the Eddington rate, so that most of
the BH's mass is acquired during episodes of ``bright mode''
accretion. However, the radio jets are assumed to couple very
efficiently with the hot halo gas, and to provide a heating term that
can partially or completely offset cooling during the ``hot flow''
mode (we assume that the jets cannot couple efficiently to the cold,
dense gas in the infall-limited or cold flow regime). 

\subsection{Multi-phase Gas Partitioning}
\label{sec:gaspart}
In this section, we describe in detail the updates to the model
ingredients that are explored in this paper. These include
partitioning of the cold gas in galactic disks into an ionized (\HII),
atomic (\HI), and molecular (\Htwo) component, an option to include
metal enhanced winds, and a set of new \Htwo-based star formation
recipes.

At each timestep, we compute the scale radius of the cold gas disc
using the angular momentum based approach described above, and assume
that the \emph{total} (\HI+\Htwo+\HII) radial cold gas distribution is
described by an exponential with scale radius $r_{\rm gas}$. We do not
attempt to track the scale radius of the stellar disk separately, but
make the simple assumption that $r_* = \chi_{\rm gas} r_{\rm gas}$,
with $\chi_{\rm gas}$ fixed to match observed stellar scale lengths at
$z=0$. \citet{Bigiel:2012} showed that this is a fairly good
representation, on average, of the disks of nearby spirals
\citep[see also][]{Kravtsov:2013}.
We then divide the gas disk into radial annuli and compute the
fraction of molecular gas, $f_{\rm H2}(r) \equiv \Sigma_{\rm
  H_2}(r)/[\Sigma_{\rm H_2}(r)+\Sigma_{\rm HI}(r)]$, in each annulus,
as described below. We use a fifth order Runga-Kutta integration
scheme to compute the integrated mass of \HI\ and \Htwo\ in the disk,
and the integrated SFR, at each timestep.

\subsubsection{Ionized gas associated with galaxies}
\label{sec:ionizedgas}

Most previous semi-analytic models have neglected the ionized gas
associated with galaxies, which may be ionized either by an external
background or by the radiation field from stars within the
galaxy. Here we investigate a simple analytic model motivated by the
work of \citet{Gnedin:2012}. We assume that some fraction of the total
cold gas in the galaxy, $f_{\rm ion, int}$, is ionized by the galaxy's
own stars. In addition, a slab of gas on each side of the disk is
ionized by the external background radiation field. Assuming that all
gas with a surface density below some critical value $\Sigma_{\rm
  HII}$ is ionized, we have
\[ f_{\rm ion} = \frac{\Sigma_{\rm HII}}{\Sigma_0}
\left[1 + \ln \left(\frac{\Sigma_0}{\Sigma_{\rm HII}} \right) + 0.5
  \left(\ln \left(\frac{\Sigma_0}{\Sigma_{\rm HII}}\right) \right)^2
  \right] \] 
where $\Sigma_0 \equiv m_{\rm cold}/(2 \pi r_{\rm gas})^2$ is the
central surface density of the cold gas ($m_{\rm cold}$ is the mass of
all cold gas in the disk and $r_{\rm gas}$ is the scale radius of the
gas disk). We typically assume $f_{\rm ion, int} = 0.2$ (as in the
Milky Way) and $\Sigma_{\rm HII} = 0.4\, \msunpcsq$, as in
\citet{Gnedin:2012}. Applying this model within our SAM gives remarkably
good agreement with the ionized fractions as a function of circular
velocity shown in Fig.~2 of \citet{Gnedin:2012}, obtained from
hydrodynamic simulations with time dependent and spatially variable 3D
radiative transfer of ionizing radiation from local sources and the
cosmic background.

\subsubsection{Molecular gas: pressure based partitioning}
\label{sec:molec:BR}
We consider three approaches for computing the molecular gas fractions
in galaxies. The first is based on the empirical pressure-based recipe
presented by \citet{BR:06}, and will be referred to as the BR
recipe. \citet{BR:06} found a power-law relation between the disc
mid-plane pressure and the ratio between molecular and atomic
hydrogen, i.e.,
\begin{equation}
R_{\mathrm{H}_2} = \left(\frac{\Sigma_{\mathrm{H}_2}}{\Sigma_{\mathrm{HI}}}\right) = \left(\frac{P_m}{P_0}\right)^\alpha
\label{eq:blitz2006}
\end{equation}
where $\Sigma_{\mathrm{H}_2}$ and $\Sigma_{\mathrm{HI}}$ are the
\Htwo\ and \HI\ surface density, $P_0$ and $\alpha_{\rm BR}$ are free
parameters that are obtained from a fit to the observational data, and
$P_m$ is the mid-plane pressure acting on the galactic disc.  We
adopted $\log P_0/k_B = 4.23$ cm$^3$ K and $\alpha_{\rm BR}=0.8$ based
on observations from \citet{Leroy:2008}.

We estimate the hydrostatic pressure as a function of the distance
from the center of the disk $r$ as
\begin{equation}
P(r) = \frac{\pi}{2} G \Sigma_{\rm gas}(r)[\Sigma_{\rm gas}(r) + 
f_{\sigma}(r) \Sigma_*(r)]
\end{equation}
where $G$ is the gravitational constant, $\Sigma_{\rm gas}$ is the
cold gas surface density, $\Sigma_*$ is the stellar surface density,
and $f_{\sigma}$ is the ratio of the vertical velocity dispersions of
the gas and stars: $f_{\sigma}(r) = \frac{\sigma_{\rm
    gas}}{\sigma_{*}}$.  Following \citet{fu:10}, we adopt
$f_{\sigma}(r) = 0.1 \sqrt{\Sigma_{*,0}/\Sigma_*}$, where $\Sigma_{*,
  0} \equiv m_*/(2 \pi r_*^2)$, based on empirical scalings for nearby
disk galaxies.

\subsubsection{Molecular gas: metallicity based partitioning}
\label{sec:molec:met}

Gnedin and Kravtsov \citep{Gnedin:2010,GK:11} performed
high-resolution ``zoom-in'' cosmological simulations with the Adaptive
Refinement Tree (ART) code of \citet{Kravtsov:1997}, including
gravity, hydrodynamics, non-equilibrium chemistry, and simplified
on-the-fly radiative transfer. These simulations are therefore able to
follow the formation of molecular hydrogen through primordial channels
and on dust grains, as well as dissociation of molecular hydrogen and
self- and dust- shielding. 

\citet{GK:11} presented a fitting function based on their simulations,
which effectively parameterizes the fraction of molecular hydrogen as
a function of the dust-to-gas ratio relative to the Milky Way, $D_{\rm
  MW}$, the UV ionizing background relative to the Milky Way $U_{\rm
  MW}$, and the neutral gas surface density $\Sigma_{HI+H_2}$. The
fraction of molecular hydrogen is given by
\[ f_{H_2} = \left[1+\frac{\tilde{\Sigma}_*}{\Sigma_{HI+H_2}}\right]^{-2} \]
where
\begin{eqnarray*}
\tilde{\Sigma}_*  & = &  20\, {\rm M_\odot pc^{-2}} \frac{\Lambda^{4/7}}{D_{\rm MW}} 
\frac{1}{\sqrt{1+U_{\rm MW} D_{\rm MW}^2}} \\
\Lambda & = & \ln(1+g D_{\rm MW}^{3/7}(U_{\rm MW}/15)^{4/7})\\
g & = & \frac{1+\alpha s + s^2}{1+s}\\
s &  = & \frac{0.04}{D_*+D_{\rm MW}}\\
\alpha &  = & 5 \frac{U_{\rm MW}/2}{1+(U_{\rm MW}/2)^2}\\
D_* & = & 1.5 \times 10^{-3} \, \ln(1+(3U_{\rm MW})^{1.7})
\end{eqnarray*}

We take the dust-to-gas ratio to be proportional to the cold gas phase
metallicity in solar units $D_{\rm MW} = Z/Z_{\odot}$. The local UV
background relative to the MW is assumed to scale in proportion with
the global SFR of the galaxy in the previous time step relative to the
MW SFR, $U_{\rm MW}= \frac{SFR}{SFR_{\rm MW}}$, where we choose
$SFR_{\rm MW} = 1.0\,\rm{M}_\odot\,\rm{yr}^{-1}$
\citep{Murray:2010,Robitaille:2010}. We refer to this as our
`fiducial' GK model.  We also investigate the results of keeping
$U_{\rm MW}$ fixed to the Milky Way value, which we refer to as the
GKFUV model (fixed UV field).

An alternate approach based on similar physical processes was
presented in a series of papers by Krumholz and collaborators
\citep{Krumholz:2008,Krumholz:2009,KMT:09}. \citet{KMT:09} developed
an analytic model for the molecular fraction in galaxies, based on the
ansatz that the interplay between the interstellar radiation field and
molecular self-shielding determines the molecular fraction. They
presented a fitting function:
\[ f_{H_2} = 1 - \left[1+\left(\frac{3}{4}\frac{s}{1+\delta}\right)^{-5}\right]^{-1/5} \]
where $s= \ln(1+0.6\chi)/(0.04 \Sigma_{\rm comp, 0}Z')$, $\chi =
0.77(1+ 3.1Z'^{0.365})$, $\delta = 0.0712(0.1s^{-1} + 0.675)^{-2.8}$,
$\Sigma_{\rm comp,0}=\Sigma_{\rm comp}/(1 M_{\odot} {\rm pc}^{-2})$,
and $Z'\equiv Z/Z_{\odot}$. The quantity denoted $\Sigma_{\rm comp}$
is the surface density within a $\sim 100$ pc sized atomic-molecular
cloud complex. \citet{KMT:09} suggest using a ``clumping factor'' to
apply the model to simulations with spatial resolution coarser than
100 pc; i.e. $\Sigma_{\rm comp} \rightarrow c \Sigma_{\rm HI+H_2}$,
where $c \geq 1$, and $\Sigma_{\rm HI+H_2}$ is the neutral gas surface
density on some larger scale. The appropriate value of $c$ depends on
this spatial scale, where $c\rightarrow 1$ as the scale approaches 100
pc. We refer to this as the KMT gas partitioning recipe.

Both \citet{GK:11} and \citet{KMT:09} note that the fitting functions,
as well as perhaps (in the case of KMT) the underlying assumptions of
the model, begin to break down at metallicities lower than about
1/50th of the Solar value.

The KMT and GK fitting functions above characterize the formation of
\Htwo\ on dust grains, which is the dominant mechanism once the gas is
enriched to more than a few hundredths of Solar metallicity. Other
channels for the formation of \Htwo\ in primordial gas must be
responsible for producing the molecular hydrogen out of which the
first stars were formed. Hydrodynamic simulations containing detailed
chemical networks and analytic calculations have shown that \Htwo\ can
form in metal-free gas in dark matter halos above a critical mass
$M_{\rm crit} \sim 10^5\, \rm{M}_\odot$
\citep[e.g.,][]{Nakamura:2001,Glover:2013}.  This gas can then form
``Pop III'' stars which can enrich the surrounding ISM to $\rm{Z}_{\rm
  III} \sim 10^{-3}\,\rm{Z}_\odot$
\citep{Schneider:2002,Greif:2010,Wise:2012}. These processes take
place in halos much smaller than our resolution limit. We represent
them by setting a ``floor'' to the molecular hydrogen fraction in our
halos, $f_{\rm H2,floor}$. In addition, we ``pre-enrich'' the initial
hot gas in halos, and the gas accreted onto halos due to cosmological
infall, to a metallicity of $\rm{Z}_{\rm pre-enrich}$. We adopt
typical values of $f_{\rm H2,floor} = 10^{-4}$ and $\rm{Z}_{\rm
  pre-enrich}=10^{-3}\, \rm{Z}_\odot$ \citep{Haiman:1996,Bromm:2004}.  We
explore the sensitivity of our results to these parameters in
Section~\ref{sec:trace}. Note that observations of resolved stars in
the halo of our Galaxy and local dwarfs have revealed stars with
metallicities below $\rm{Z}\sim10^{-3}\, \rm{Z}_\odot$
\citep{Tolstoy:2009,Schorck:2009,Kirby:2011}, precluding much higher
values for $\rm{Z}_{\rm pre-enrich}$.

\subsection{Star Formation Recipes}
\label{sec:sam:sfrecipes}

\begin{figure} 
\begin{center}
\includegraphics[width=0.45\textwidth]{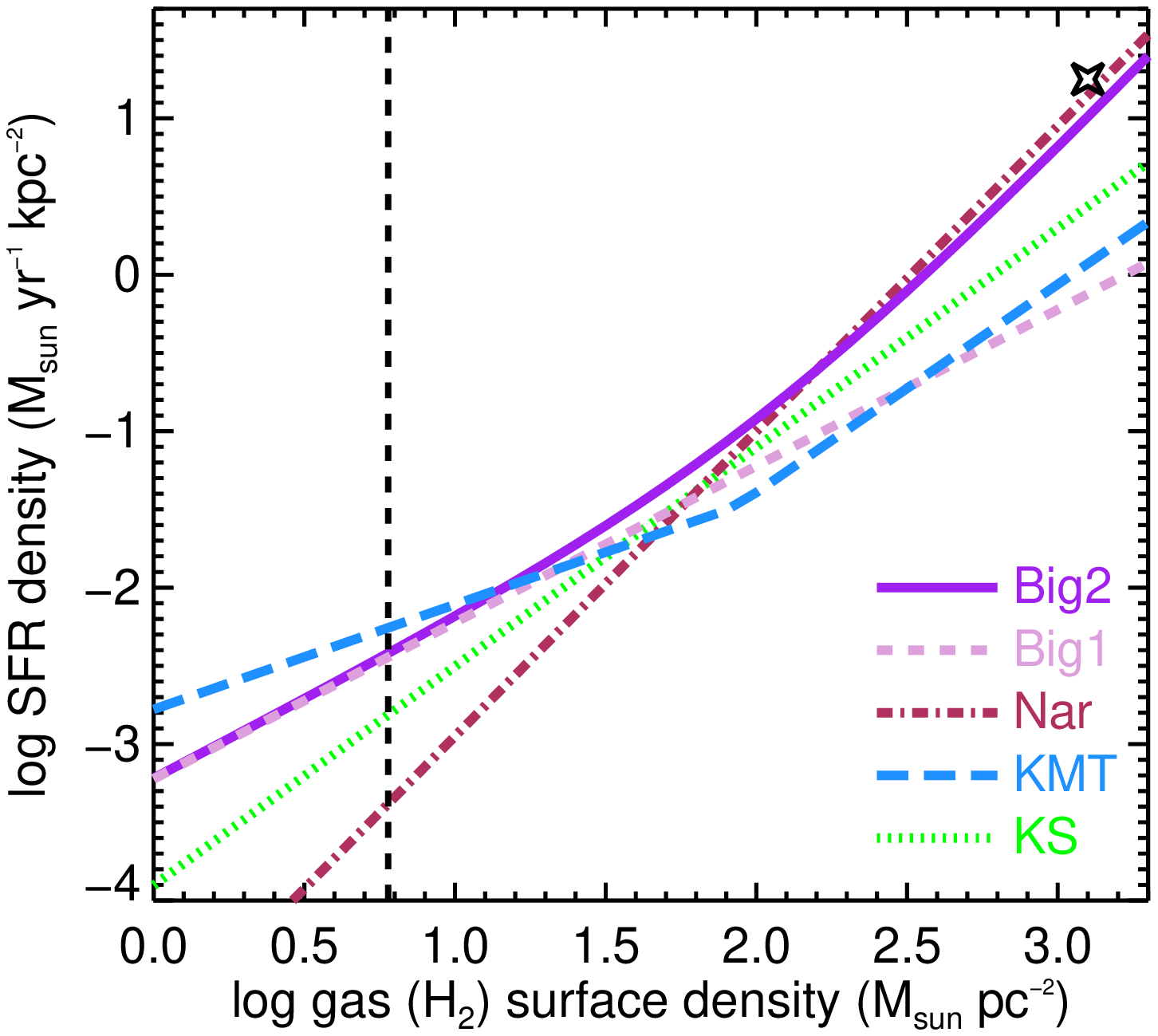}
\end{center}
\caption{Empirically based star formation recipes used as input in our
  models, and from the literature. The solid purple line shows the
  two-slope \Htwo-based recipe (Big2); the dashed lavender line shows
  the single slope \Htwo-based recipe (Big1); the dot-dashed dark red
  line shows the recipe based on the analysis of
  \protect\citet{Narayanan:2012}; the dotted green line shows the (\HI
  + \Htwo) based ``classic'' KS recipe. The vertical dashed line
  shows the critical total gas surface density used in our models that
  implement the classic KS recipe ($\Sigma_{\rm crit}$). The star
  symbol shows the relation derived by \protect\citet{Sharon:2013} for
  an extreme starburst galaxy. Note that the
  \protect\citet{Narayanan:2012} results are shown for reference only;
  we do not show the results of incorporating this recipe in our
  models here.
\label{fig:H2kennobs}}
\end{figure}

\subsubsection{The Kennicutt-Schmidt (KS) Recipe}

The KS recipe \citep{kennicutt:98} assumes that the surface density of
star formation in a galaxy is a function of the \emph{total} surface
density of the cold neutral gas (atomic and molecular), above some
threshold surface density $\Sigma_{\rm crit}$.

The star formation rate density (per unit area) for $\Sigma_{\rm
  gas}>\Sigma_{\rm crit}$ is given by:
\begin{equation}
\Sigma_{\rm SFR} = A_{\rm SF} \, {\Sigma_{\rm gas}}^{N_{\rm SF}} ,
\label{eqn:ks}
\end{equation}
where $\Sigma_{\rm SFR}=0$ for $\Sigma_{\rm gas}<\Sigma_{\rm
  crit}$. This recipe is the same one used in most of our previously
published SAMs (S08, S12), and is similar to recipes commonly adopted
in many other SAMs. The values of the free parameters are given in
Table~\ref{tab:samparam}.

\subsubsection{Molecular Hydrogen-based Recipes}

In the same spirit as the KS recipe, we use empirical
relationships from observations to motivate our \Htwo-based
recipes. \citet{Bigiel:2008} found, based on observations of spiral
galaxies from the THINGS survey,
that the star-formation
rate surface density can be directly related to the surface density of
molecular gas, i.e.
\begin{equation}
\Sigma_{\rm SFR} = \left(\frac{A_{\rm SF}}{10\, M_\odot {\rm
  pc}^{-2} }\right) \, {\Sigma_{\rm H_2}}^{N_{\rm SF}}
\label{eqn:big1}
\end{equation}
with $N_{\rm SF} \simeq 1$ \citep[see
  also][]{Bigiel:2011,Leroy:2013}. Observations of higher density
environments suggest that above some critical \Htwo\ surface density,
the slope of the relation described in Eqn.~\ref{eqn:big1} steepens
\citep{Narayanan:2012}. We therefore also consider a two-part scaling
law given by:
\begin{equation}
\Sigma_{\rm SFR} = A_{\rm SF} \, \left(\frac{\Sigma_{\rm H_2}}{10\, M_\odot {\rm
  pc}^{-2}}\right) \left(1+ \frac{\Sigma_{H_2}}{\Sigma_{\rm H_2,
    crit}}\right)^{N_{\rm SF}}
\label{eqn:big2}
\end{equation}
The values of the parameters $A_{\rm SF}$, $N_{\rm SF}$, and
$\Sigma_{\rm H_2, crit}$ are given in Table~\ref{tab:samparam}.

A star formation relation that changes slope above a critical density
is also expected based on theoretical grounds. \citet{KMT:09} adopt
the star formation relation:
\begin{equation}
\Sigma_{\rm SFR} = A_{\rm SF} \Sigma_{H_2}
\left(\frac{\Sigma_{\rm gas}}{\Sigma_{\rm crit}}\right)^{N_{\rm SF}}
\label{eqn:kmt}
\end{equation}
where $N_{\rm SF} = -0.33$ for $\Sigma_{\rm gas}/\Sigma_{\rm crit}<1$
and $N_{\rm SF} = 0.33$ for $\Sigma_{\rm gas}/\Sigma_{\rm
  crit}>1$. KMT adopt $A_{\rm SF}=1/2.6$ Gyr$^{-1}$ and $\Sigma_{\rm
  crit}=85\, \msun {\rm pc}^{-2}$. We adopt the same parameter values
in our ``KMT'' model.

A comparison of various star formation relations in the literature,
and used in this work, is shown in Fig.~\ref{fig:H2kennobs}.

\subsection{Metal Enhanced Winds}
\label{sec:metalwind}

Most of our previous models have assumed that metals are ejected from
galaxies with the same efficiency as the gas, i.e. with the same mass
loading factor $\eta \equiv \dot{m}_{\rm out}/\dot{m}_*$. However,
since metals are produced by the same massive stars and supernovae
that are believed to drive galactic outflows, it is possible that
metals are preferentially ejected (i.e., have a higher effective mass
loading factor than the gas averaged over the whole disk).
Because two of our recipes for gas partitioning depend on the gas
metallicity, the dispersal of metals in our models has a potentially
important impact on our results. We therefore include an optional
treatment of metal-enhanced winds in our models.

We base our parameterization of metal-enhanced winds on the approach
used in \citet{Krumholz:2012}, in part because we want to be able to
compare our results with theirs. The fraction of metals that is
ejected is parameterized by:
\[ \zeta = \zeta_{\rm lo} \exp(-M_h/M_{\rm ret}) \]
where both $\zeta_{\rm lo}$ and $M_{\rm ret}$ are free parameters, and
$M_h$ is the virial mass of the halo. The modified equation for the evolution
of the mass in metals in the cold gas phase is then:
\[ \dot{M}_Z = y(1-R)(1-\zeta)\dot{m}_* + Z_{\rm hot} \dot{m}_{\rm inf} - 
Z_{\rm cold} \dot{m}_{\rm out} \] where $R$ is the recycled fraction,
$y$ is the chemical yield, $Z_{\rm hot}$ is the metallicity of the hot
gas, $Z_{\rm cold}$ is the metallicity of the cold gas, and
$\dot{m}_*$, $\dot{m}_{\rm inf}$, and $\dot{m}_{\rm out}$ are the star
formation rate, inflow rate of gas from the hot halo into the disk,
and the outflow rate of gas from the disk, respectively.

\subsection{Calibrating the Free Parameters}
\label{sec:results:norm}
\begin{table*}
\centering
\caption{Summary of Model Parameters}
\begin{tabular}{llcc}
\hline \hline
parameter  & description & section defined & value\\
\hline \hline
$\epsilon_{\rm SN}$ & supernova feedback efficiency & \ref{sec:oldsam} & 1.5\\
$\alpha_{\rm SN}$ & supernova feedback slope & \ref{sec:oldsam} & -2.2\\
$y$ & chemical yield & \ref{sec:oldsam} & 1.6 Z$_\odot$ \\
$\kappa_{\rm AGN}$ & radio mode AGN feedback & S08 \S2.11, Eqn. 20 & $3.8 \times 10^{-3}$ \\
$\chi_{\rm gas} $ & ratio of stellar to gas scale length & \ref{sec:gaspart} & 0.59 \\
\hline
$\Sigma_{\rm HII}$ & critical density for ionized gas & \ref{sec:ionizedgas} & 0.4 \msunpcsq \\
$f_{\rm ion, int}$ & internal fraction of ionized gas & \ref{sec:ionizedgas} & 0.2 \\
\hline
$P_0$ & pressure scaling in BR recipe & \ref{sec:molec:BR} & 4.23 $k_B\, \rm{cm}^3$ K\\
$\alpha_{\rm BR}$ & slope in BR recipe & \ref{sec:molec:BR} & 0.8 \\
\hline
$c$ & clumping factor in KMT recipe & \ref{sec:molec:met} & 5\\
\hline
$f_{\rm H2, floor}$ & primordial \Htwo\ fraction & \ref{sec:molec:met} & $10^{-4}$ \\
$Z_{\rm pre-enrich}$ & metallicity due to Pop III stars & \ref{sec:molec:met} & $10^{-3}$ \\
\hline
$\zeta_{\rm lo}$ & metal enhanced winds normalization & \ref{sec:metalwind} & 0.1\\
$M_{\rm ret}$ & metal enhanced winds mass scale & \ref{sec:metalwind} & 0.9\\
\hline
$A_{\rm SF}$ (KS) & star formation relation normalization & \ref{sec:sam:sfrecipes}, Eqn.~\ref{eqn:ks} &  $1.1 \times 10^{-4}$ \\
$N_{\rm SF}$ (KS) & star formation relation slope & \ref{sec:sam:sfrecipes}, Eqn.~\ref{eqn:ks} &  1.4 \\
$\Sigma_{\rm crit}$ (KS) & critical density for SF & \ref{sec:sam:sfrecipes}, Eqn.~\ref{eqn:ks} &  6 \msunpcsq \\
$A_{\rm SF}$ (Big1, Big2) & star formation relation normalization & \ref{sec:sam:sfrecipes}, Eqn.~\ref{eqn:big1}, \ref{eqn:big2} &  $4.0 \times 10^{-3}$ \\
$N_{\rm SF}$ (Big1, Big2) & star formation relation slope & \ref{sec:sam:sfrecipes}, Eqn.~\ref{eqn:big1}, \ref{eqn:big2} &  1.0 \\
$\Sigma_{\rm H2, crit}$ & critical \Htwo\ density & \ref{sec:sam:sfrecipes}, Eqn.~\ref{eqn:big2} & 70 \msunpcsq \\
\hline \hline
\end{tabular}
\label{tab:samparam}
\end{table*}

\begin{table*}
\centering
\caption{Summary of Model Variants}
\begin{tabular}{lccc}
\hline \hline
model  & \HI/\Htwo\ partitioning & SF law & metal-enhanced winds\\
\hline \hline
KS ``fiducial'' & none & KS  & N \\
BR ``fiducial'' & BR & Big2  & N \\
GK ``fiducial'' & GK, $U_{\rm MW} \propto {\rm SFR}$ & Big2 & N\\
GK+Big1 & GK & Big1 & N \\
GKFUV & GK, $U_{\rm MW}=1$ & Big2  & N \\
BR+Big1 & BR & Big1  & N \\
KMT+Big1 & KMT & Big1  & N\\
KMT & KMT & KMT  & N\\
KMT+MEW & KMT & KMT  & Y\\
\hline \hline
\end{tabular}
\label{tab:models}
\end{table*}

As in any cosmological simulation, we must parameterize the sub-grid
physics in our models. In keeping with common practice, we choose the
values of the free parameters by tuning to a subset of observations in
the local universe. In this subsection we summarize the values of the
free parameters used here (see Table~\ref{tab:samparam}) and the
observations we used to constrain them. The parameters are the same as
those used in the models presented in \citet{Popping:2014}.

We assume values for the cosmological parameters consistent with the
five year WMAP results (WMAP5): $\Omega_m$ = 0.28, $\Omega_{\Lambda}$
= 0.72, $H_0$ = 70.0, $\sigma_8$ = 0.81, and $n_s=0.96$
\citep{komatsu:09}.  We note that these values are generally
consistent with those obtained from the analysis of the seven-year
WMAP data release \citep{komatsu:10}. The adopted baryon fraction is
0.1658.  We assume a recycled fraction of $R=0.43$, as appropriate for
a Chabrier stellar Initial Mass Function \citep{chabrier:03}.

As discussed in S08 \citep[see also][]{White:2014}, in our models the
supernova feedback parameters mainly control the low-mass end of the
stellar mass function ($m_* \lesssim M_{\rm char}$, where $M_{\rm
  char}$ is the characteristic mass of the ``knee'' in the Schechter
function describing the stellar mass function), or equivalently, the
fraction of baryons that is turned into stars in halos with $M_h
\lesssim 10^{12} \msun$. On the other side, the efficiency of the
``radio mode'' AGN feedback (one can think of this schematically as
the efficiency with which radio jets couple to and heat the hot
intragroup and intracluster medium) controls the number density of
massive galaxies $m_* \gtrsim M_{\rm char}$, or the fraction of
baryons that are able to turn into stars in massive halos ($M_h
\gtrsim 10^{12} \msun$). We tune the parameters controlling supernova
feedback and AGN feedback to reproduce the observed stellar mass
function at $z=0$ in our traditional KS model (see S08 for
details). These parameters are then kept fixed as we explore the
effects of varying the modeling of gas partitioning and star
formation.

The parameters of the star formation recipe mainly control the
fraction of cold gas in galaxies, and do not strongly affect the $z=0$
stellar mass function \citep{White:2014}. We require the parameters
characterizing our star formation recipes to lie within the
observational uncertainties from recent empirical constraints, and
tune them within these limits to match the \emph{total} gas fractions
as a function of galaxy stellar mass at $z=0$ (see PST14).

The chemical yield $y$ could in principle be obtained from stellar
evolution models, but these model yields are uncertain by a factor of
$\sim 2$, and the single-element instantaneous recycling approach to
chemical evolution that we are using here is somewhat crude, so we
instead treat the chemical yield as a free parameter (though we
restrict it to be in the expected range). We tune our yield to match
the normalization of the observed \emph{stellar} metallicity vs. mass
relation of \citet{gallazzi:05}.

We take the parameter values for the BR gas partitioning recipe from
the observational results of \citet{Leroy:2008}. We implement the GK
recipe as it is given in \citet{GK:11}, with no tunable
parameters. The KMT recipe has one free parameter, the clumping factor
of the gas. We adopt $c=5$, following \citet{Krumholz:2012}.

Our new models predict the fraction of cold gas in different phases:
ionized, atomic, and molecular. We showed the predictions of our two
fiducial models (GK and BR) for the fraction of \HI\ and \Htwo\ as a
function of galaxy internal stellar density and stellar mass in
PST14. It is encouraging that our new models reproduce these observed
scalings for nearby galaxies quite well without any additional
tuning. We emphasize that the only new free parameters in the gas
partioning recipe have been taken directly from observations (in the
case of the BR recipe) or from numerical simulations (in the case of
KMT and GK).

\section{Results}
\label{sec:results}

\subsection{Effects of Varying Model Ingredients, Parameter Values, and Resolution}
\label{sec:trace}

In this subsection we explore the sensitivity of our model results to
our new model ingredients and parameter values related to gas
partitioning and star formation, as well as to our numerical
resolution. To illustrate these effects, we show the properties of the
largest progenitor galaxy as a function of time (or redshift) in a set
of halos with masses at $z=0$ ranging from $\log M_h/M_\odot =
10.0$--11.5 (except in one case, where we show a more massive halo
with $\log M_h/M_\odot = 12.5$). The variations that we explore here
mainly affect lower mass halos, and produce no significant differences
for galaxies in halos more massive than those shown. We use the same
merger trees for each model, and fix all model properties that are
chosen from random distributions to their average values.  Each panel
shows the average over $60$ different realizations of halos with the
specified final mass. For each experiment we show the stellar mass,
total neutral cold gas mass (\HI+\Htwo), SFR, \Htwo\ fraction, and gas
phase metallicity. To facilitate comparison, the stellar mass, gas
mass, and SFR are normalized by dividing by the root halo mass at
$z=0$.

\begin{figure*} 
\begin{center}
\includegraphics[width=0.75\hsize]{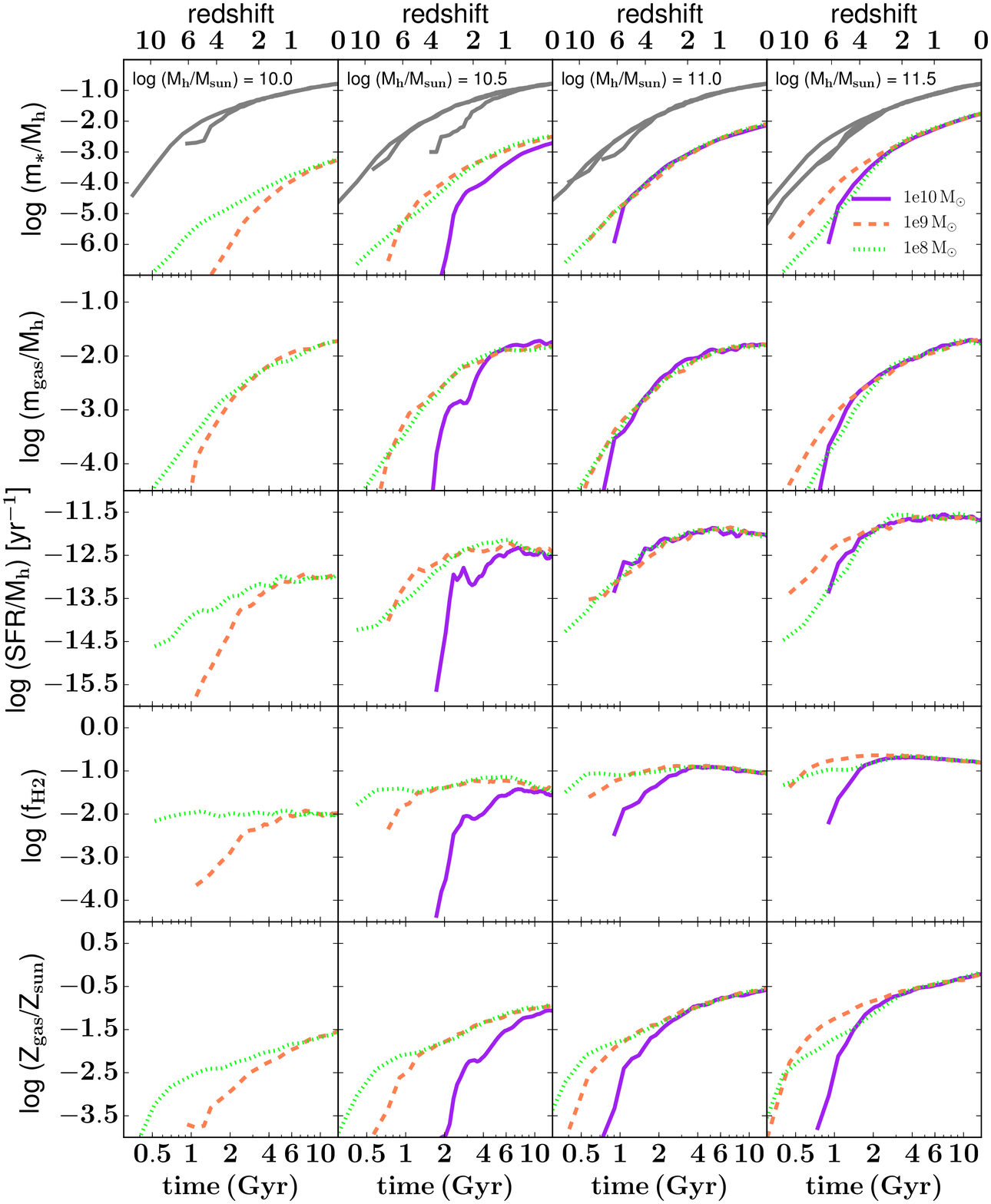}
\end{center}
\caption{From top to bottom, colored lines show the stellar mass, cold
  neutral gas mass (\HI+\Htwo), and SFR normalized by the mass of the
  root halo at $z=0$ for the largest progenitor galaxy as a function
  of cosmic time (redshift). The \Htwo\ fraction ($f_{\rm H2} \equiv
  m_{\rm H2}/(m_{\rm H2}+m_{\rm HI}$) and gas phase metallicity (in
  solar units) are also shown. Gray lines show the maximum baryon
  fraction in the halo, $f_b\, M_h(t)$, where $M_h(t)$ is the mass of
  the largest progenitor halo at time $t$ and $f_b$ is the universal
  baryon fraction (different gray lines correspond to different
  resolutions; lower resolution runs cannot resolve the halo mass
  accretion history as far back in time). In this experiment, we test
  the dependence of our results on the mass resolution of our merger
  trees, varying the mass resolution by two orders of
  magnitude. Results are shown for a mass resolution of
  $10^{10},\ \msun$ (solid purple), $10^{9}\, \msun$ (dashed orange),
  and $10^{8} \, \msun$ (dotted green). The mass accretion histories
  are well-resolved when the mass resolution is at least 1/100 the
  mass of the root halo. 
\label{fig:trace_resolution}}
\end{figure*}

As a first basic check, we test the impact of varying the mass
resolution of our merger trees (Fig.~\ref{fig:trace_resolution}). Note
that what we mean by the `mass resolution' here is the mass of the
smallest halos that are tracked in the merger tree. This is not
equivalent to the particle mass in an $N$-body simulation, but rather
to the smallest halo mass that can be robustly identified. We see from
this test that in order to robustly reconstruct the whole halo mass
accretion history back to $z\sim 10$, we require a minimum halo mass
of $\sim 1/100$ of the root mass at the output redshift. Accordingly,
we impose this condition on all halos in the runs used in this
work. It is reassuring to see that, once the halo mass accretion
history is well resolved, our SAM predictions converge extremely well
(note that we do not retune the free parameters when we change the
mass resolution).

\begin{figure*} 
\begin{center}
\includegraphics[width=0.75\hsize]{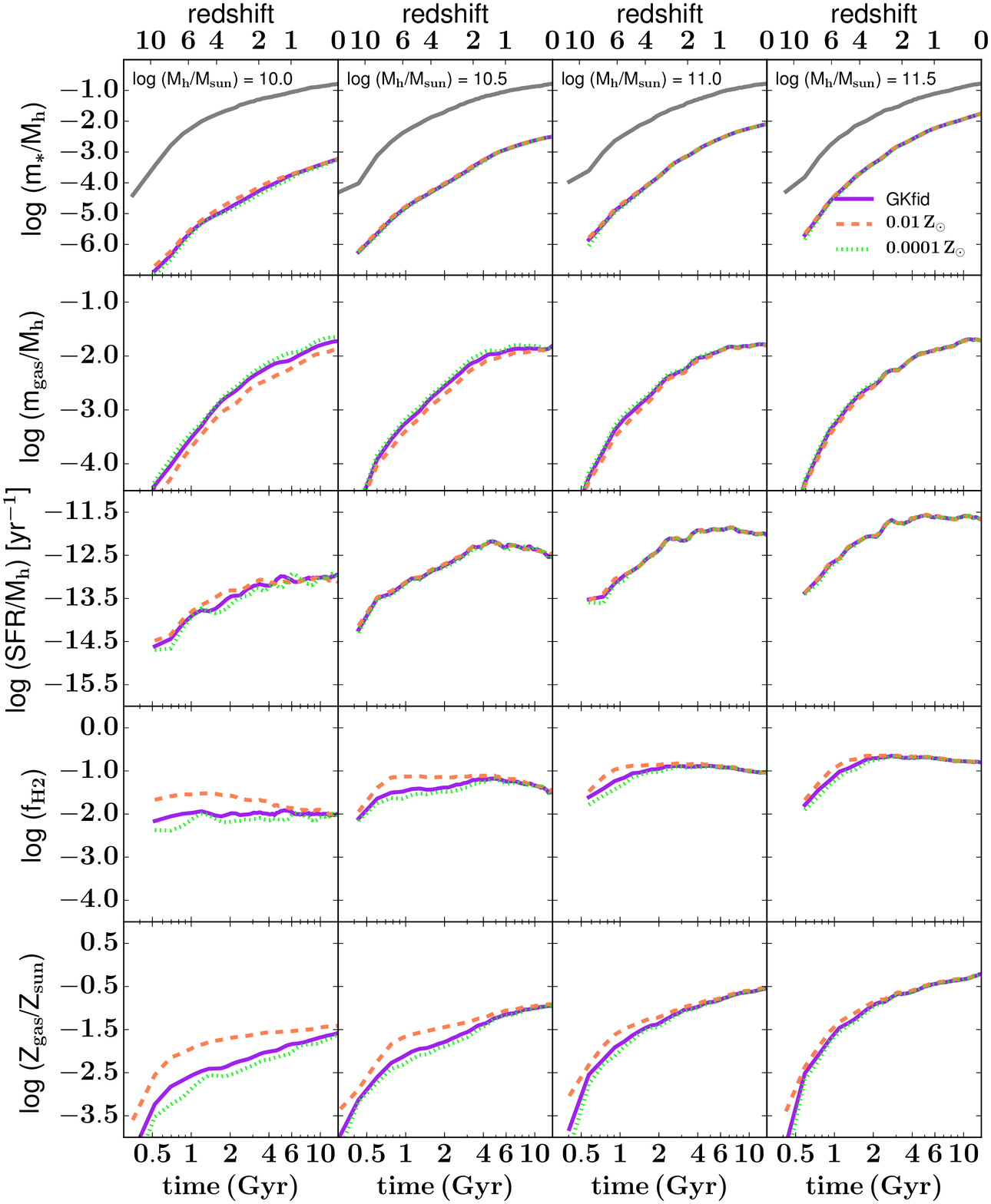}
\end{center}
\caption{Same as Fig.~\ref{fig:trace_resolution}, except here we
  compare the results of our fiducial GK model with different values
  for the ``pre-enriched'' gas metallicity ($Z_{\rm pre-enrich}$), as
  shown in the key. Our results are quite insensitive to the value of
  this parameter within a reasonable range. 
\label{fig:trace_Zpre}}
\end{figure*}

In Fig.~\ref{fig:trace_Zpre} we test for possible sensitivity to the
values of two parameters that we introduce to simulate the formation
of stars in primordial gas, the metallicity of the ``pre-enriched''
gas $Z_{\rm pre-enrich}$ (most relevant for the metallicity-based
recipes), and the primordial molecular hydrogen fraction $f_{\rm H2,
  floor}$ (most relevant for the BR recipe). Leaving all other
settings of the fiducial GK model fixed, we vary $Z_{\rm pre-enrich}$
from its fiducial value of $10^{-3}$ by one order of magnitude
downwards, to $10^{-4}$, and upwards to $0.01$. Overall, the impact of
even such extreme variations is fairly minor. The most noticable
impact is on the stellar and gas phase metallicities. The metallicity
builds up earlier in models with higher values of $Z_{\rm
  pre-enrich}$, as expected. As a result, the \Htwo\ fraction is
higher at earlier times in the model with higher $Z_{\rm pre-enrich}$,
leading to higher star formation efficiency (SFE) and slightly lower
gas fractions. Similarly, we varied the value of $f_{\rm H2, floor}$
from its fiducial value of $10^{-4}$ up and down by an order of
magnitude. This has no discernable effect on our results, and we
therefore omit the corresponding figure.

\begin{figure*} 
\begin{center}
\includegraphics[width=0.75\hsize]{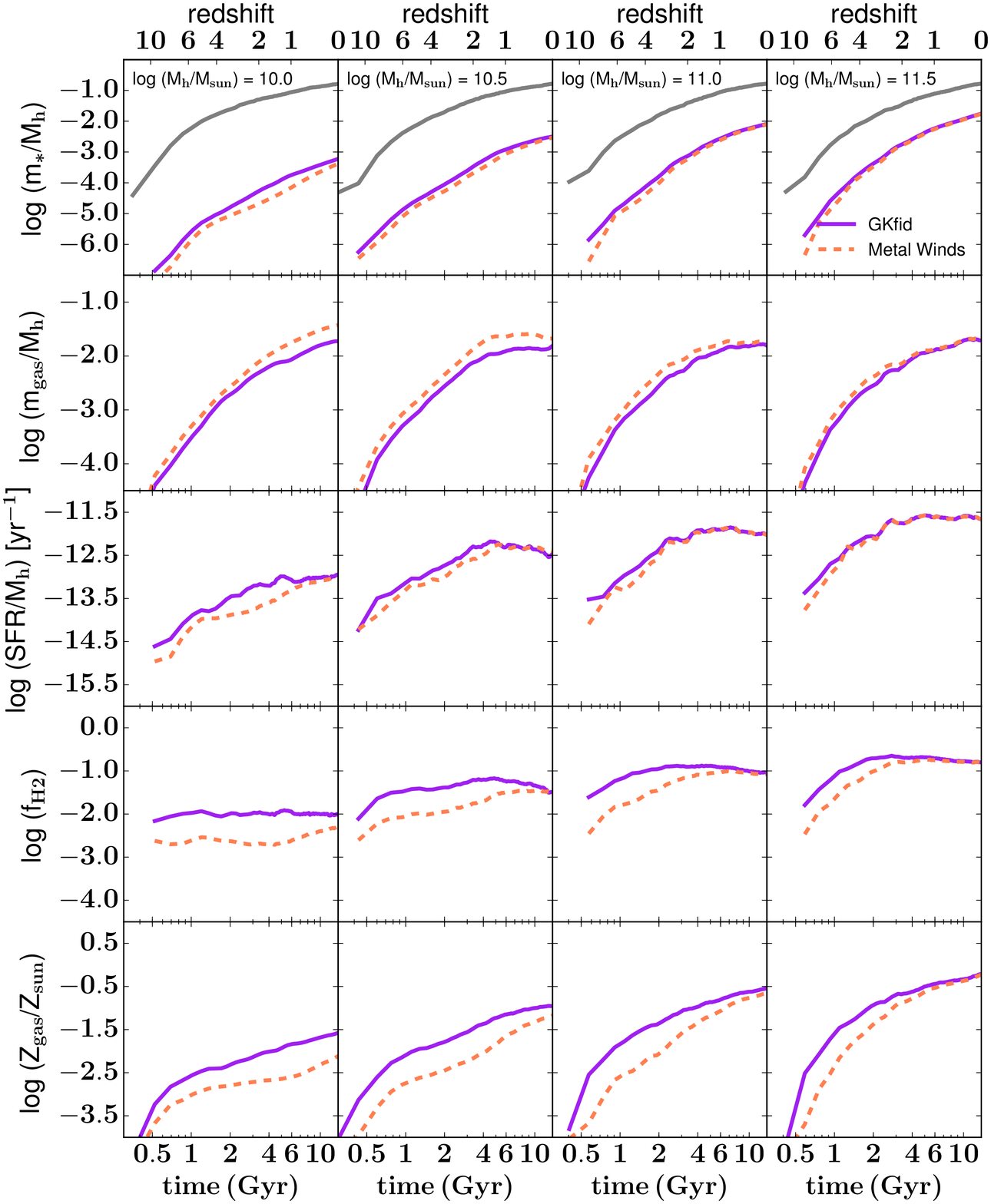}
\end{center}
\caption{Same as Fig.~\ref{fig:trace_resolution}, except here we
  compare the results of our fiducial GK model with and without metal
  enhanced winds. Metal enhanced winds can delay enrichment and, to a
  lesser extent, \Htwo\ and star formation in low mass halos.
\label{fig:trace_metalwind}}
\end{figure*}

In a related experiment, we run our (otherwise) fiducial GK model with
metal-enriched winds, described in Section~\ref{sec:metalwind}. Here,
the metallicity of stellar driven outflows can be metal-enhanced
relative to the ISM by a factor that depends on the halo mass. As seen
in Fig.~\ref{fig:trace_metalwind}, we find that metal enhanced winds
can significantly delay the formation of \Htwo\ and stars in very
low-mass halos ($\log M_h/M_\odot = 10.0$), and cause the build-up of
slightly more cold gas in low-mass halos ($\log M_h/M_\odot \lesssim
10.5$); note however that these halos host galaxies that are well
below the detection limits of most surveys except in the very nearby
Universe ($m_{\rm *} \simeq 10^7$--$10^8 \msun$). Metal-enriched winds
can also delay the build-up of metal-enriched gas even
in more massive halos ($\log M_h/M_\odot \lesssim 11.5$). 

\begin{figure*} 
\begin{center}
\includegraphics[width=0.75\hsize]{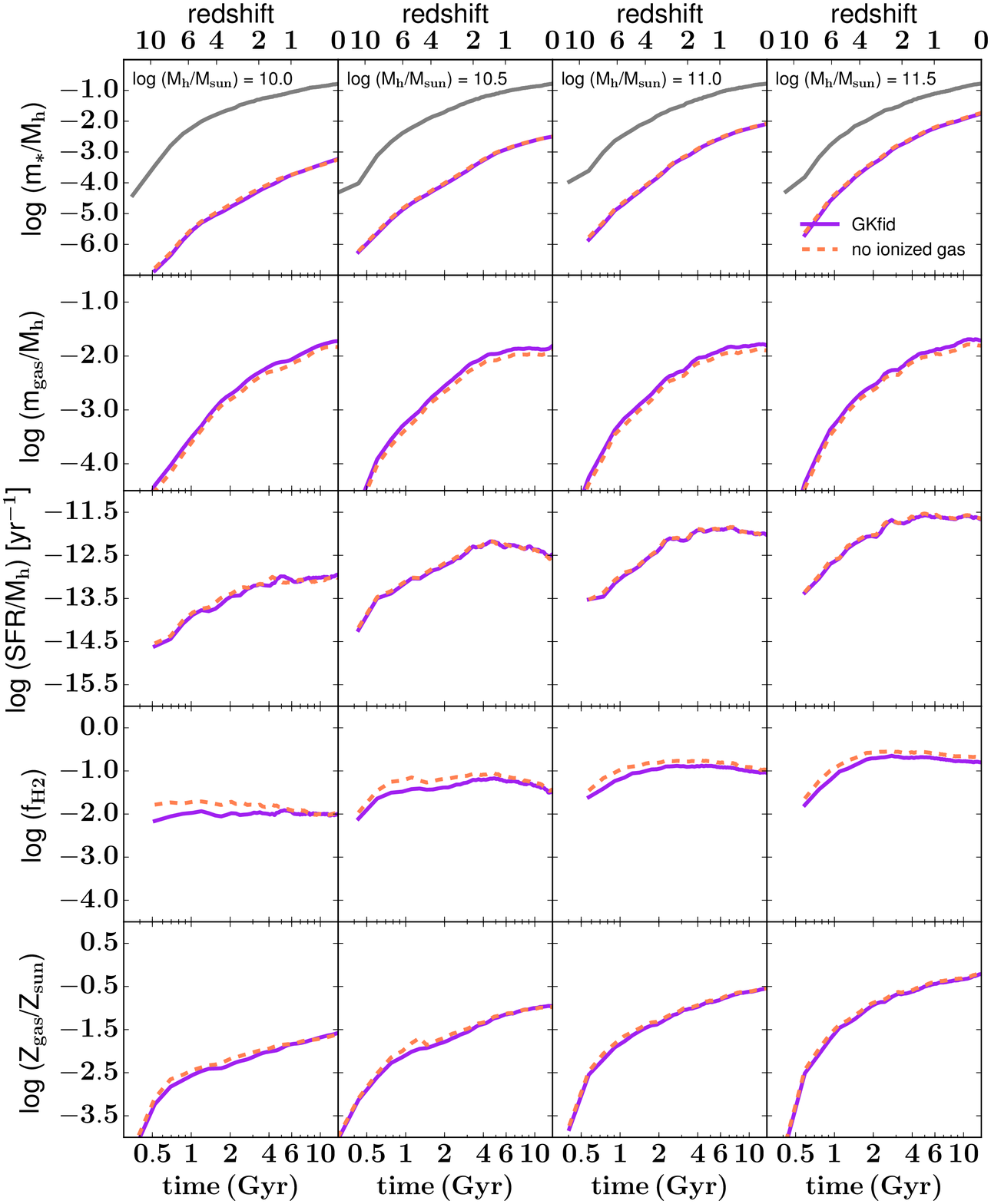}
\end{center}
\caption{Same as Fig.~\ref{fig:trace_resolution}, except here we
  compare the GK model with and without the partioning of ionized gas
  (\HII) into a separate reservoir. Our results are nearly unchanged
  whether or not we include the ionized gas component.
\label{fig:trace_iongas}}
\end{figure*}

A new ingredient we have introduced into our models is the tracking of
gas that is photoionized either by an external radiation field or by
internal sources. This gas is not eligible to form \Htwo\ or stars. In
Fig.~\ref{fig:trace_iongas} we show the galaxy properties in the
fiducial GK model with and without tracking of \HII. Although our
model predicts that galaxies contain a substantial amount of
\HII\ (see Fig.~2 in \citealt{Popping:2014}), partitioning this gas
into a separate reservoir has a very weak effect on our results. The
only noticable effect is slightly lower \Htwo fractions at high
redshift, particularly in the lowest mass halos.

\begin{figure*} 
\begin{center}
\includegraphics[width=0.75\hsize]{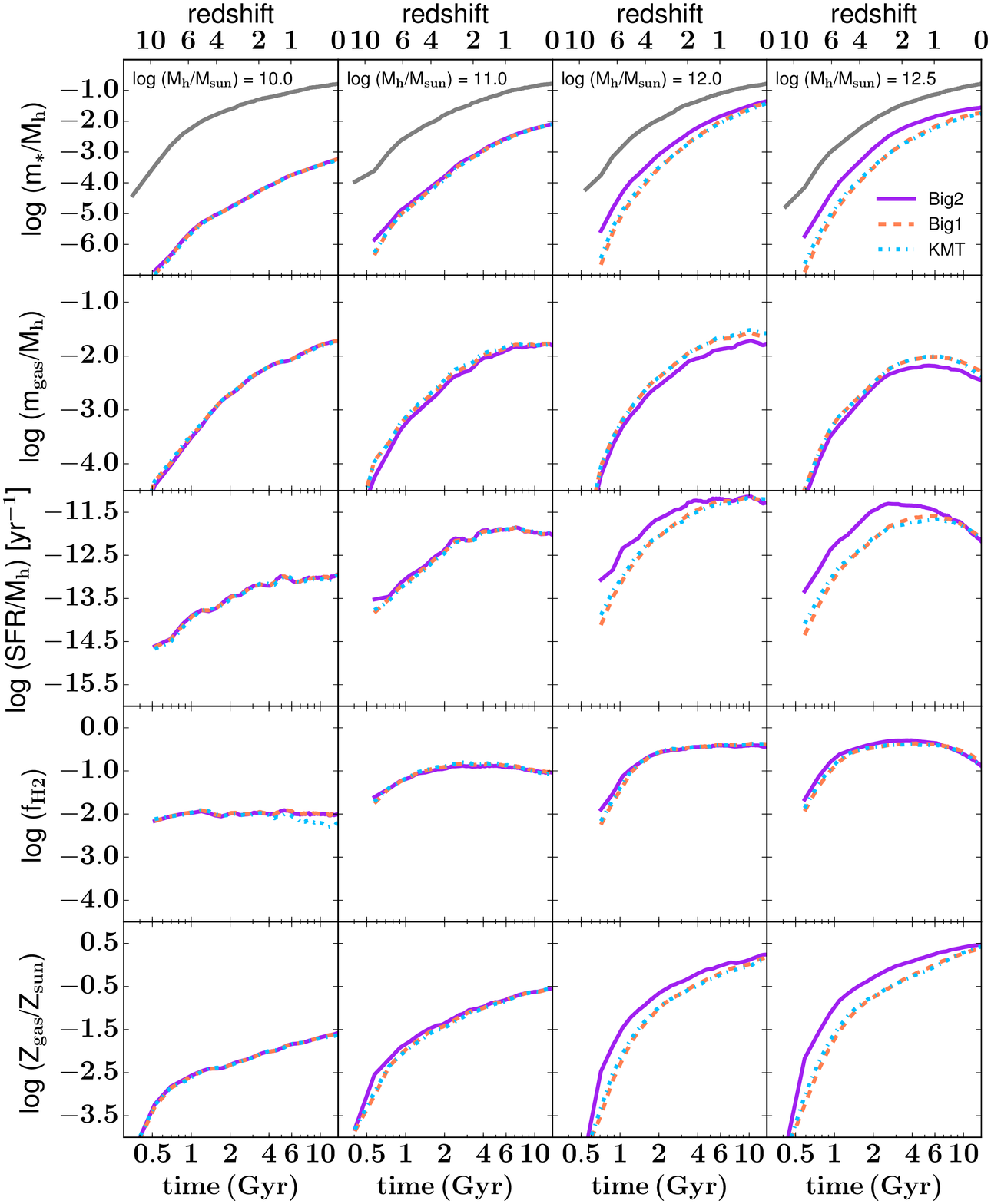}
\end{center}
\caption{Same as Fig.~\ref{fig:trace_resolution}, except here we
  compare different recipes for converting molecular gas into stars
  within our fiducial GK models: Big2, Big1, and KMT. Note the
  different range of halo masses from the other plots. Here, the
  strongest effect seen is the more efficient production of stars and
  earlier enrichment in massive halos in the Big2 model. This is owing
  to the non-linear dependence of the star formation efficiency on
  \Htwo\ density at high densities in this model.
\label{fig:trace_SF}}
\end{figure*}

In the next experiment, shown in Fig.~\ref{fig:trace_SF}, we
investigate several different recipes for converting cold molecular
gas into stars within the fiducial GK model. We consider two variants
of the empirical recipe based on the observations by
\citet{Bigiel:2008}. In addition, we consider the recipe proposed by
KMT based on theoretical arguments (see
Section~\ref{sec:sam:sfrecipes} for details). Big1 refers to
Eqn.~\ref{eqn:big1} and Big2 to Eqn.~\ref{eqn:big2}. In contrast to
most of our other experiments, these variations have almost no
discernable effect on the low mass halos. However, the Big2 recipe
leads to significantly earlier build-up of stellar mass, more
efficient star formation at high redshift, and earlier metal
enrichment in \emph{massive} halos ($\log M_h/M_\odot \gtrsim
12.0$). This is owing to the non-linear dependence of the star
formation efficiency on \Htwo\ density at high densities in this
model, and the positive correlation between halo mass and galaxy
surface density in our models.

\begin{figure*} 
\begin{center}
\includegraphics[width=0.75\hsize]{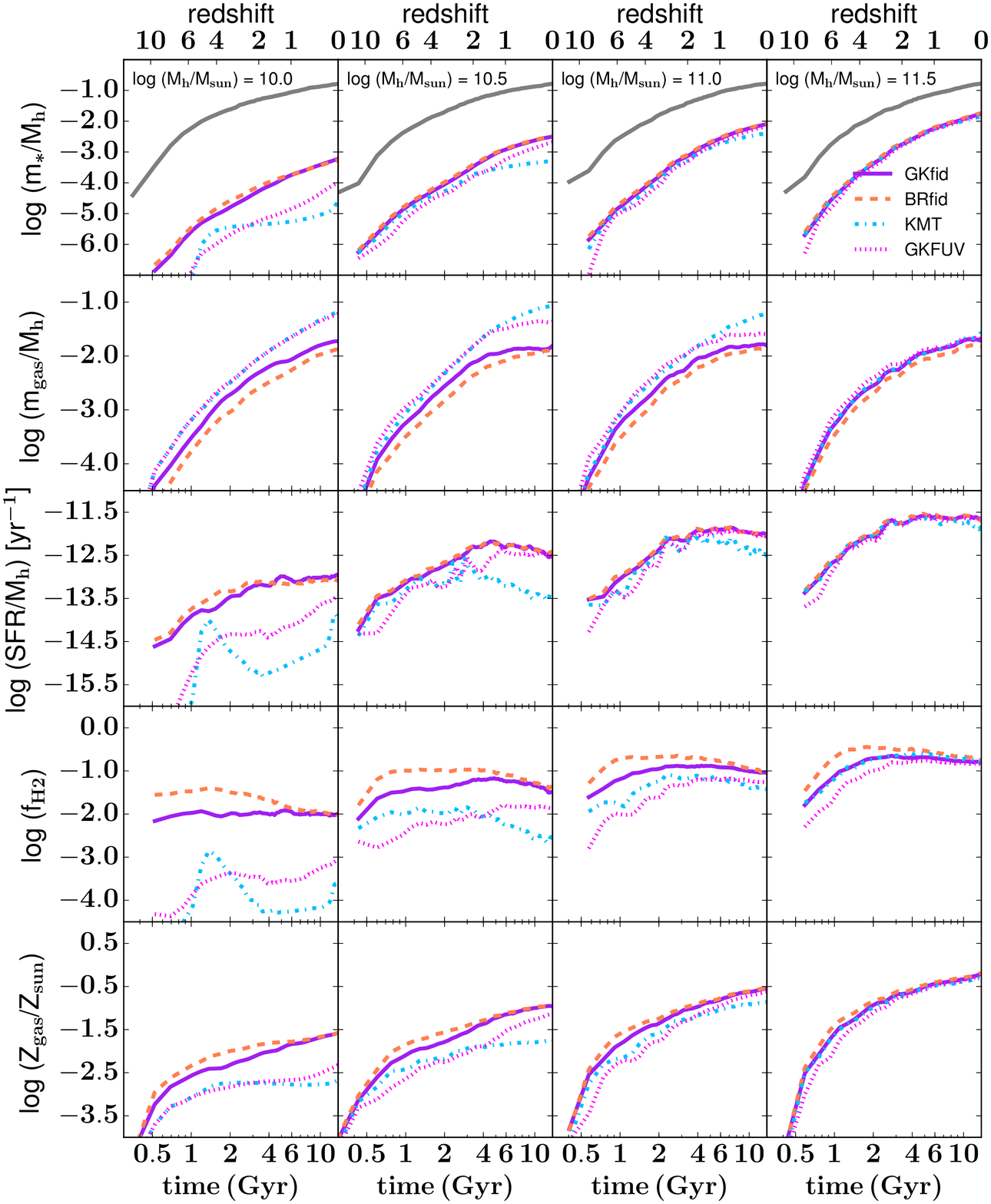}
\end{center}
\caption{Same as Fig.~\ref{fig:trace_resolution}, except here we
  compare our fiducial GK and BR models with the KMT recipe for gas
  partioning and the GK recipe for gas partitioning with a fixed UV
  background field. The KMT recipe and the GKFUV recipes, which both
  neglect the dependence of \Htwo\ formation on the local UV radiation
  field, both predict much lower \Htwo\ fractions in low mass halos,
  especially at high redshift, leading to later stellar mass assembly,
  slightly higher overall gas fractions, lower SFE, and later metal
  enrichment. 
\label{fig:trace_gasrecipes}}
\end{figure*}

Next we experiment with changing the recipe for partitioning gas into
\Htwo (Fig. ~\ref{fig:trace_gasrecipes}). All other ingredients are
the same as our fiducial GK models. We show the pressure-based BR
model as well as two alternate metallicity based models. Recall that
in the fiducial GK model, $f_{\rm H2}$ depends on the total gas
density and metallicity as well as the local UV radiation field (which
we scale with the global galaxy SFR). In the GKFUV model, we remove
the UV radiation field dependence by using the Milky Way value in all
galaxies. In the KMT model, $f_{\rm H2}$ depends only on total gas
density and metallicity, and has different dependencies on these
quantities than the GK model. The results of this experiment are quite
interesting. The predictions of the BR and fiducial GK models are
quite similar, although the GK model tends to predict higher gas
masses, lower \Htwo\ fractions, and lower metallicities at early times
in the two lowest mass halo bins. The KMT and GKFUV models also
produce similar results, as expected based on the findings of
\citet{Krumholz:2011}, who also showed the two models to be very
similar. However, the KMT and GKFUV models predict significantly
suppressed \Htwo\ formation, leading to lower star formation rates and
stellar masses, reduced chemical enrichment, and higher gas fractions
in the two lowest halo mass bins. This is because galaxies in low-mass
halos tend to have lower metallicities but also lower SFR. Therefore
in our fiducial GK models, the lower metallicity, which makes
\Htwo\ formation less efficient, is mitigated by the lower SFR, which
leads to weaker photo-dissociation and relatively higher
\Htwo\ fractions.

\begin{figure*} 
\begin{center}
\includegraphics[width=0.75\hsize]{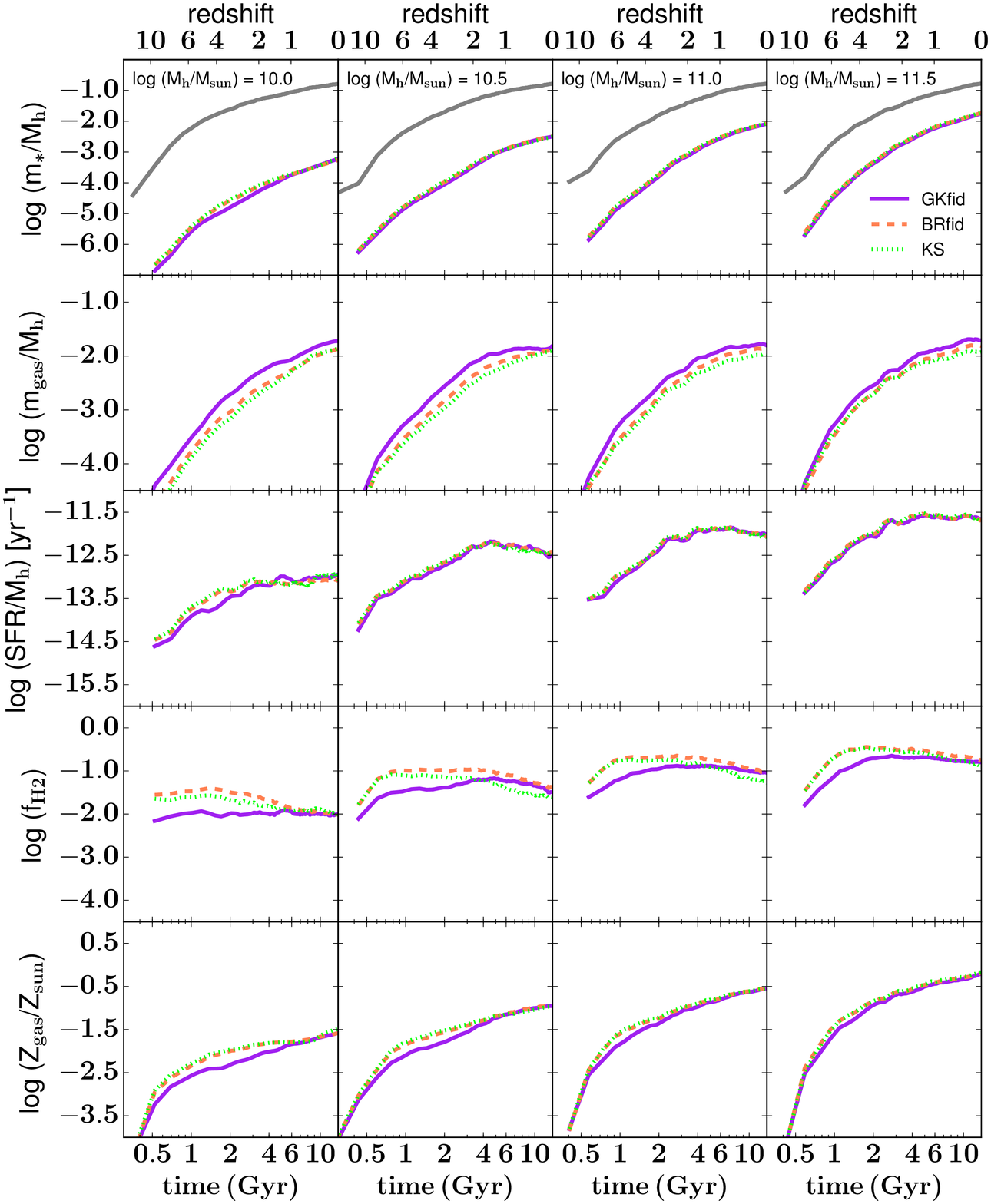}
\end{center}
\caption{Same as Fig.~\ref{fig:trace_resolution}, except here we
  compare the three ``fiducial'' models, GK, BR, and KS. We find
  remarkably similar results among all three models. The largest
  differences are in the predicted overall gas fraction and the
  \Htwo\ fraction at high redshift in the lowest mass halos. 
\label{fig:trace_recipes}}
\end{figure*}

In our penultimate experiment (Fig.~\ref{fig:trace_recipes}), we
compare our two new fiducial multiphase gas recipes with the
``classic'' Kennicutt-Schmidt recipe (see
Section~\ref{sec:sam:sfrecipes}), in which all cold gas above a fixed
surface density is eligible for star formation. This KS recipe has
been used in many previous SAMs \citep[e.g. S08, S12,
][]{Porter:2014}. The build-up of stellar mass is almost identical in
all three models. However, the cold gas mass is highest in the GK
model and tends to be lowest in the KS model. The \Htwo\ fraction is
also lower at early times in the lowest halo mass bin in the GK
model. Note that the \Htwo\ fraction shown for the KS model has been
computed using the BR recipe in post-processing, but this has no
impact on the star formation in the model. Overall, the degree of
similarity between the three model results is quite surprising, given
the rather different physical premises on which they are based. We
discuss possible reasons for this in \S\ref{sec:discussion}.

\begin{figure*} 
\begin{center}
\includegraphics[width=0.75\hsize]{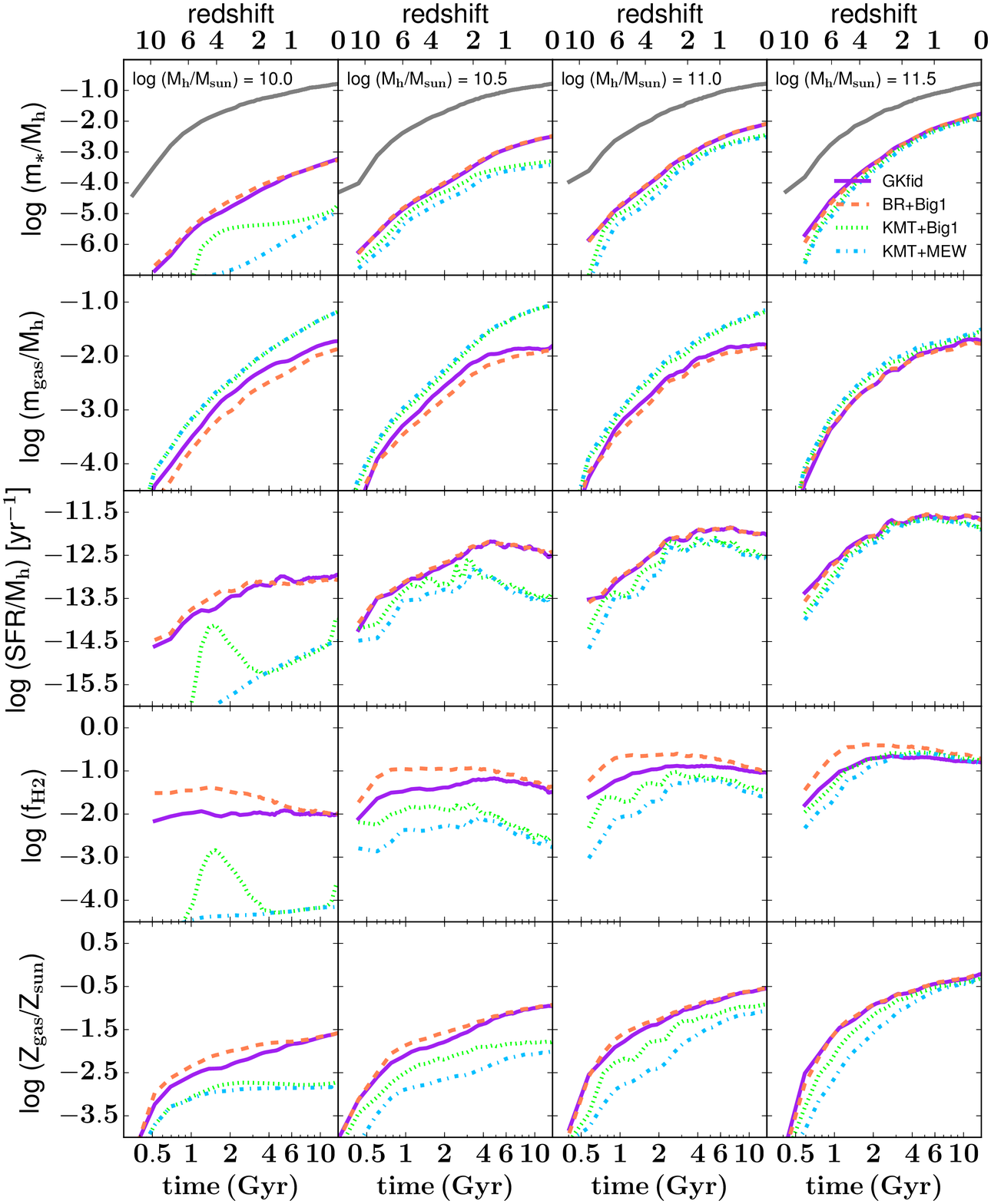}
\end{center}
\caption{Same as Fig.~\ref{fig:trace_resolution}, except here we
  compare the results of our fiducial GK model with combinations of
  model ingredients that are similar to those used in several models
  in the literature. Models that neglect the effect of a varying UV
  background predict later star formation, higher cold gas
  fractions, lower \Htwo\ fractions, and later chemical enrichment in
  low-mass halos. Metal-enhanced winds, when coupled with
  metallicity-dependent gas partition recipes, further delay star
  formation and enrichment.
\label{fig:trace_literature}}
\end{figure*}

In our final experiment, shown in Fig.~\ref{fig:trace_literature}, we
compare the evolution in our fiducial GK model with variants that
include combinations of recipes that are similar to those used in
several published models from the literature. For example, the BR+Big1
model treats gas partitioning and conversion of \Htwo\ to stars using
similar recipes to the BR model of \citet{lagos_sflaw:11} and the
``Bigiel + \Htwo\ prescription 2'' of \citet{fu:12}. The KMT+Big1
contains similar ingredients to the ``Krumholz + \Htwo\ prescription
1'' of \citet{fu:12}. In the KMT+MEW model, we use the KMT recipes for
both gas partioning and star formation, as well as including metal
enhanced winds, as in the models of \citet{Krumholz:2012}. Note that
the KMT model of \citet{lagos_sflaw:11} adopts the KMT recipes for
both gas partioning and star formation, but does not adopt
metal-enhanced winds, so does not correspond exactly to any of the
cases shown here. However, adopting these choices in our models yields
results very similar to the KMT+Big1 model shown. We emphasize that
many other aspects of our models differ from those used by other SAMs
in the literature, so these may not correspond to the actual
predictions of those models. This exercise is intended to shed some
light on the effect of choosing different recipes for gas partitioning
and conversion of \Htwo\ into stars in a controlled environment where
all other aspects of the models are held fixed. Most other SAMs to
date that have attempted to track multi-phase gas with a
metallicity-based approach have done so using the KMT recipe. Our
experiment shows that this may result in more delayed star formation
and enrichment in low-mass halos than our fiducial models
predict. \citet{Krumholz:2012} additionally adopt strongly halo mass
dependent metal-enhanced winds. These two effects together lead to
strong suppression of star formation in low-mass galaxies,
particularly at early times.

\subsection{Star Formation Relations}
\label{sec:sfrel}

\begin{figure*} 
\begin{center}
\includegraphics[width=0.45\textwidth]{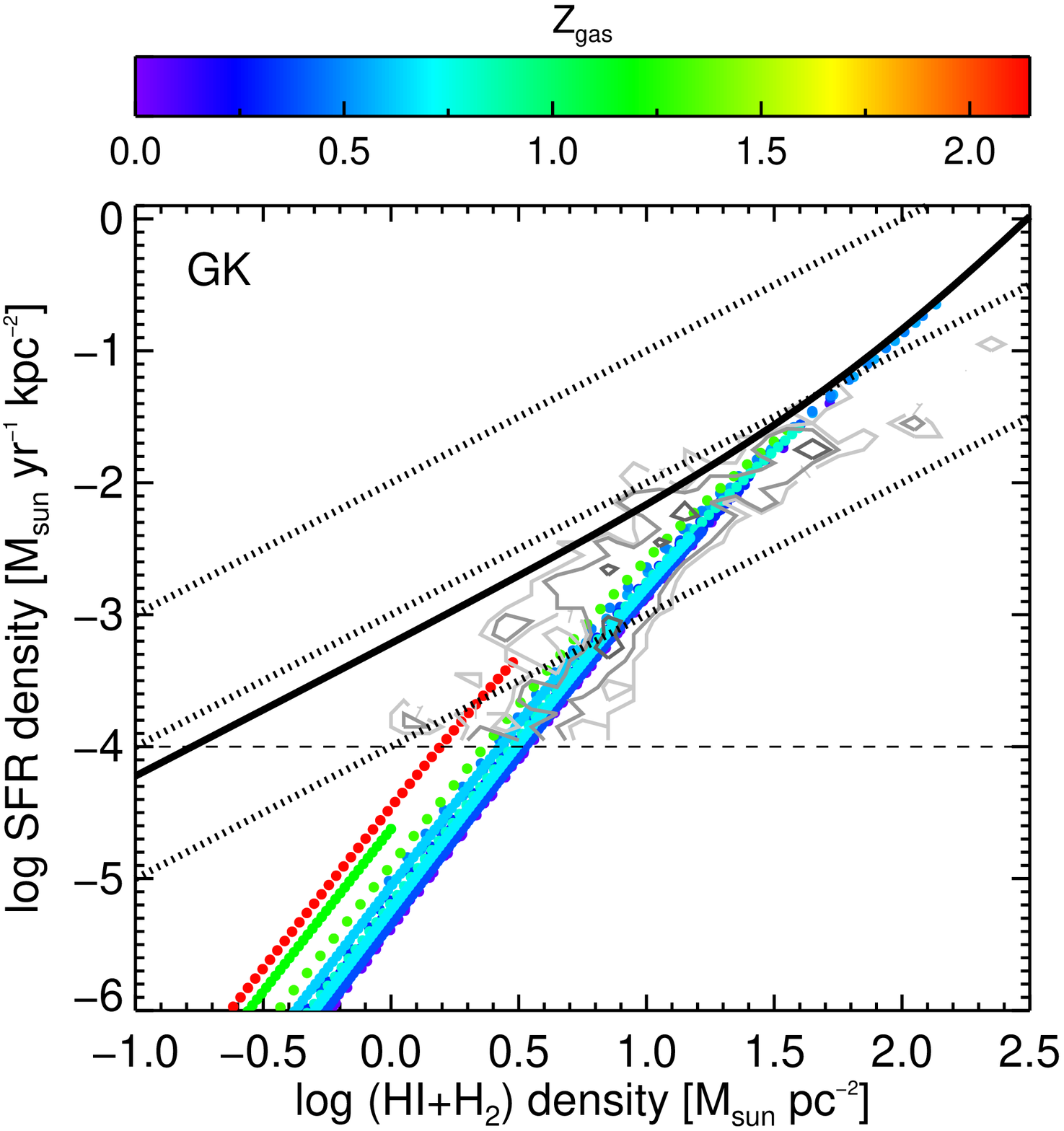}
\includegraphics[width=0.45\textwidth]{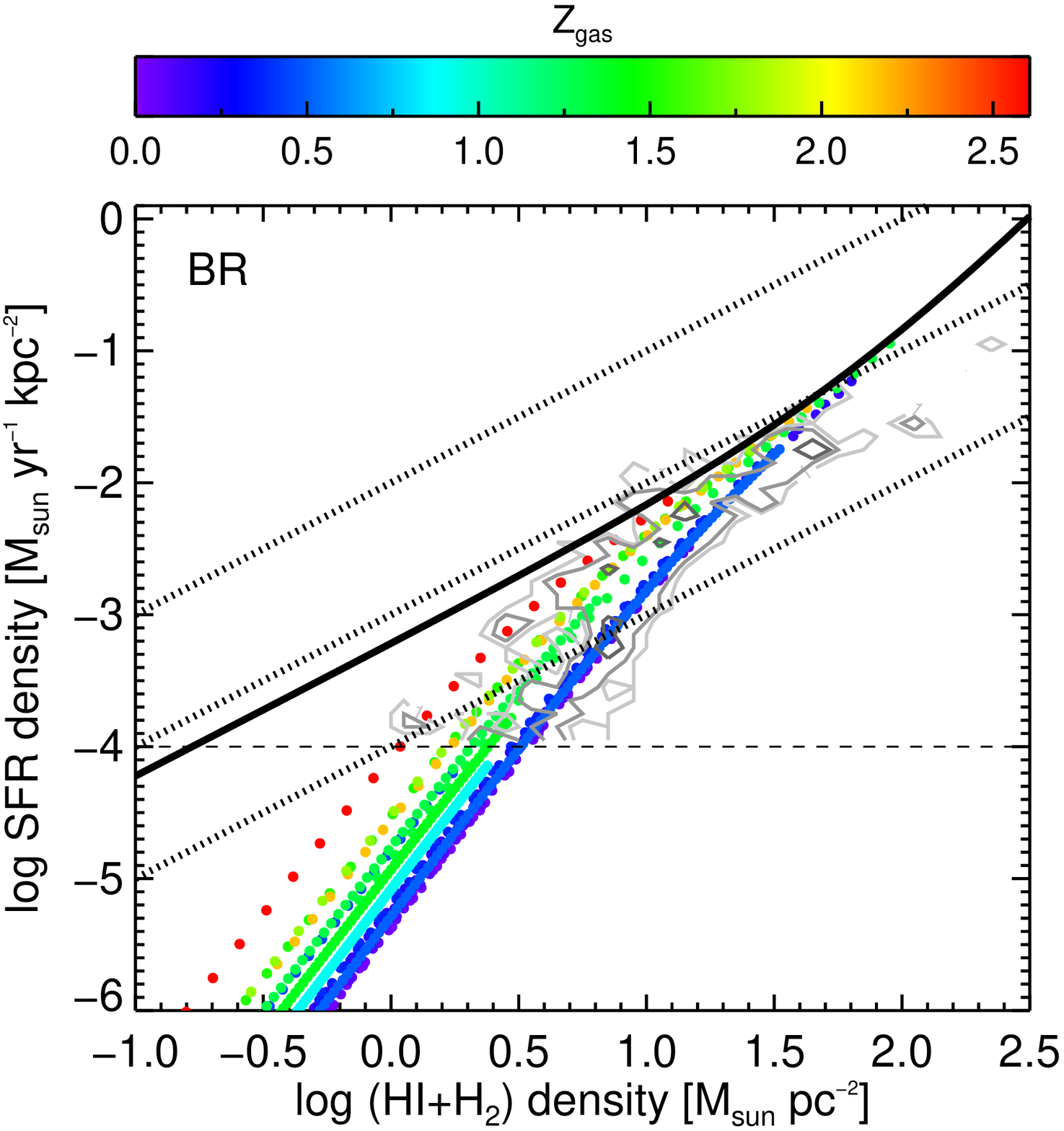}
\end{center}
\caption{Relation between total cold neutral gas density (\HI + \Htwo)
  and star formation rate density (SFRD). The solid black line shows
  the input \Htwo-based star formation recipe (Big2). Slanted dotted
  lines show a star formation efficiency of 1\%, 10\%, and 100\% per
  $10^8$ yr. Gray contours show observational estimates from the 13
  Spiral galaxies from the THINGS+Heracles sample presented in
  \protect\citet{Leroy:2008}. Colored dots show a selection of 25
  galaxies in our fiducial GK (left) and BR (right) models (at $z=0$),
  with a stellar mass range chosen to match the THINGS sample. Each
  point shows the value in an annulus with radius 500 pc. The points
  are color-coded with the average gas phase metallicity of the
  galaxy.
Note that \emph{in both models}, galaxies with lower metallicity gas
have lower SFRD for a given total gas density, because a smaller
fraction of the gas is predicted to be in the form of \Htwo. In the GK
model, a direct dependence of \Htwo\ fraction on metallicity is
assumed. In the BR model, a dependence on disk midplane pressure is
assumed, but this quantity turns out to be highly correlated with
metallicity in our models.
\label{fig:kenn_prof}}
\end{figure*}

Fig.~\ref{fig:kenn_prof} shows the relationship between the total
neutral cold gas surface density $\Sigma_{HI+H_2}$ and star formation
rate density $\Sigma_{\rm SFR}$ in our two new fiducial models with
multiphase gas partitioning. In our previous generation of models,
galaxies had a deterministic relation between $\Sigma_{HI+H_2}$ and
$\Sigma_{\rm SFR}$ given by the assumed KS relation (as plotted in
Fig.~\ref{fig:H2kennobs}), and $\Sigma_{\rm SFR}$ was set to zero
below the critical gas surface density $\Sigma_{\rm crit}$ (also shown
in Fig.~\ref{fig:H2kennobs}). In our new models, neutral gas is
`partitioned' into \HI\ and \Htwo, and only \Htwo\ is allowed to
participate in star formation. Therefore the value of $\Sigma_{\rm
  SFR}$ at a given $\Sigma_{HI+H_2}$ has a ``second parameter''
dependence. This second parameter is metallicity in the case of the GK
(and KMT, not shown) recipes and disk mid-plane pressure (stellar
surface density $\Sigma_*$, to first order) in the BR
recipe. Fig.~\ref{fig:kenn_prof} shows the the average metallicity of
the cold gas in each galaxy, where each dot shows one annulus with
radius 500 pc. In the GK model, galaxies with higher metallicity have
a higher $\Sigma_{\rm SFR}$ for a given $\Sigma_{HI+H_2}$, as
expected. However, we also see a similar dependence on metallicity in
the BR model, although in this case it is not directly input into the
model. The reason for this apparent dependence on metallicity is
simply that $\Sigma_*$ and gas phase metallicity are highly
correlated. The curvature in $\Sigma_{\rm SFR}$-$\Sigma_{HI+H_2}$ at
low gas surface densities in both models is in good agreement with
observations of nearby spiral galaxies \citep{Leroy:2008,Bigiel:2008}.

\subsection{Evolution of Galaxy Populations}
\label{sec:results:pops}

\subsubsection{Stellar Mass Functions and Stellar Fractions}
\begin{figure*} 
\begin{center}
\includegraphics[width=0.95\hsize]{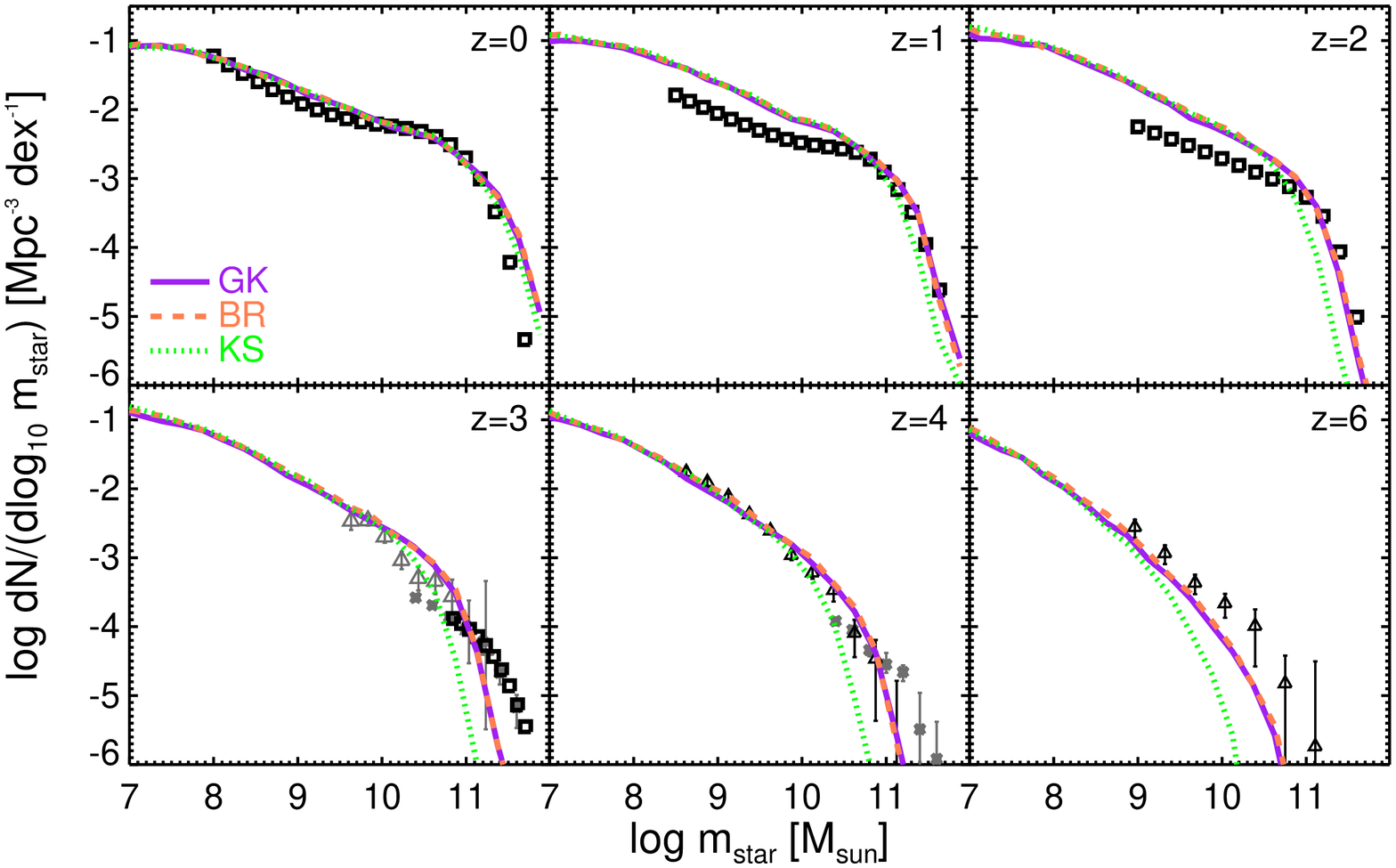}
\end{center}
\caption{Stellar mass function evolution with redshift. Symbols show
  observational estimates as follows. In the $z=0.1$, 1, 2, and 3
  panels, black square symbols show a double-Schechter fit to a
  compilation of observational estimates. Observations included in the
  fit are: $z=0.1$ -- \protect\citet{Baldry:2008},
  \protect\citet{Moustakas:2013}; $z=1$ and $z=2$ panels --
  \protect\citet{Tomczak:2014}, \protect\citet{Muzzin:2013}, $z=3$
  panels -- \protect\citet{Muzzin:2013}. The fits shown at $z=1$,
  $z=2$ and $z=3$ are interpolated to these redshifts from adjacent
  redshift bins in the original published results. In the $z=3$ panel
  we also show estimates from
  \protect\citet[][triangles]{Santini:2012} and
  \protect\citet[][crosses]{Caputi:2011}. In the $z=4$ panel we show
  estimates from \protect\citet[][triangles]{Duncan:2014} and
  \protect\citet[][crosses]{Caputi:2011}. In the $z=6$ panel we show
  the estimates from \protect\citet[][triangles]{Duncan:2014}.
The purple solid line shows the results of the fiducial GK model, the
orange dashed line shows the fiducial BR model, and the green dotted
line shows the KS model.
\label{fig:smf}}
\end{figure*}

\begin{figure*} 
\begin{center}
\includegraphics[width=0.95\hsize]{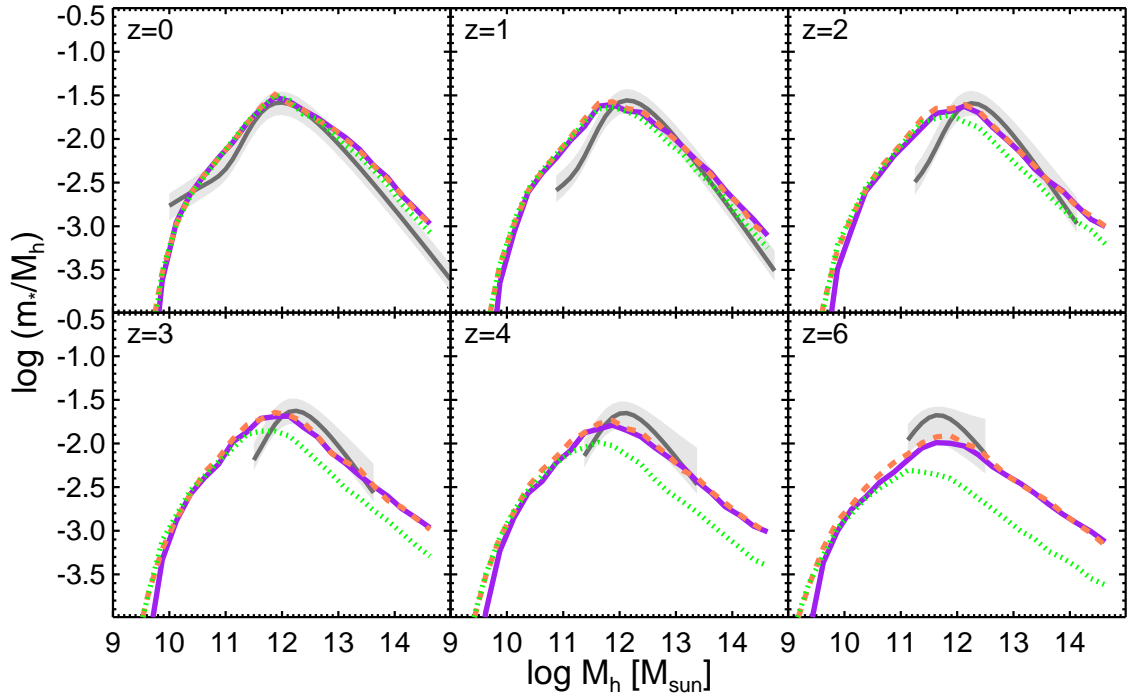}
\end{center}
\caption{The stellar mass of central galaxies divided by the total
  mass of their dark matter halo, in redshift bins from $z=0$--6. Dark
  gray lines and shaded areas show constraints from halo abundance
  matching from \protect\citet{Behroozi:2013}. The purple solid line
  shows the results of the GK model, the orange dashed line shows the
  BR model, and the green dotted line shows the KS model.
\label{fig:fstar_comp}}
\end{figure*}

\begin{figure*} 
\begin{center}
\includegraphics[width=0.75\hsize]{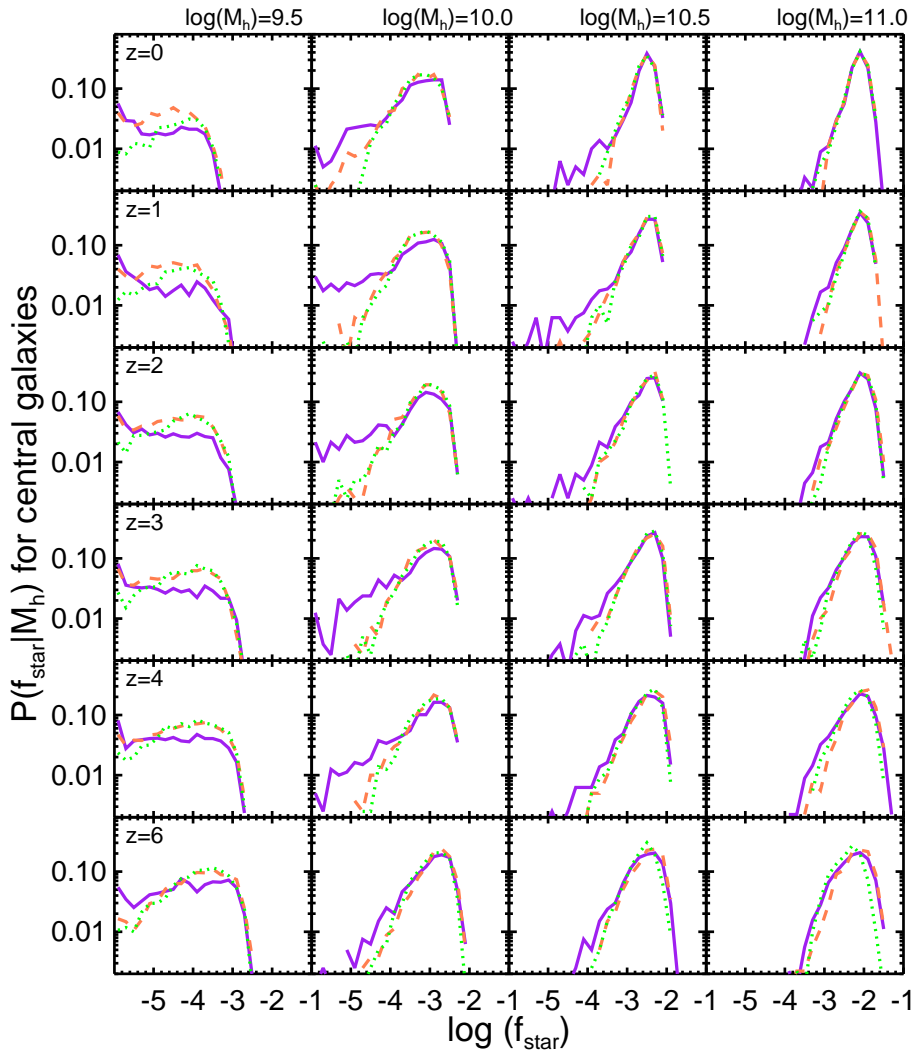}
\end{center}
\caption{Distribution functions for the stellar fraction ($f_{\rm
    star} \equiv m_{\rm star}/M_h$) of central galaxies in bins of
  halo mass ($\log M_h = 9.25-9.75$, $9.75-10.25$, $10.25-10.75$,
  $10.75-11.25$) and redshift as indicated on the panels.  The purple
  solid line shows the results of the GK model, the orange dashed line
  shows the BR model, and the green dotted line shows the KS model.
\label{fig:fstarhist_central}}
\end{figure*}

\begin{figure*} 
\begin{center}
\includegraphics[width=0.75\hsize]{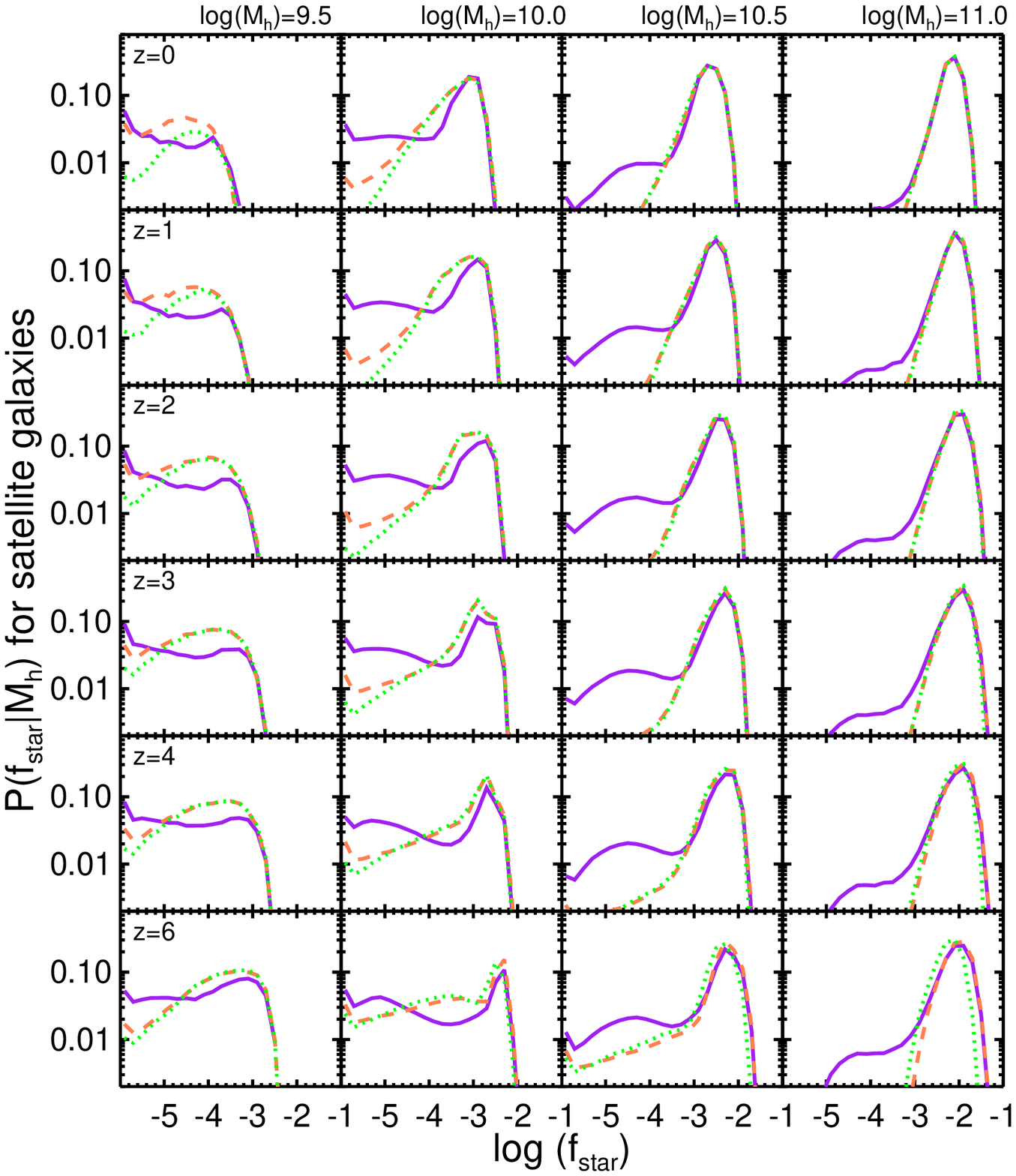}
\end{center}
\caption{The same as Fig.~\protect\ref{fig:fstarhist_central}, but for
  satellite galaxies (sub-halos). The halo mass here is the mass of
  the halo when it first becomes a sub-halo.
\label{fig:fstarhist_sat}}
\end{figure*}

In this sub-section we examine the evolution of galaxy populations,
which are directly comparable with observations. We consider three
main model variants: the fiducial versions of the GK, BR and KS models
(see Table~\ref{tab:models}). The KS model is the star formation
recipe used in previously published Santa Cruz SAMs
\citep[e.g.][]{s08,Somerville:2012,Porter:2014}. The GK and BR models
are the same as the models used in \citet{Popping:2014} and
\citet{Berry:2014}. Selected results for other model variants are
shown in the Appendix.

In Fig.~\ref{fig:smf} we present predictions for the stellar mass
function of galaxies from $z=0$ to $z=6$. We compare these predictions
with a compilation of observations as described in the figure
caption. The similarity of the three model predictions is striking,
particularly on the low mass end where we might have expected the
differences to be largest. The most noticable differences are instead
at high masses, particularly at redshifts $z \gtrsim 2$. Here, the KS
model produces significantly lower number densities of massive
galaxies at $z\gtrsim 1$, with the deviation growing with increasing
redshift. It is important to note that we have not accounted for the
expected errors in the observational estimates of the stellar masses
when comparing to the models --- this would tend to lead to an
apparent increase in the number of massive galaxies due to Eddington
bias \citep[see][e.g.]{Lu:2014}. However, even if this effect were
included, the earlier formation of massive galaxies predicted by the
GK and BR models is clearly in better accord with recent
observations. All three models suffer from the familiar excess of
low-mass galaxies at $z\sim 0.5$--2 which, as we have already
discussed, is a widespread problem in both semi-analytic models and
numerical hydrodynamic simulations
\citep{fontanot:09,weinmann:12,White:2014,Somerville_Dave:2014}. One
of the important conclusions of this paper is that \emph{varying the
  star formation efficiency according to physically motivated recipes
  does not appear to be able to cure this problem within the current
  model framework}. A possibly related problem is that over this
redshift range, the predicted gas fractions in low mass galaxies in
these same models may be too low \citep[][Popping et al. in
  prep]{White:2014,Somerville_Dave:2014,Popping:2014}\footnote{An
  important caveat, however, is that the gas mass estimates that lead
  to this conclusion are based on indirect methods. It remains to be
  seen whether this is conformed by direct observations of cold gas in
  these low-mass galaxies.}.

Fig.~\ref{fig:fstar_comp} shows a related quantity, the stellar
fraction (stellar mass divided by halo mass; $f_{\rm star} \equiv
m_*/M_h$) over the same redshift range. Our model predictions are now
compared with constraints from (sub)-halo abundance matching from
\citet{Behroozi:2013}.  The conclusions are similar to the ones above,
unsurprisingly since the $f_{\rm star}$ constraints are derived from
observational estimates of stellar mass functions (though not exactly
the same ones plotted in our Fig.~\ref{fig:smf}). The median stellar
fractions are nearly identical in the three models in low mass halos
($\log M_h/M_{\odot} \lesssim 11$) and are very similar in the GK and
BR model over the whole range of halo masses. The median value of
$f_{\rm star}$ in the KS model is much lower in massive halos than in
the other two models, and the difference increases with redshift to
about 0.4--0.5 dex at $z=6$.

Although the median values of $f_{\rm star}$ are similar in the three
models, the distributions differ significantly for low mass host
halos. Fig.~\ref{fig:fstarhist_central} and \ref{fig:fstarhist_sat}
show the distribution of $f_{\rm star}$ in halo mass and redshift
bins, for central and satellite galaxies respectively. For all models
and at all epochs, the distribution of $f_{\rm star}$ becomes broader
and more skewed with decreasing halo mass. For massive halos, the
width of the distribution becomes slightly narrower with increasing
time, while for low mass halos, a more noticable tail towards lower
values of $f_{\rm star}$ develops with time. In the two intermediate
halo mass bins, this tail is more prominent in the GK models than in
the other two models.  The predicted broadening in $f_{\rm star}$ has
potentially important implications for empirical halo-based models,
which generally assume a narrow and fixed scatter in $f_{\rm
  star}(M_h)$. We show the results for these two types of galaxies
separately because a) central and satellite galaxies are treated
differently in SAMs. For example, in our models, satellite galaxies
are not allowed to accrete new gas from the IGM; b) our predictions
for differences between stellar fractions for satellites and centrals
can provide useful input to empirical models such as Halo Occupation
Distribution (HOD) and abundance matching models. Many such models do
not distinguish between satellite and central galaxies.

\subsubsection{Star Formation rates and gas depletion times}
\begin{figure*} 
\begin{center}
\includegraphics[width=0.95\hsize]{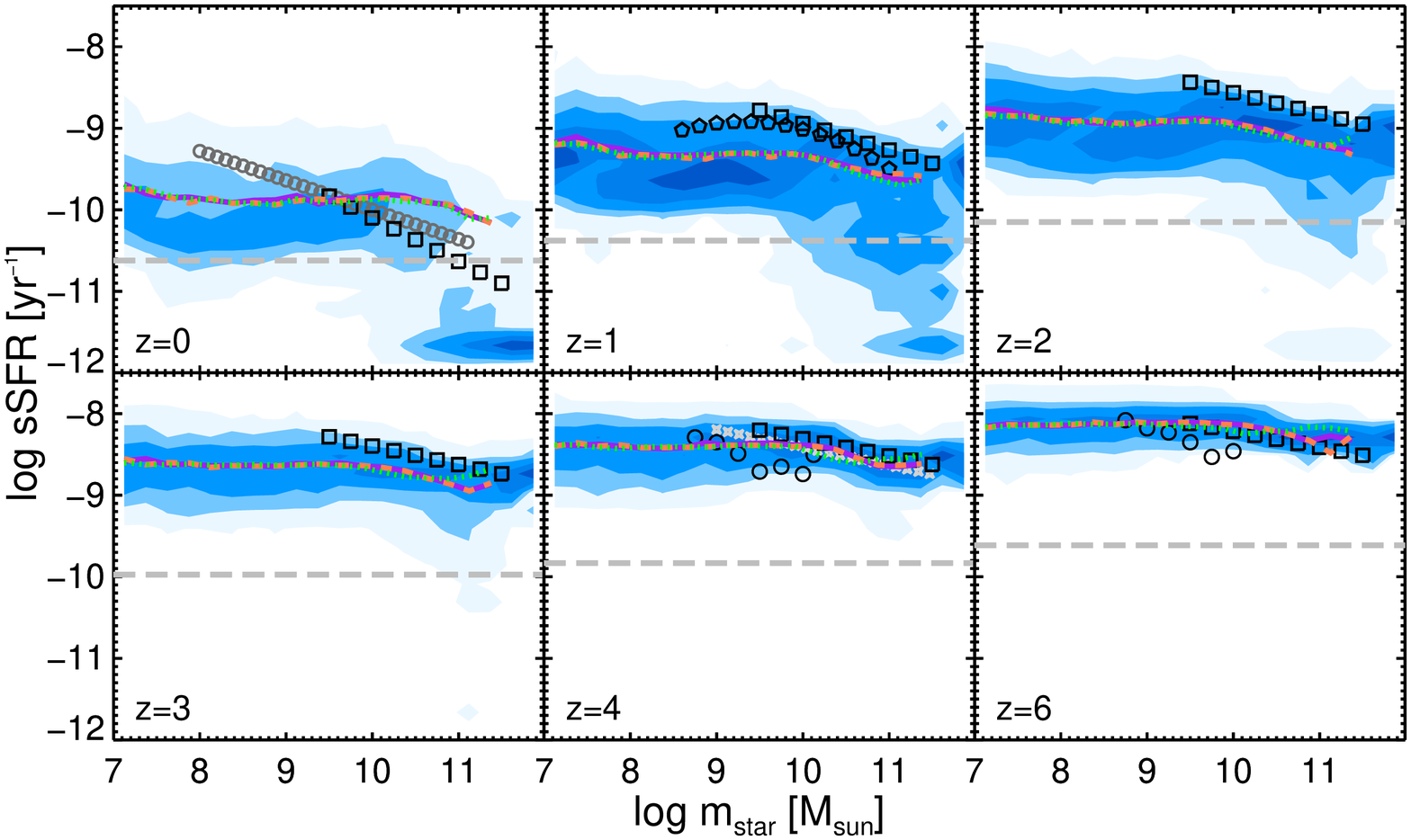}
\end{center}
\caption{Mean specific star formation rate (sSFR $\equiv
  \dot{m}_*/m_*$), as a function of stellar mass for our three
  fiducial models (purple solid: GK; orange dashed: BR; green dotted:
  KS), in redshift bins from $z=0$--6.  The blue contours show the
  conditional sSFR in the GK model. The horizontal gray line shows the
  sSFR corresponding to $1/(3 t_H)$, where $t_H$ is the Hubble time at
  that redshift. Only galaxies with sSFR $>1/(3 t_H)$ are included in
  the mean. Symbols show a compilation of observations for star
  forming galaxies as follows: $z=0.1$ -- \protect\citet[][open
    circles]{Salim:2007}; $z=1$ -- \protect\citet[][pentagons,
    interpolated in redshift from the published
    results]{Whitaker:2014}; $z=4$--
  \protect\citet[][crosses]{Steinhardt:2014}; $z=4$ and $z=6$ --
  \protect\citet[][circles]{Salmon:2014}; all panels -- fit to data
  compilation from \protect\citet[][squares]{Speagle:2014}.
\label{fig:ssfrz}}
\end{figure*}

\begin{figure*} 
\begin{center}
\includegraphics[width=0.75\hsize]{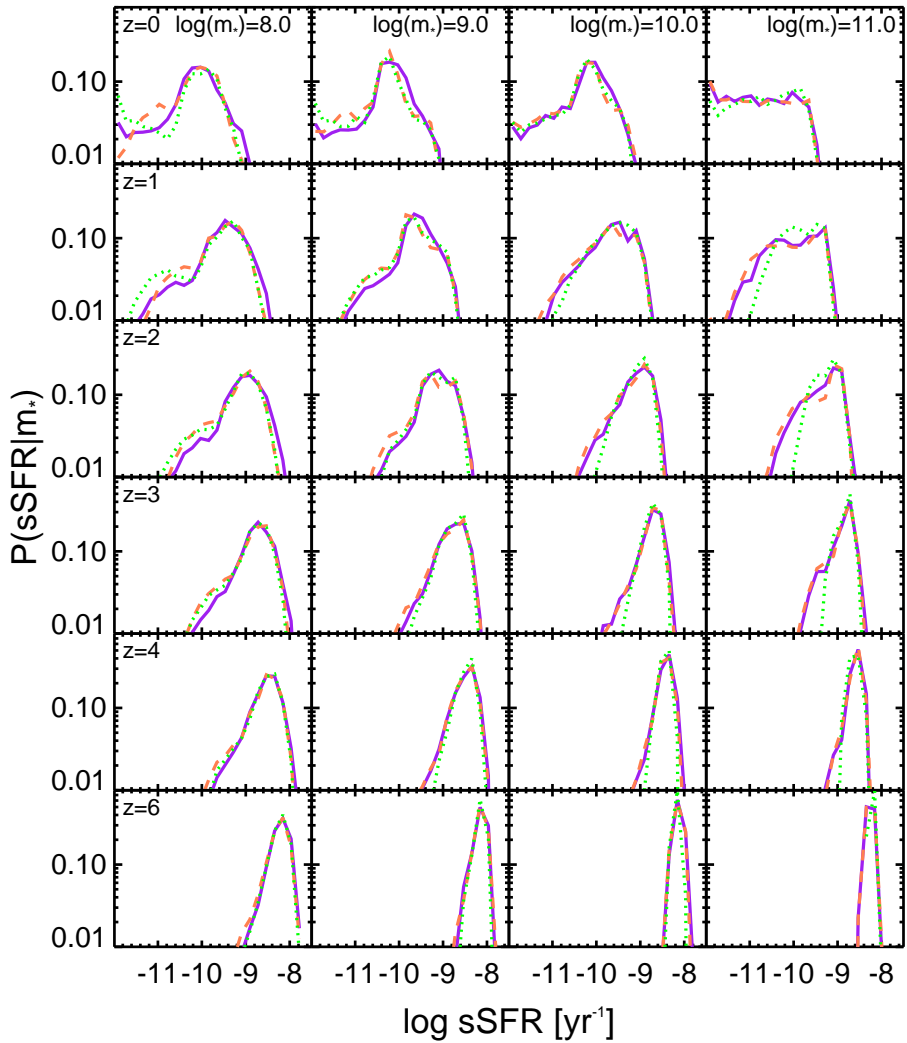}
\end{center}
\caption{The conditional probability distributions of sSFR in stellar
  mass bins at different redshifts, for our three fiducial models
  (purple solid: GK; orange dashed: BR; green dotted: KS).
\label{fig:ssfrdist}}
\end{figure*}

\begin{figure*} 
\begin{center}
\includegraphics[width=\hsize]{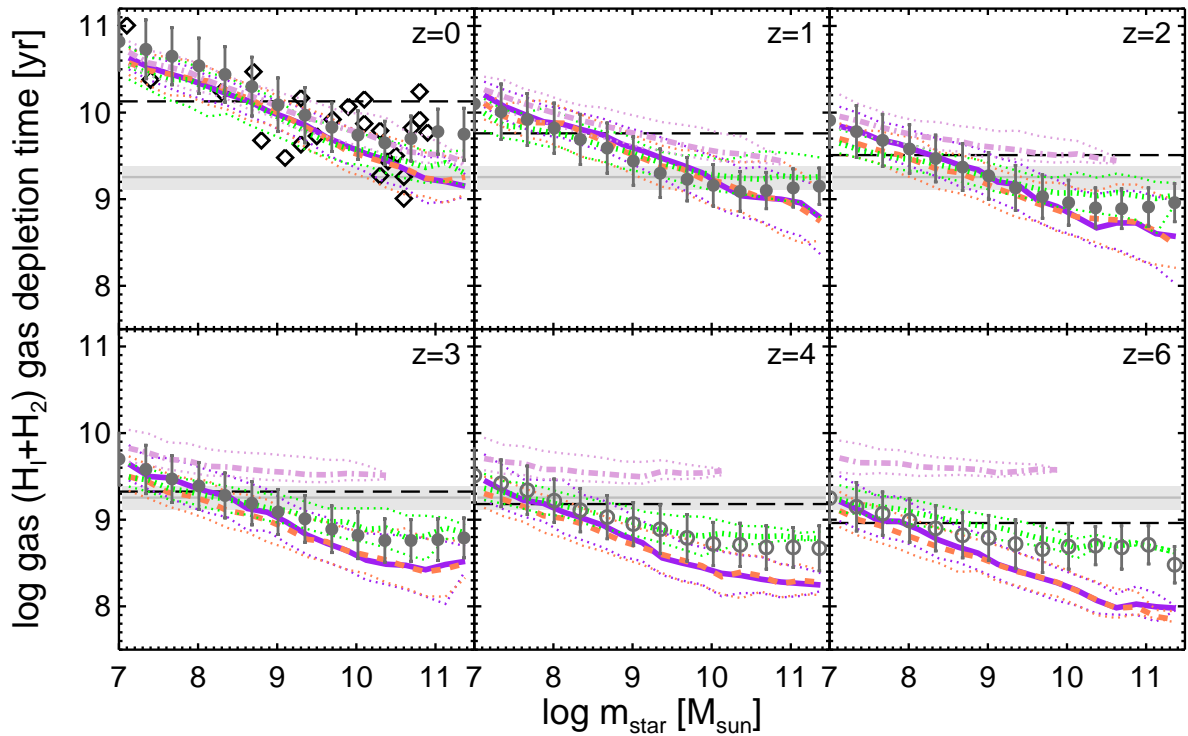}
\end{center}
\caption{Total gas depletion time, defined as the total cold neutral
  gas mass (\HI + \Htwo) divided by the star formation rate. The
  purple solid lines show the predictions of the fiducial GK model,
  the orange dashed lines show the BR model, and the green dotted
  lines show the KS model. The lavender dot-dashed lines show the
  GK+Big1 model. The 16th, 50th, and 84th percentiles are shown for
  central star forming galaxies in the models. The horizontal black
  dashed line shows the age of the Universe at that redshift. The open
  diamonds in the $z=0$ panel show observational estimates for
  galaxies in the THINGS+Heracles sample from
  \protect\citet{Leroy:2008}. The horizontal gray line shows the
  average \emph{molecular} gas depletion time estimated by
  \protect\citet{Leroy:2013} for nearby galaxies; the shaded gray area
  indicates the uncertainty in the measurement due to the uncertain
  conversion factor between CO and \Htwo. The gray circles show the
  estimates obtained via the empirical method of
  \protect\citet{Popping:2014b}. These are plotted with open symbols
  at $z>3$ to indicate that the estimates are quite speculative in
  this redshift regime (see text for more details). All three models
  reproduce the observed trend of decreasing depletion time with
  increasing stellar mass, but different underlying physics are
  responsible for the trends in the different models.
\label{fig:tdep}}
\end{figure*}

\begin{figure*} 
\begin{center}
\includegraphics[width=\hsize]{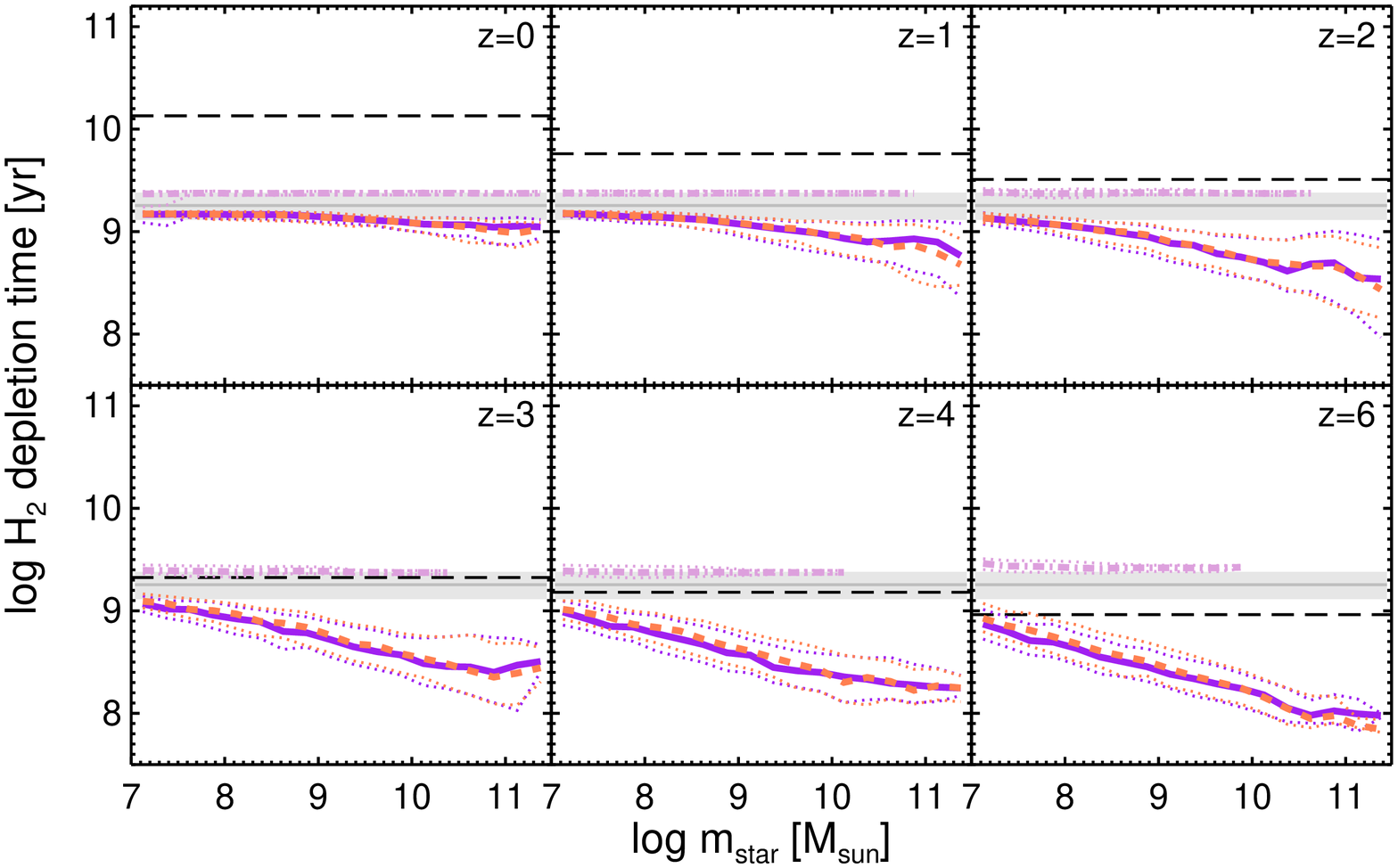}
\end{center}
\caption{\Htwo\ depletion time, defined as the \Htwo\ mass divided by
  the star formation rate. Models shown (colored lines) are as in
  Fig.~\protect\ref{fig:tdep}. The horizontal black dashed line shows
  the age of the Universe at that redshift. The horizontal gray line
  shows the average \emph{molecular} gas depletion time estimated by
  \protect\citet{Leroy:2013} for nearby galaxies; the shaded gray area
  indicates the uncertainty in the measurement due to the uncertain
  conversion factor between CO and \Htwo. Our models predict that at
  high redshift, molecular gas depletion times were significantly
  shorter, and there is a much stronger trend between galaxy stellar
  mass and \Htwo\ depletion time.
\label{fig:tdep_h2}}
\end{figure*}

Fig.~\ref{fig:ssfrz} shows the specific star formation rate (sSFR
$\equiv \dot{m}_*/m_{\rm star}$) as a function of stellar mass over
the redshift range $z=0$--6. Our model predictions are compared with a
compilation of observations as described in the figure caption. We
have selected only ``star forming'' galaxies using the criterion
sSFR $> 1/(3 t_H(z))$, where $t_H(z)$ is the Hubble time at the
galaxy's redshift. This has been shown to produce similar results to
commonly used observational methods for selecting star forming
galaxies \citep[e.g.][]{Lang:2014}. Our models agree well with
the observed slope and normalization of the Star Forming Main Sequence
(SFMS) at $z\sim 6$--4, but then the normalization of the model SFMS
drops below the observationally estimated one between $z\sim
3$--0.5. At $z\sim 0$, the predicted SFMS has approximately the
correct normalization for massive galaxies (here the precise value may
be impacted by the details of the selection of ``star forming'' versus
quiescent galaxies), but the slope is much shallower than the
observations suggest. This is another facet of the ``dwarf galaxy
conundrum'' discussed in \citet{White:2014}, and again is common to
most cosmological models of galaxy formation
\citep{Somerville_Dave:2014}. Our results show that this relation is
extremely robust to changing the star formation recipe in models. 

Fig.~\ref{fig:tdep} shows the total gas depletion time $t_{\rm dep}
\equiv (m_{\rm HI} + m_{\rm H_2})/\dot{m}_*$ in the fiducial GK and BR
models, and in the GK+Big1 model. Here we compute the depletion time
using only the SFR due to the `disc' mode of SF, i.e. not including
star formation due to merger-triggered bursts, but the plot looks very
similar when the burst mode is included. Observations of nearby
galaxies show that $t_{\rm dep}$ increases with decreasing stellar
mass, i.e. the conversion of cold gas into stars is less efficient in
low mass galaxies \citep[e.g.][]{Leroy:2008}. A similar trend is
indicated by the empirical estimates of total gas depletion time from
\citet{Popping:2014b}, also shown in Fig.~\ref{fig:tdep} for
comparison. The empirical estimates are based on a SFR-halo mass
relation inferred from abundance matching, and an indirect estimate of
the \HI\ and \Htwo\ masses from inverting the SFR density. These
estimates rely on a number of assumptions (e.g., that disk cold gas
radial profiles are well-represented by exponentials), and on the
observed relationship between size and stellar mass in disk-dominated
galaxies. In \citet{Popping:2014b}, the empirical predictions are
shown only up to $z\sim 3$, because it is not known whether these
assumptions and empirical relations hold at higher redshift. Here we
show the results of extrapolating the same method to $z\sim 6$, but
these should be considered highly uncertain.

Our three fiducial models reproduce the same qualitative trends
indicated by the observations and by the empirical predictions. First,
depletion times are longer in lower-mass galaxies. In detail, the
physics that is responsible for this trend is different in the three
fiducial models. Low mass galaxies have lower gas and stellar surface
density on average. In the KS model, gas below the critical surface
density is not allowed to make stars, and low mass galaxies tend to
have a larger fraction of their gas below this threshold. Low mass
galaxies also tend to have lower gas-phase metallicities, and in the
GK model, this results in less efficient formation of \Htwo\ and thus
of stars. In the BR model, the lower SFE in low-mass galaxies is due
to their lower stellar surface density, which leads to a lower disc
mid-plane pressure, and again a lower \Htwo\ fraction. Second, in all
models, $t_{\rm dep}$ at a given stellar mass was lower in the past
and increases with cosmic time. The GK and BR models show more
pronounced evolution and shorter depletion times (higher SFE) at high
redshift, particularly in massive galaxies. This is due to the steeper
dependence of the SFR density on gas density adopted in our GK and BR
models (slope $N_{\rm SF} =2$ instead of $1.4$). High-redshift
galaxies contain higher surface density gas overall, and so the
results are more sensitive to the slope of the SF relation at high gas
densities. This result explains the less efficient formation of stars
in massive galaxies at high redshift in the KS model relative to the
GK and BR models, seen in Fig.~\ref{fig:smf} and
\ref{fig:fstar_comp}. Note that already by $z\sim 3$, and increasingly
so at higher redshift, the predictions using the Big1 recipe for SF
are inconsistent with the empirical constraints. This suggests that
the assumption of a constant \Htwo\ depletion time in galactic disks
(which is inherent in the Big1 recipe) may not be universally
applicable.

Fig.~\ref{fig:tdep_h2} shows the \Htwo\ depletion time ($t_{\rm dep,
  H_2} \equiv m_{\rm H_2}/\dot{m}_*$) in the GK, BR, and GK+Big1
models (the KS model is not shown, as we do not track
\Htwo\ self-consistently in this model). This figure, in combination
with results shown in PST14, shows that the trends seen in our models
in Fig.~\ref{fig:tdep} are due to a combination of two factors: a) at
a given redshift, more massive galaxies have larger fractions of their
cold gas in the form of \Htwo, and at fixed mass, higher redshift
galaxies have higher \Htwo\ fractions; b) the \Htwo\ depletion time is
also shorter in more massive galaxies and at high redshift.

\subsubsection{Mass-metallicity relations}

\begin{figure*} 
\begin{center}
\includegraphics[width=0.95\hsize]{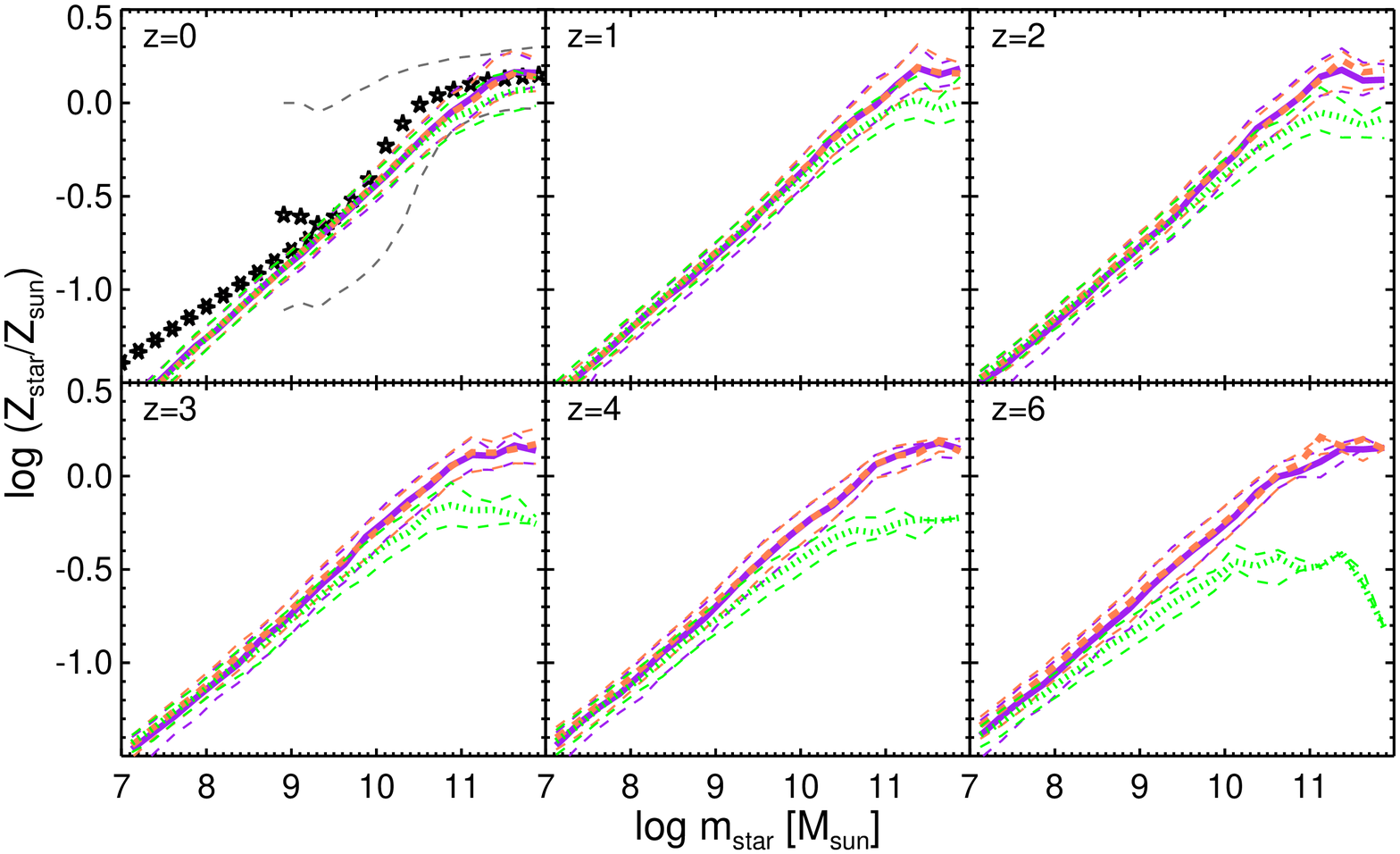}
\end{center}
\caption{Stellar mass vs. stellar metallicity.  Five-pointed star
  symbols and dashed lines show observational estimates from
  \protect\citet{gallazzi:05}, and six-pointed stars show the fit to
  the observed MZR estimated from Local Group dwarf galaxies by
  \protect\citet{Kirby:2013}. The purple lines show the 16, 50, and
  84th percentiles for our fiducial GK model, the orange line shows
  the BR model, and the green line shows the KS model.
\label{fig:massmetstar}}
\end{figure*}

\begin{figure*} 
\begin{center}
\includegraphics[width=0.95\hsize]{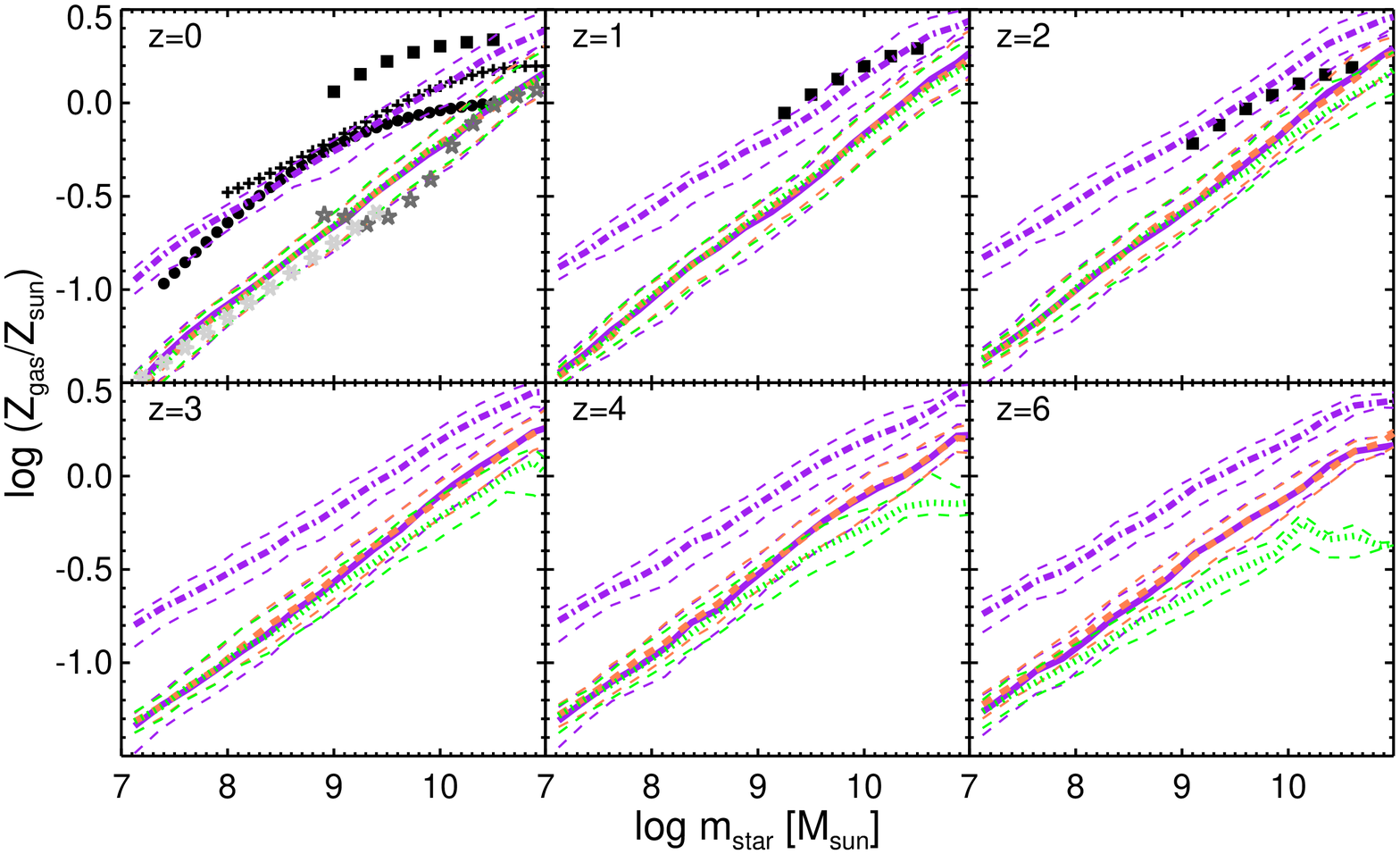}
\end{center}
\caption{Stellar mass vs. cold gas phase metallicity. Black symbols
  show observational estimates of the gas-phase metallicity: $z=0.1$
  -- \protect\citet[][pluses]{Peeples:2014}; \protect\citet[][filled
    circles]{Andrews:2013}. In all panels, the filled squares show the
  compilation of observational estimates from
  \protect\citet{Zahid:2013}. For comparison, we also show the
  observational estimates of \emph{stellar metallicity}, as in
  Fig.~\ref{fig:massmetstar}, with gray star symbols. Colored lines
  show model predictions, as in Fig.~\protect\ref{fig:massmetstar}.
  The dot-dashed purple line shows the GK model with an approximate
  correction for varying [$\alpha$/Fe] (see text). Taken at
  face value, our predicted gas-phase MZR appears to be much steeper
  than the observational estimates. However, properly accounting for
  varying abundance ratios of $\alpha$ versus Fe elements may at least
  partially remove this tension (see text).
\label{fig:massmetgas}}
\end{figure*}

Fig.~\ref{fig:massmetstar} and Fig.~\ref{fig:massmetgas} show the
mass-metallicity relation (MZR) for stellar and cold gas phase
metallicities, respectively. Recall that the chemical yield parameter
in our models has been adjusted to approximately reproduce the
normalization of the stellar MZR measured by \citet{gallazzi:05}. Our
models naturally predict a slope for the stellar MZR that is in fairly
good agreement with observations \citep{Woo:2008,Kirby:2013} down to
very low stellar masses ($m_{\rm star} \sim 10^{7} \msun$). The three
fiducial models make very similar predictions for the stellar phase
MZR, except that massive galaxies become enriched much earlier in our
two new models (GK and BR) than in the KS model. This is owing to the
steeper slope of our Big2 star formation recipe at high gas densities
(see discussion in \S\ref{sec:trace}, especially
Fig.~\ref{fig:trace_SF}, and above). It is also interesting to note
that our models predict a much smaller dispersion in the stellar MZR
than the observational dispersion estimated by \citet{gallazzi:05}.

In Fig.~\ref{fig:massmetgas}, the model results shown are for central,
star forming galaxies selected using the same criteria described
above. This is because the observational estimates of gas-phase
metallicity are based on emission line diagnostics from \HII\ regions,
which are only detectable in star forming galaxies. A compilation of
observational estimates for the gas phase MZR is shown. We have
converted the observed values of $12+\log ({\rm O/H})$ to $Z/Z_\odot$
assuming $12+\log ({\rm O/H})=8.76$ for the Sun \citep{Caffau:2011},
and $Z_g/Z_\odot = ({\rm O/H})/({\rm O/H})_\odot$. Note that the
observed MZR is uncertain by a factor of 2--3 as different calibration
methods produce different zero-points and, to some extent, different
slopes \citep{Kewley:2008}. The predictions of our three models are
again quite similar, except that the KS model again produces later
enrichment of massive galaxies, so the MZR is shallower at $z\sim
3$--6. All three models produce a cold gas phase MZR that is, taken at
face value, considerably steeper than the observed gas phase
MZR. Moreover, the predicted \emph{evolution} of gas phase metallicity
in our models is quite different from that implied by current
observations. The models predict that the gas phase metallicity for
galaxies of fixed stellar mass declines slightly with decreasing
redshift, while observations indicate an increase of almost a factor
of two between $z\sim 2.2$ and $z\sim 0$. This discrepancy was shown
previously by \citet{White:2014} for our KS models; we see here that
the results are qualitatively similar for our new fiducial GK and BR
models. We discuss possible reasons that our models reproduce the
stellar MZR fairly well but seem to fail to reproduce the observed gas
phase MZR in \S\ref{sec:discussion:mzr}.

\subsubsection{Evolution of Global Quantities}

\begin{figure*} 
\includegraphics[width=0.3\textwidth]{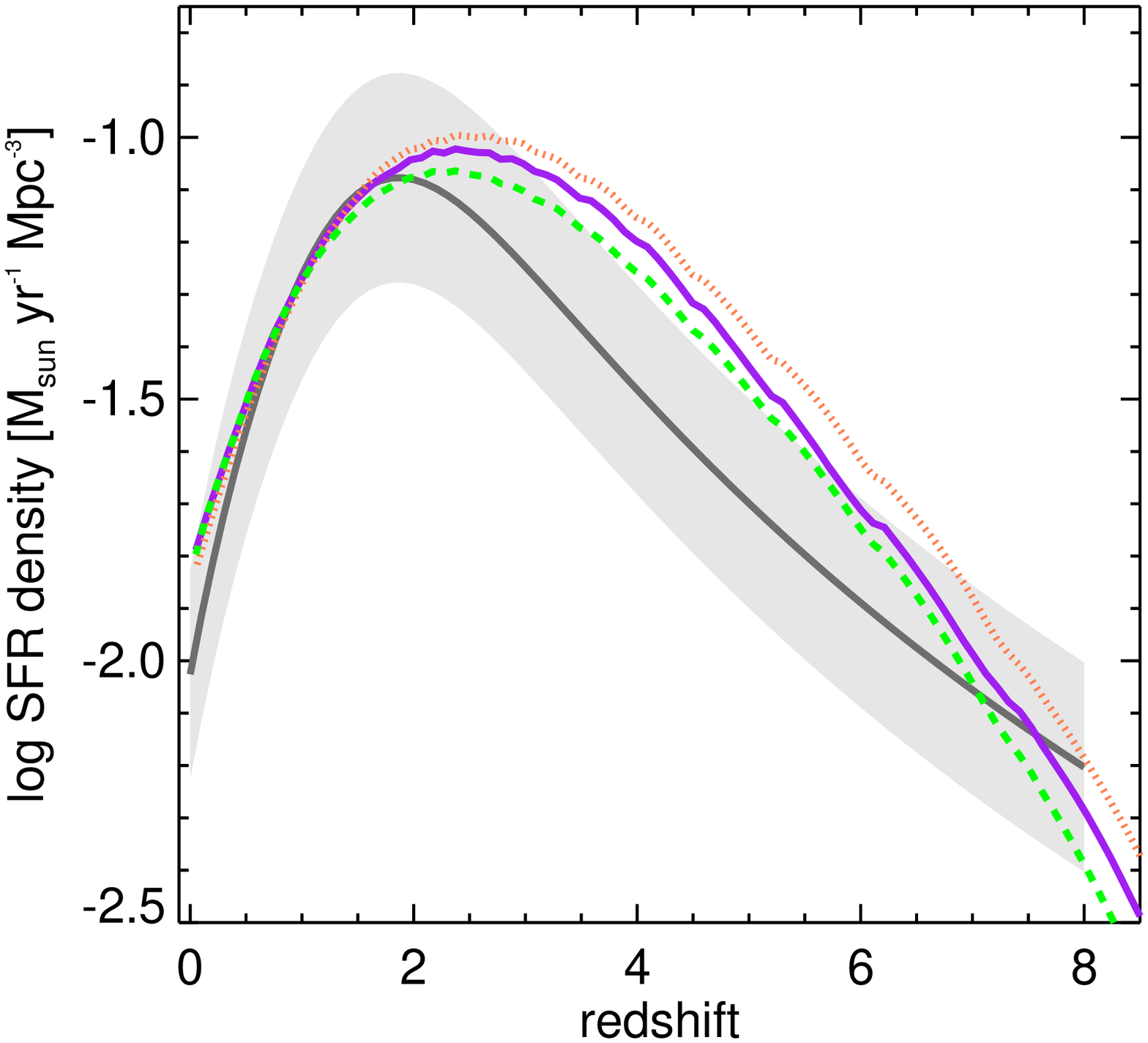}
\includegraphics[width=0.3\textwidth]{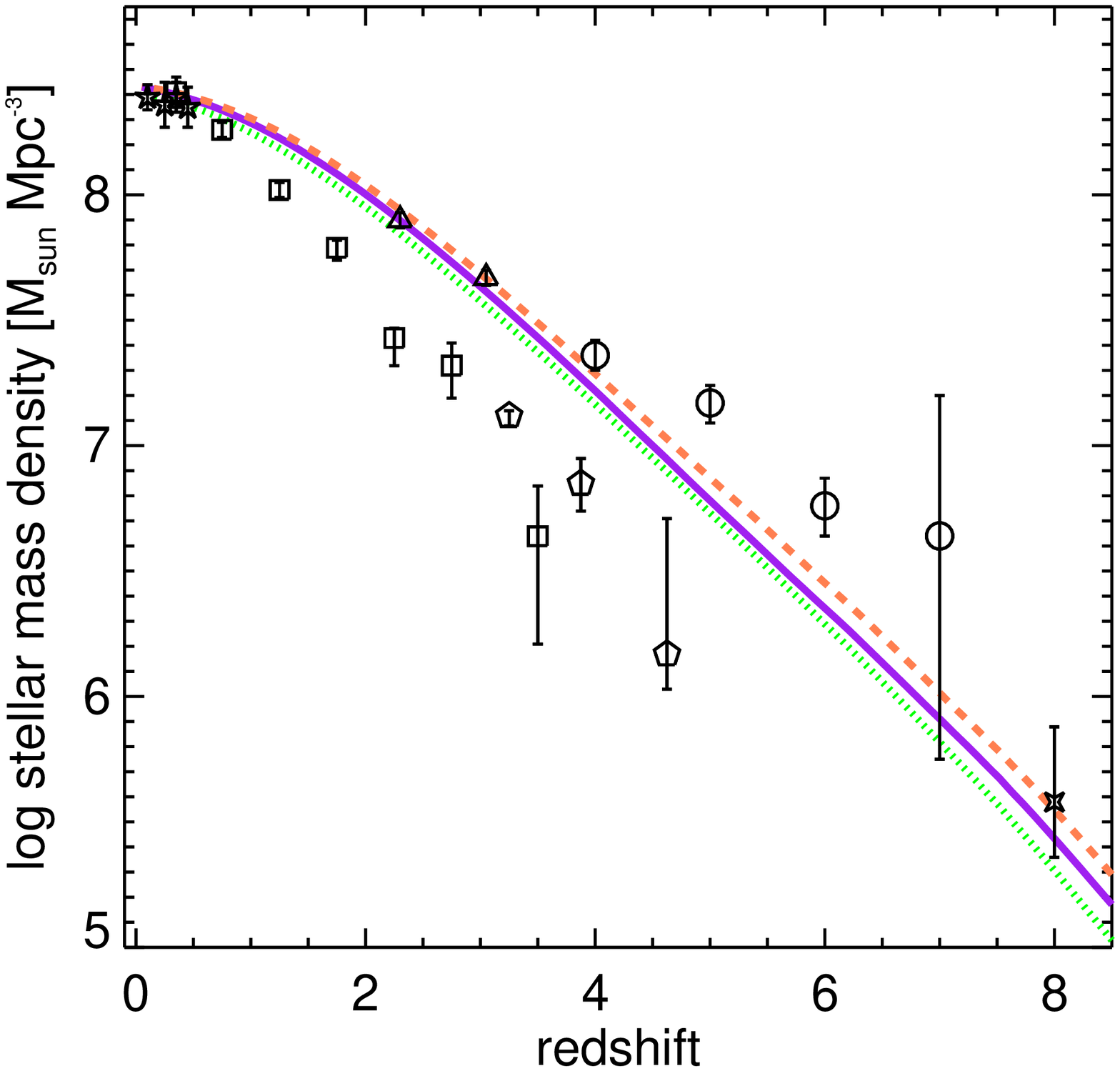}
\includegraphics[width=0.3\textwidth]{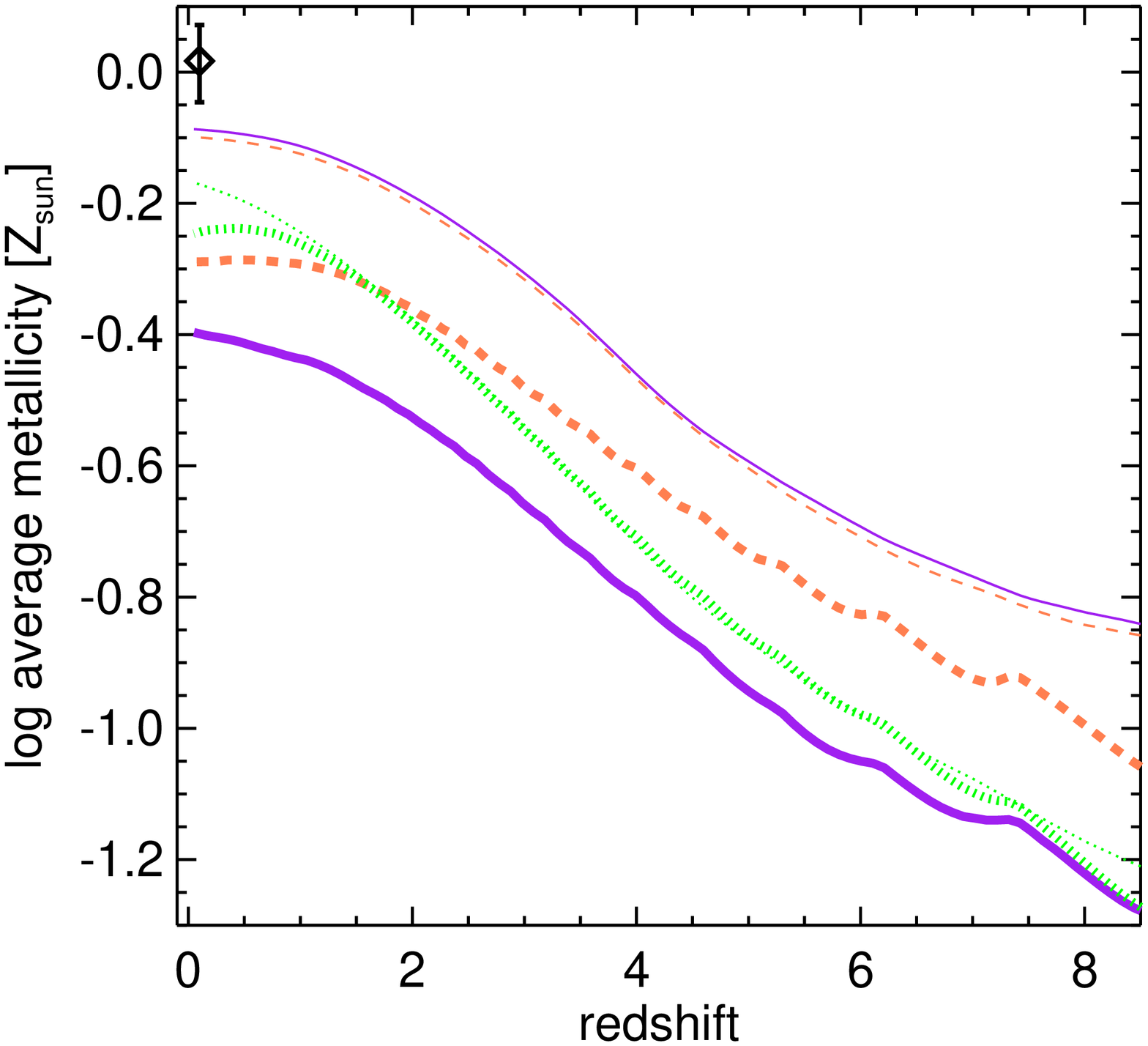}
\caption{Global history of star formation, stellar mass assembly, and
  metallicity as a function of redshift. The solid purple line shows
  the results of our fiducial GK model, the dashed orange line shows
  the BR model, and the dotted green line shows the KS model. Left:
  global star formation rate density; gray lines show the fit to the
  compilation of observational estimates from
  \protect\citet{Madau:2014}. Middle panel: global stellar mass
  density; symbols show selected observational estimates taken from
  Table~2 of \protect\citet{Madau:2014}. Right: mean mass-weighted
  metallicity; thick lines show the metallicity of the cold gas
  component and thin lines show the stellar component. The diamond
  symbol at $z\sim 0$ is the mean stellar metallicity derived by
  \citet{gallazzi:05}.
\label{fig:history}}
\end{figure*}

In Fig.~\ref{fig:history} we show our model predictions for the
evolution of the global SFR density, global stellar mass density, and
average metallicity of cold gas and stars over cosmic time. The figure
shows that the three models predict almost identical global SFR
densities at low redshift, while at $z\gtrsim 2$, the BR model
produces the highest SFR density, and the KS model the lowest, with
the GK model in between. We also see that our model predictions are in
reasonably good agreement with observations at $z\lesssim 2$ and
$z\gtrsim 6$, though this is in part a fortuitous cancellation --- the
models overproduce galaxies with low SFR but underproduce ones with
high SFR. Moreover, the observational estimates of the SFR density
from \citet{Madau:2014} are integrated only down to 0.03 $L_*$, while
our theoretical predictions are integrated over all galaxies. This
same behavior is echoed in the build-up of the global stellar mass
density. The largest difference in the three models appears in the
evolution of the stellar and cold gas phase metallicity. The models
differ in the normalization and evolution of the mean metallicity for
gas and stars. In addition, in the KS model, the mean metallicity of
cold gas and stars is very similar, while there is a much larger
difference between the stellar and gas phase metallicities of gas and
stars in the BR model and GK models. As one can see in
Fig.~\ref{fig:massmetstar} and \ref{fig:massmetgas}, at stellar masses
above $m_{\rm star} \sim 10^7\, \msun$, the models make similar
predictions for the relative metallicities of gas and stars. The
differences seen in Fig.~\ref{fig:history} are entirely due to very
low mass halos.

\section{Discussion}
\label{sec:discussion}

\subsection{Interpreting our results: the equilibrium model}

One of the main conclusions of our work is that modifying the recipes
for how cold gas is converted into stars has very little effect on the
properties of low-mass galaxies ($m_{*} \lesssim M_{\rm char}$, where
$M_{\rm char}$ is the ``knee'' in the stellar mass function). Instead,
modifying the star formation recipe mainly changes the ratio of cold
gas to stars in galaxies.  Similar conclusions were reached in the
study by \citet{White:2014}, in which more extreme (though in some
cases less physically motivated) modifications to the SF recipe
in similar models were made. This is because in our models, star
formation in low-mass galaxies is strongly self-regulated: if star
formation is made less efficient, less gas is ejected by stellar
winds, leading to more efficient star formation, and vice versa.

A number of recent works have pointed out this property of
self-regulation, which is a rather generic feature of modern galaxy
formation models, arising from the broadly adopted hypothesis of
efficient stellar-driven feedback
\citep{Schaye:2010,Haas:2013a,White:2014,Somerville_Dave:2014}. A
useful analytic framework for understanding the behavior of the fairly
complex intertwined suite of physical processes at play in
state-of-the-art galaxy formation simulations is the ``equilibrium
model'', sometimes called the ``bathtub model''
\citep[e.g.][]{Dave:2012,Dekel:2013,Dekel:2014}. The basic assumption
in this model is that due to self-regulation, on some timescale
$t_{\rm eq}$, galaxies establish an equilibrium state in which the
rate of change of their cold gas reservoir is small,
i.e. $\dot{m}_{\rm cold} \simeq 0$. Once equilibrium is established,
the star formation rate is balanced by global inflows and outflows,
$\dot{m}_* = \dot{m}_{\rm in} /(1 + \eta)$, where $\dot{m}_{\rm in}$
is the rate at which gas flows into the galaxy due to cosmological
accretion and $\eta \equiv \dot{m}_{\rm out}/\dot{m}_*$ is the mass
loading factor of a large-scale stellar driven outflow. The time for a
galaxy to come into equilibrium (or to re-establish equilibrium after
a disruption) is
\begin{equation}
t_{\rm eq} = \frac{\tau_{\rm SF}}{1 + \eta} 
\end{equation}
where $\tau_{\rm SF} \equiv m_{\rm cold}/\dot{m}_*$ is the star
formation (or gas depletion) timescale \citep[][hereafter
  DFO12]{Dave:2012}.

In simulations, $\eta$ is a fairly strong inverse function of galaxy
mass, while $\tau_{\rm SF}$ is a weaker function of galaxy mass
(DFO12).  Therefore low mass galaxies come into equilibrium
earlier. This helps us to understand why changing our star formation
recipe had less impact on low-mass vs. massive galaxies. In addition,
it explains why high-mass galaxies were affected only at high
redshifts -- at $z\gtrsim 2$, these galaxies have not yet come into
equilibrium. This work makes the interesting prediction that
observations of massive galaxies at very high redshift ($z\gtrsim 2$)
will place strong constraints on the physics of star formation in cold
dense gas, while constraints on the low-mass end of the galaxy stellar
mass function at high-$z$ will mainly constrain the physics of
outflows\footnote{All of this discussion implicitly assumes that
  ``preventative'' feedback --- physical processes that could prevent
  gas from accreting into galaxies or becoming available for star
  formation --- is sub-dominant. At very low masses, preventative
  feedback due to photo-ionization squelching likely becomes
  important. Preventative feedback due to AGN heating and winds, and
  virial shock heating, is probably dominant in massive galaxies at
  late times ($z\lesssim 2$). See DFO12 and
  \citet{Somerville_Dave:2014} for a more complete discussion. }.

\subsection{Mass-metallicity relations for gas and stars}
\label{sec:discussion:mzr}
It is puzzling that our models reproduce the stellar MZR fairly well
but predict a much steeper gas phase MZR than observations appear
to indicate. This has been seen in other models as well --- models that
invoke a weaker dependence of mass outflow rate on galaxy circular
velocity, and normalize their yield parameter to the observed
\emph{gas phase} MZR, produce better agreement with the observed gas
phase MZR but then fail to reproduce the stellar MZR
\citep{Lu:2014}. \citet{Peeples:2013} combined empirical star
formation histories derived from the observed SFMS with the observed
relation between SFR, stellar mass and gas phase metallicity (assumed
to be universal) and predicted the stellar MZR, finding fairly good
agreement with observations. However, \citet{Munoz:2014} did a more
realistic calculation using a similar approach, but accounting for
stochasticity in the SF histories and quenching, and found more
significant tension between the gas and stellar phase MZR.

We can see by comparing Fig.~\ref{fig:massmetstar} and
\ref{fig:massmetgas} that our models predict that the stellar
metallicity in galaxies is about a factor of 1.5--1.7 lower than the
gas phase metallicity, with weak trends on stellar mass and
redshift. We can also see from the figures in \S\ref{sec:trace} that
the gas phase metallicity tends to increase rapidly and monotonically
with time in our models. The stellar metallicity is effectively a
mass-weighted average over the chemical enrichment history of the
galaxy, so it makes sense that the stellar metallicities are slightly,
but not enormously, lower than the gas phase metallicities at any
given time. Taken at face value, the observational results --- which
imply that $Z_{\rm gas}/Z_{\rm star}$ is as high as a factor of $\sim
10$ or more, and is a fairly strong function of stellar mass --- may
be difficult to reproduce in cosmological models without invoking
accretion of highly metal pre-enriched gas.

An alternative explanation is that the normalization, and possibly the
slope, of the gas and stellar phase MZRs are not accurately calibrated
to the same system. Indeed, some gas phase metallicity indicators do
yield a gas MZR normalization and slope that is more consistent with
the stellar MZR \citep{Kewley:2008,Munoz:2014}. Another potential
issue is that we have plotted \emph{stellar mass} weighted
metallicities, while the observed stellar metallicities are
\emph{luminosity} weighted. However, other studies have found that the
luminosity weighted stellar MZR does not differ significantly in slope
from the stellar mass weighted MZR \citep{Trager:2009,Peeples:2013}.

Another issue is that different observational probes are sensitive to
different chemical elements. The stellar metallicities derived by
\citet{gallazzi:05} are sensitive to a combination of Fe and Mg, and
the stellar MZR derived by \citet{Kirby:2013} measures
[Fe/H]. \citet{gallazzi:05} quote their results in terms of $Z_{\rm
  star}/Z_\odot$ and claim that their results are independent of
$\alpha/{\rm Fe}$ (A. Gallazzi, priv. comm.). The observational gas
phase MZRs are primarily sensitive to $\alpha$ elements and are quoted
in terms of oxygen abundance ($12 + \log({\rm O}/{\rm H})$). We have
assumed that $Z_{\rm gas}/Z_{\odot} = ({\rm O}/{\rm H})/({\rm O}/{\rm
  H})_\odot$ and $Z_{\rm star}/Z_\odot = ({\rm Fe}/{\rm H})/({\rm
  Fe}/{\rm H})_\odot$ for the \citet{Kirby:2013} observations. This is
equivalent to assuming that all the stars in our model galaxies have
$(\alpha/{\rm Fe}) = (\alpha/{\rm Fe})_\odot$. However, $(\alpha/{\rm
  Fe})$ is known to differ significantly from the Solar value in stars
in our own Galaxy \citep[e.g.][]{Stoll:2013}, in nearby dwarf galaxies
\citep[][and references therein]{Tolstoy:2009}, and in giant
ellipticals \citep{Thomas:2005,Trager:2000}. 

In the models presented here, we track the total metallicity assuming
a constant yield, and we also adopt the instantaneous recycling
approximation. Metals are produced in direct proportion to the
formation of stars, so enrichment in our models probably most closely
traces $\alpha$ elements, but we normalized our yield parameter to
observations that are also sensitive to Fe (see above). This simple
version of single-element chemical enrichment with the instantaneous
recycling approximation is the standard approach adopted in
semi-analytic models. However, physical processes in the models are
actually dependent on different elements in potentially significant
ways. For example, the (also widely adopted in SAMs) cooling tables of
\citet{sutherland:93} used to model the cooling rate of hot halo gas
are parameterized via [Fe/H] and adopt assumed [Fe/H]-dependent
abundance ratios. However, \Htwo\ formation is more closely tied to
$\alpha$ elements such as oxygen and carbon, which are primary
coolants in the interstellar medium \citep[e.g.][]{Glover:2014}. One
conclusion of the work presented here is that if we wish to include
more realistic physics in our models, it is important to track
multiple chemical elements and their production via different channels
(stars of various masses, Type II SNae and prompt and delayed Type Ia
SNae) and on different timescales. Several groups have developed SAMs
that include more detailed chemical evolution models
\citep{Arrigoni:2010,Yates:2013}. We have integrated the more
sophisticated chemical evolution models presented in
\citet{Arrigoni:2010} within our new SAMs, including the new
metallicity-dependent \Htwo\ formation and \Htwo-based star formation
recipes as described here, and plan to investigate the predicted
metallicities of cold gas and stars and their evolution in these
models in a future work (Peeples et al. in prep).

In the meantime, we can perform an empirical correction for the
variation of $\alpha/{\rm Fe}$ to see if this is a plausible
explanation for the discrepancy. If we assume that the metallicity
tracked in our models is actually Fe, and apply the empirical relation
presented by \citet{Stoll:2013}\footnote{$[{\rm Fe}/{\rm H}] = -0.34 +
  1.25 [{\rm O}/{\rm H}]$}
to our model galaxies to ``convert'' to [{\rm O}/{\rm H}], we find
much better agreement between our predicted gas phase MZR and at least
some calibrations of the observed MZR (see
Fig.~\ref{fig:massmetgas}). Note that, conceptually following
\citet{Munoz:2014}, we are effectively assuming that a relationship
between [Fe/H] and [O/H] derived for individual stars in the Milky Way
holds for the \emph{average} stellar and gas metallicities in galaxies
with a variety of star formation histories --- an assumption that may
well not be valid. However, it suggests that properly accounting for
non-Solar $\alpha/{\rm Fe}$ and its possible trends with other
galaxy properties (such as stellar mass and metallicity) may at least
partially relax the tension between the stellar and gas phase MZR seen
in our models and others.

\subsection{Caveats and limitations of our models}
Understanding how the ``small scale'' processes of star formation and
stellar feedback interact with cosmological scale processes such as
galactic scale inflows and outflows to shape the observable properties
of galaxies is currently one of the major unsolved problems in the
study of galaxy formation and evolution. The models presented here
neglect a large number of processes that are thought to be important
in influencing how efficiently gas can form molecules, and in turn how
stars form within molecular gas. For example, we do not consider the
possible impact of the local shear field, and non-axisymmetric
perturbations such as spiral arms and bars. Nor do we attempt to model
the ``local'' effects of stellar feedback (through stellar winds,
supernovae, and H$_{\rm II}$ regions) on star formation. We instead
assume that the efficiency of converting \emph{molecular gas} into
stars is roughly constant, as suggested by observations of nearby
spiral galaxies \citep{Bigiel:2008,Bigiel:2011}. As pointed out by
\citet{Krumholz:2012b} and many others, the efficiency of forming
stars \emph{within} GMCs is surprisingly low, only about 1\% per free
fall time. Our picture is that local feedback processes are
responsible for setting that efficiency, and that when we smooth over
several 100 pc regions of the ISM, it averages out to a nearly
universal value.

However, there have been some recent studies that suggest that the
galaxy-averaged value of this molecular star formation efficiency
(often expressed as a depletion time, $t_{\rm dep, H_2} \equiv
m_{H_2}/\dot{m}_*$) may vary significantly from galaxy to galaxy, and
may have a strong dependence on global galaxy
properties. \citet{Saintonge:2011} find that in the COLD GASS sample,
$t_{\rm dep, H_2}$ is weakly correlated with the galaxy stellar mass
and stellar surface density, and rather strongly (anti-) correlated
with sSFR. However, \citet{Saintonge:2011} adopted a constant
(Galactic) value for the conversion factor between CO and
\Htwo\ ($\alpha_{\rm CO}$). \citet{Leroy:2013} found similar
correlations in their sample of 30 disk galaxies from the HERACLES
survey, but found that most of the correlation disappeared when they
applied a theoretically motivated dependence of $\alpha_{\rm CO}$ on
the dust-to-gas ratio. They found smaller residual variations in
$t_{\rm dep, H_2}$, mainly associated with nuclear gas
concentrations. This suggests that there may be different values of
$t_{\rm dep, H_2}$ in undisturbed disk galaxies and in galaxies
experiencing mergers and interactions \citep[see
  also][]{Daddi:2010,Genzel:2010}, possibly due to a super-linear
dependence of SFRD on \Htwo\ density, as assumed in our fiducial
models. We do include an enhancement of SFE in mergers in our models,
but the treatment is based on hydrodynamic simulations of binary
mergers with a rather outdated treatment of sub-grid physics, so this
is clearly an area that should be explored with state-of-the-art
high-resolution hydrodynamic simulations.

Another limitation of our approach is that although we compute the
\Htwo\ fraction and $\Sigma_{\rm SFR}$ in radial annuli in each disk,
and integrate over the disk to obtain the global properties, we do not
store the information on the stellar, gas, and metal content of each
annulus over different timesteps. Therefore we assume that both the
gaseous and stellar disks have radial exponential profiles, and adopt
a simple fixed factor relating the size of the gaseous and stellar
disk. While this may be a reasonable approximation on average, it may
miss important trends. We also use the global values of the gas phase
metallicity and SFR (which we use as a proxy for the UV radiation
field) in the GK recipe, instead of the local values of these
quantities in annuli. In future work, we plan to construct more
detailed models of disks in which all of these quantities are tracked
as a function of radius, along the lines of work by \citet{fu:10} and
\citet{Dutton:2010}.

\subsection{Comparison with previous work}

Several other groups have carried out studies similar to this one.
\citet{lagos_sflaw:11} considered two models without gas
partitioning. The first unpartitioned model adopts the original SF
relations implemented in the \citet{Baugh:2005} and \citet{Bower:2006}
GALFORM models, in which the SFR was assumed to be proportional to the
total cold gas mass divided by a timescale $\tau_*$. In the
\citet{Baugh:2005} models, $\tau_*$ was scaled with the galaxy
circular velocity to a power, and in the \citet{Bower:2006} models,
also with the galaxy dynamical time. The second unpartitioned model
used a Kennicutt-Schmidt SF recipe similar to the one we have adopted
here. They also considered two models with gas partitioning: one with
the pressure-based BR recipe, and one with the metallicity-based KMT
recipe. In their BR model, they adopted SF relations similar to our
Big1 and Big2 recipes. In the KMT model, they used the SF relation
given by KMT. They did not attempt to separate the effects of the
different gas partitioning recipes versus \Htwo-based SF recipes. We
focus on the results of the \citet{Bower:2006} variant of the GALFORM
models, which are more similar to our models than the
\citet{Baugh:2005} variant. \citet{lagos_sflaw:11} do not show stellar
mass function predictions, but find that the rest frame K-band
luminosity function at $z=0$, 1, and 2 shows little change between the
six different star formation recipes they explored, consistent with
our results. \citet{lagos_sflaw:11} emphasize differences in the SFR
distributions as a function of stellar mass, in particular the
prominence of a passive population in models with different star
formation recipes. We find very small differences in the SFR
distributions between our different models, and note that the
prominence and location of a passive population will certainly also be
very sensitive to the treatment of AGN feedback. It is also
interesting to note that \citet{lagos_cosmic:11} end up favoring their
BR models and strongly disfavor the KMT recipe because they find that
it does not reproduce the observed \HI\ and \Htwo\ gas scaling
relations. However, we showed in PST14 that both our BR and GK models
do about equally well at reproducing available observations of cold
neutral gas in galaxies. \citet{lagos_sflaw:11} do not show
metallicity predictions.

\citet{fu:12} conducted a similar study, investigating four SF recipes
and separately studying two gas partitioning recipes, BR and KMT,
within the framework of the models developed by \citet{fu:10} based on
the MPA Millennium SAMs \citep{Croton:2006,Guo:2011}. Their ``Bigiel''
SF recipe is similar to our Big1 recipe, but depends on the
\Htwo\ fraction, steepening below a critical value. Their
``Kennicutt'' recipe is similar to our KS recipe. Their ``Genzel''
recipe contains a linear scaling of $\Sigma_{\rm SFR}$ with
$\Sigma_{H2}$, but also scales as the inverse galaxy dynamical time
(which is redshift dependent). They also consider the KMT SF
recipe. Again their results are quite consistent with ours. The
stellar and \Htwo\ mass functions are almost identical for all models,
while the largest difference between models is in the
\HI\ content. This is the same conclusion that we reach based on PST14
and this study. All of their models underproduce massive galaxies at
high redshift ($z\gtrsim 2$), as we found with similar SF
prescriptions to the ones they adopted. \citet{fu:12} do not show
their $m_{\rm star}$-SFR relation, but they show that the different
star formation recipes can lead to substantial deviation between model
predictions for the cosmic SFR density at high redshift ($z\gtrsim
3$--4), as we also find. Interestingly, \citet{fu:12} find that their
gas phase MZR is \emph{too shallow} compared with observations --- the
opposite problem to the one we encounter in our models. This is
probably due to their adopted scaling for stellar driven winds. They
assume that the mass outflow rate $\dot{m}_{\rm out} \propto
\dot{m}_*$, i.e., a fixed mass loading factor, while we assume that
the mass loading factor $\dot{m}_{\rm out}/\dot{m}_*$ scales
approximately with inverse circular velocity squared. This dependence
of MZR slope on wind scaling parameters is well known
\citep{Peeples:2011}. They also find that the \emph{redshift
  evolution} of the MZR is quite sensitive to the SF recipe adopted;
in particular, recipes in which SFE scales with galaxy dynamical time
predict very weak or no evolution in the MZR, while their models
without dynamical time scalings predict stronger evolution. None of
our models contain an explicit scaling with dynamical time, but the
non-linear slope of our Big2 and KS recipes has a similar effect,
consistent with the weak evolution in the MZR seen in our
models. 

\citet{Krumholz:2012} implemented an updated version of the KMT recipe
within simplified semi-analytic models that only follow the mass
accretion history of the main branch, and do not track the full merger
trees. In their fiducial model they additionally assume a constant
mass loading factor for stellar-driven winds (no dependence on galaxy
mass or circular velocity) and strongly metal-enhanced winds. When we
implement similar ingredients in our models (except that we retain our
``energy driven'' stellar wind scalings), we obtain qualitatively
similar results. Namely, a metallicity-dependent formation efficiency
for \Htwo\ and self-consistent \Htwo-based star formation recipe can
significantly suppress and delay star formation and metal enrichment
in very low-mass halos.  We find that this effect is considerably
stronger in our KMT+MEW models than in our fiducial GK models. This is
because a) in our fiducial GK model, the effect of a varying UV
background partially compensates for the metallicity dependence of
\Htwo-formation; and b) the metal enhanced winds delay enrichment of
the cold gas, further suppressing star formation (see
\S\ref{sec:trace}). However, we stress that a noticable effect is seen
only in halos with virial mass $M_h \lesssim 10^{10.5}\, \msun$ and at
early times ($z\gtrsim 1$).  Therefore, although \citet{Krumholz:2012}
do not show their predicted stellar mass functions or stellar
fractions, it is likely that their model also still suffers from the
overprediction of low-mass galaxies ($m_{\rm star} \sim 10^{9-10}\,
\msun$) at intermediate redshifts ($0.5\lesssim z \lesssim 4$) that we
have discussed extensively above (see also Fig.~\ref{fig:smf2} and
\ref{fig:fstar_comp2}). Certainly they see the same qualitative
problems with predicted specific star formation rates being too low at
intermediate redshifts that we have described here.

Galaxies hosted by halos in the strongly affected mass range have
typical stellar masses of $m_{\rm star} \sim 10^{7-8}\, \msun$, and it
is probably not feasible to obtain complete samples of these objects
at high redshift with existing facilities. However, this strong
suppression of star formation in low mass halos could have
implications for reionization, future observations with the James Webb
Space Telescope, and stellar archaeology in local dwarf
galaxies. Moreover, \citet{Berry:2014} showed that the strong
suppression of \Htwo\ formation and star formation in low-mass halos
predicted by the KMT-like models would lead to a large population of
``barren'' halos that never experience significant star formation, and
so are filled with \HI\ \citep[see
  also][]{kuhlen:12}. \citet{Berry:2014} found that the existence of
this population is in apparent tension with observations of
\HI\ absorption systems at high redshift.

We found that the inclusion of metal-enhanced winds (MEW) produces a
steeper MZR, leading to an even greater tension with
observations. \citet{Krumholz:2012} also found that their models
produce a steeper gas phase MZR than observations, in spite of their
adopted constant mass loading factor which generally leads to
shallower predicted MZR. Interestingly, \citet{Krumholz:2012} find
significant evolution in the gas phase MZR from $z\sim 2$--0 in their
models, in better apparent agreement with observations.

\section{Summary and Conclusions}
\label{sec:conclusions}

We have presented new models of galaxy formation, set in the framework
of cosmological merger trees, which include physically motivated
recipes for the partioning of cold gas into an atomic, molecular, and
ionized phase. These models have several advantages: first, we can
make explicit predictions for the atomic and molecular gas properties
of galaxies over cosmic time, which can be directly confronted with
observations from current and upcoming facilities. The first of these
predictions, for \HI\ and \Htwo\ gas observed in emission, were
presented in \citet{Popping:2014}, and predictions for \HI\ gas
observed in absorption were presented in
\citet{Berry:2014}. \citet{Popping:2014c} extended these models to
predict sub-mm line emission from several atomic and molecular
species. These predictions may be directly compared with observations
from current and upcoming facilities such as ALMA (Popping et al. in
prep), and also used to plan future observations with these
facilities. Second, we can implement more physically motivated
\Htwo-based recipes for star formation within our models. In this
paper we focussed on the predictions for the stellar mass content,
star formation rate, and metallicity of galaxies in our new models,
and compared these with available observational estimates from $z\sim
0$--6.

We summarize our main conclusions below: 
\begin{itemize}
\item We performed a number of tests of the robustness of our models
  to resolution and various parameter values and ingredients. We find
  that our code is very well converged with respect to variation of
  the mass resolution of our merger trees. In addition, our results do
  not depend sensitively on the assumed values of the metallicity of
  pre-enriched gas or the molecular hydrogen floor (within a
  reasonable range of values).

\item Accounting for a reservoir of ionized gas (due to an internal
  and external photo-ionizing radiation field) in our models, which is
  not allowed to form molecular hydrogen or stars, does not
  significantly change our predictions.

\item We explored the effect of adopting different \Htwo-based star
  formation recipes. All the recipes we considered gave similar
  results for low-mass galaxies. Models that adopted a ``two-slope''
  recipe (Big2), which has a linear dependence of star formation rate
  density on molecular gas surface density below a critical value,
  steepening to $\Sigma_{\rm SFR} \propto {\Sigma_{\rm H2}}^2$ above a
  critical value of $\Sigma_{\rm H2}$, produced more efficient star
  formation and metal enrichment in massive galaxies at high
  redshift. Models that implement the Big2 recipe appear to be in
  better agreement with current observational estimates of the number
  density of massive galaxies at high redshift.

\item We explored the effect of different recipes for partioning gas
  into an atomic and molecular component. The metallicity-based GK
  recipe and the pressure-based BR recipe gave surprisingly similar
  results, perhaps because of the strong correlation between disk
  mid-plane pressure ($\propto \Sigma_*$, to first order) and
  metallicity in our models. The KMT and GKFUV recipes, which do not
  include a dependence on the FUV radiation background as our fiducial
  GK recipe does, predicted less efficient formation of \Htwo, less
  star formation and metal enrichment at early times, and later
  stellar mass assembly. These differences are only noticable,
  however, in very low mass halos ($\log (M_h/\msun) \lesssim 10.5$).

\item Both of our new fiducial models (GK and BR) reproduce the
  curvature in the relationship between total cold gas surface density
  and $\Sigma_{\rm SFR}$ seen in observations of nearby spiral
  galaxies. Our results illustrate the difficulty of disentangling the
  physical processes that are responsible for the scatter in this
  relationship, however, because of the strong correlations in galaxy
  properties.

\item The stellar mass function and mean stellar fractions ($f_{\rm
  star} \equiv m_*/M_h$) of galaxies in our three fiducial models (the
  ``classic'' KS model, the metallicity-based GK model and the
  pressure-based BR model) are almost identical for low-mass galaxies
  at all redshift $z\sim 0$--6. Both of the new models (GK and BR)
  predict earlier formation of massive galaxies, in better agreement
  with observational estimates than models that adopt the KS recipe.

\item Although the median values of $f_{\rm star}$ are very similar in
  low-mass halos in all three models, the models can have
  significantly differently shaped \emph{distribution functions}
  $P(f_{\rm star}|M_h)$. In particular, the GK model tends to have a
  much more pronounced tail to low values of $f_{\rm star}$ in
  low-mass halos ($\log (M_h/\msun) \lesssim 10.5$). The predicted
  broadening in $f_{\rm star}$ has potentially important implications
  for halo occupation models, which generally assume a narrow and
  fixed scatter in $f_{\rm star}(M_h)$.

\item All three fiducial models produce nearly identical predictions
  for the relationship between stellar mass and SFR at all
  redshifts. Even the distributions of sSFR at a given stellar mass
  are very similar. The KS model predicts a slightly narrower
  distribution of sSFR in high-mass galaxies at $z \gtrsim 1$ than the
  other two models.

\item All three models predict a weak dependence of the gas depletion
  time ($t_{\rm dep} \equiv (m_{\rm HI}+m_{\rm H_2})/\dot{m}_*$) on
  stellar mass, in agreement with observations of nearby normal disk
  galaxies and empirical estimates. The predicted $t_{\rm dep}$
  decreases by about 1.3 dex from $z\sim 0$ to $z\sim 6$. The KS model
  predicts milder evolution in the depletion time for massive galaxies
  to high redshift, resulting in longer depletion times in massive
  high redshift galaxies compared to the other two models.

\item All three models predict quite good agreement with the observed
  $z=0$ stellar mass versus metallicity relation (MZR) for
  \emph{stellar} metallicities, but predict a gas phase MZR that is
  much steeper than observational estimates taken at face
  value. However, this tension may \emph{perhaps} be relieved by
  properly accounting for the possible dependence of [$\alpha$/Fe] on
  galaxy properties. The KS model predicts later metal enrichment of
  massive galaxies, leading to a shallower MZR at high redshift. In
  tension with observational results, all of our models predict a
  nearly constant or slightly declining metallicity for galaxies
  selected at fixed stellar mass from $z\sim 4$--0.

\end{itemize}

\section*{Acknowledgments}

We thank Molly Peeples, Mark Krumholz, Fabian Walter, Adam Leroy, Nick
Gnedin and Joop Schaye for helpful discussions. We also thank Peter
Behroozi and Paula Santini for providing their data in electronic
form.  rss gratefully thanks the Downsbrough family for their generous
support. This work has been supported in part by grant HST AR-13270.01
from NASA. GP acknowledges funding from NOVA (Nederlandse
Onderzoekschool voor Astronomie).


\bibliographystyle{mn}
\bibliography{mn-jour,newsf}

\section*{Appendix: Results for Additional Model Variants}
\begin{figure*} 
\begin{center}
\includegraphics[width=0.75\hsize]{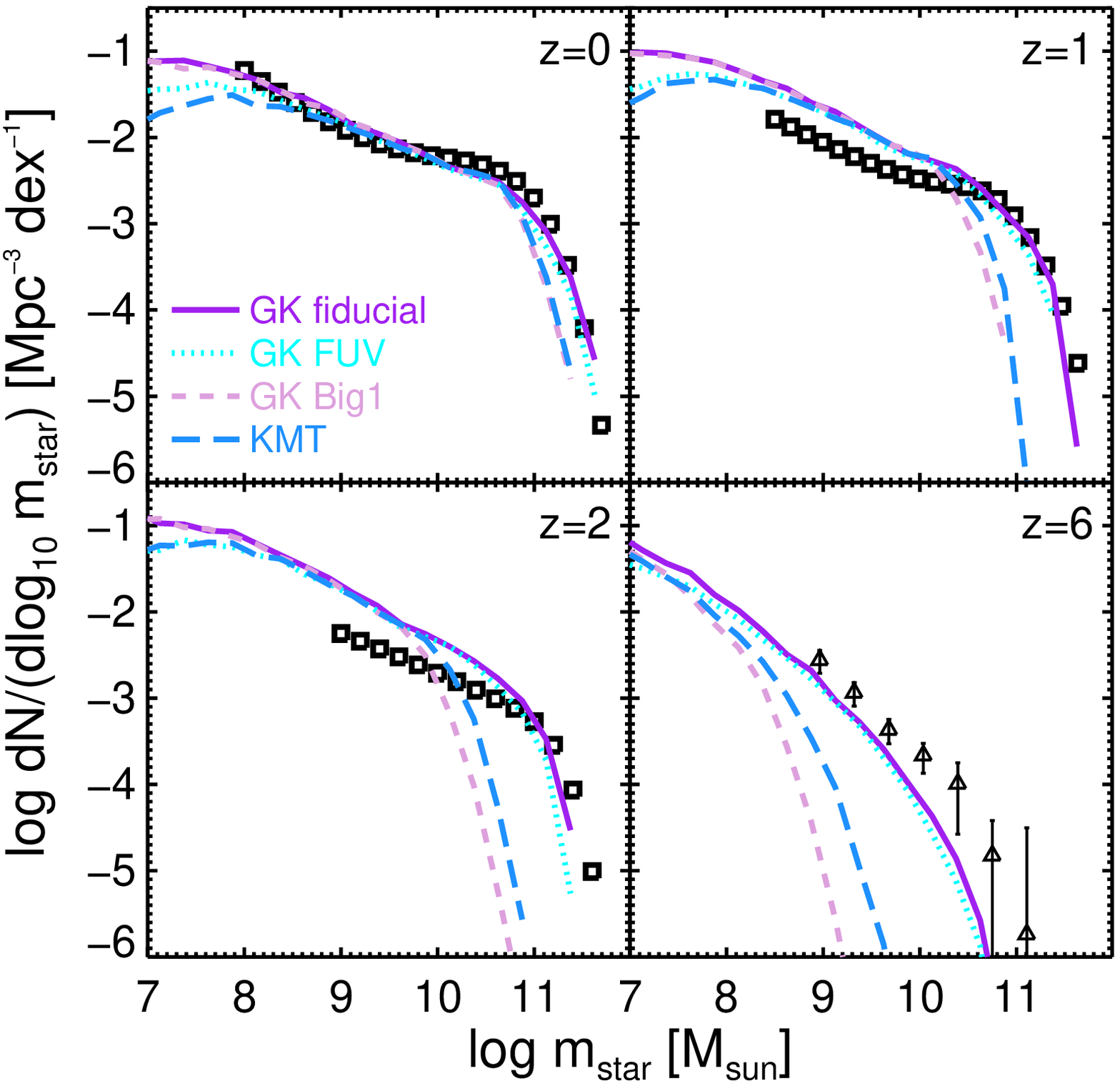}
\end{center}
\caption{Stellar mass function evolution with redshift. Symbols show
  observational estimates as detailed in
  Fig.~\protect\ref{fig:smf}. The purple solid line shows the results
  of the fiducial GK model, the cyan dotted line shows the GKFUV
  model, the dashed lavender line shows the GK Big1 model and the
  long-dashed blue line shows the KMT model (see text and
  Table~\protect\ref{tab:models}).
\label{fig:smf2}}
\end{figure*}

\begin{figure*} 
\begin{center}
\includegraphics[width=0.75\hsize]{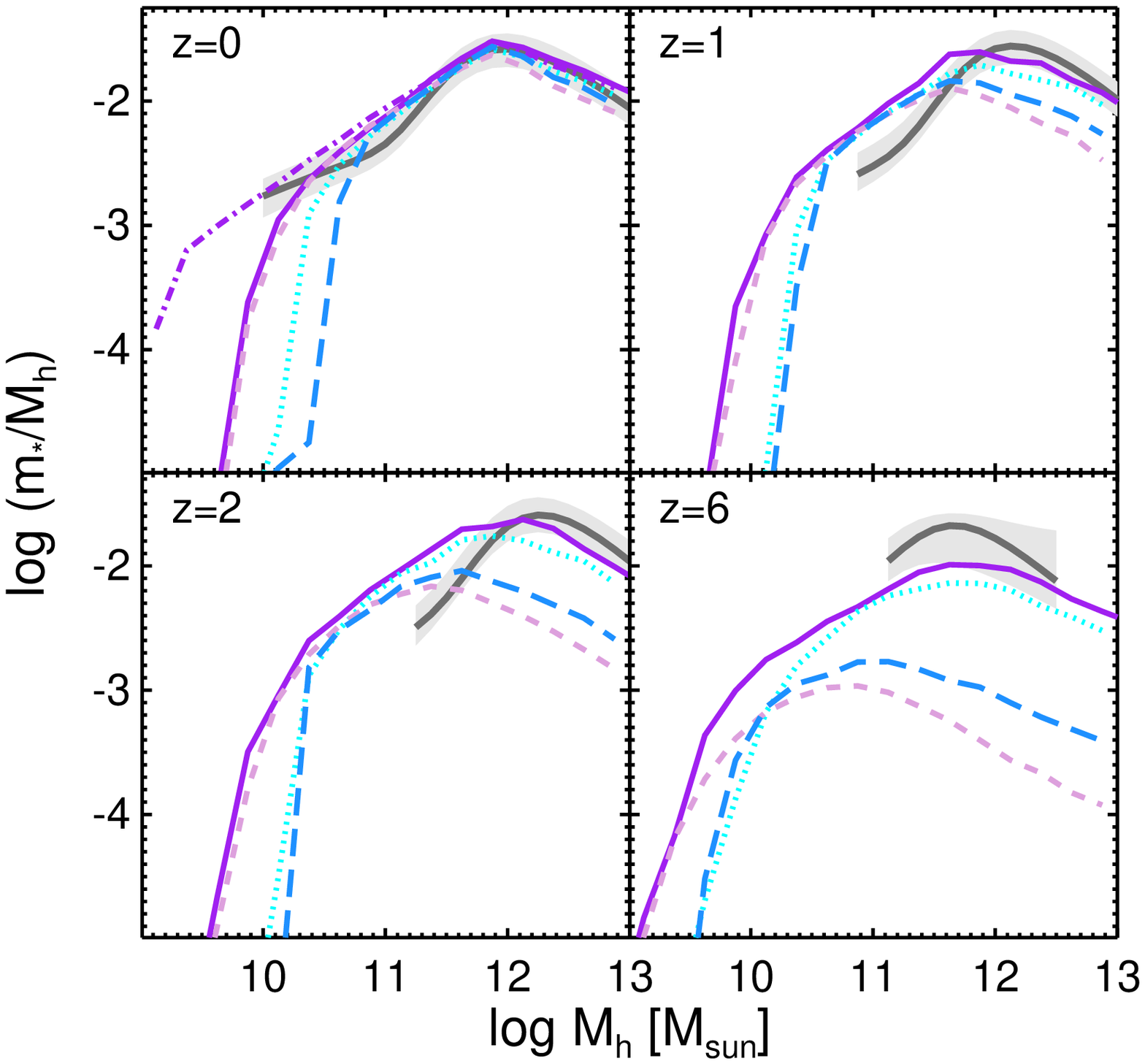}
\end{center}
\caption{The stellar mass of central galaxies divided by the total
  mass of their dark matter halo, in redshift bins from $z=0$--6. Dark
  gray solid lines and shaded areas show constraints from abundance
  matching from \protect\citet{Behroozi:2013}. Models shown are as in
  Fig.~\protect\ref{fig:smf2}. In addition, the dot-dashed purple line
  (shown in the $z=0$ panel only) shows the fiducial GK model with
  photo-ionization squelching switched off.
\label{fig:fstar_comp2}}
\end{figure*}

In this Appendix we show results for the stellar mass functions and
stellar fractions in several additional model variants, in order to
aid the interpretation of the results presented in the main text. In
Fig.~\ref{fig:smf2}, we show the stellar mass function at $z=0$, 1, 2,
and 6 for our fiducial GK model (the same one shown in
Fig.~\ref{fig:smf}), compared with the GK model with a fixed value of
the UV radiation background $U_{\rm MW}=1$ (GKFUV), the GK model for
gas partioning with the Big1 SF relation (GK+Big1), and a model with
the KMT recipes for gas partioning and a KMT SF relation (see
Table~\ref{tab:models}). Fig.~\ref{fig:fstar_comp2} shows the median
stellar fraction as a function of halo mass for central galaxies, in
the same suite of models, with the addition of the GK model with
photo-ionization ``squelching'' switched off shown in the $z=0$ panel
only.

These plots illustrate several points, which we already discussed in
\S\ref{sec:trace}. First, the metallicity and UV radiation field
dependence in the GK recipe partially counteract each other (lower
mass galaxies have lower metallicity, resulting in less efficient
\Htwo\ formation, but also a lower SFR, resulting in less efficient
\Htwo\ destruction). Recipes that do not account for the effect of a
varying UV radiation field (GKFUV and KMT) predict that
\Htwo\ formation becomes so inefficient in low mass galaxies that the
stellar mass function actually turns over at $\log (m_*/\msun) \simeq
8$. Similarly, $f_{\rm star}(M_h)$ declines sharply at $\log
(M_h/\msun) \simeq 10$. Note that although in the models shown, the
abundance of very low-mass galaxies is actually probably too small
compared with observations, we could probably adjust our recipes for
stellar feedback or photo-ionization squelching to fix this. However,
it does appear that even in these models, the excess of low-mass
galaxies ($m_{\rm star} \sim 10^{9-10}\, \msun$) at intermediate
redshift ($0.5\lesssim z \lesssim 4$) persists. This indicates that
the halo mass scale where star formation can become inefficient enough
to break the self-regulation equilibrium is smaller than the one where
the discrepancy with current observations appears.

The second point is that the differences between models seen in more
massive halos are almost solely due to the assumed scaling of the star
formation relation at large gas surface densities. The model with the
steepest dependence of $\Sigma_{\rm SFR}$ on $\Sigma_{\rm H2}$ (Big2,
with $N_{\rm SF} \rightarrow 2$ in dense gas) has the highest number
density of massive galaxies and the highest values of $f_{\rm star}$
in massive halos. The GKFUV model is almost identical to the fiducial
GK model at high masses. The KMT model is the next highest ($N_{\rm
  SF} \rightarrow 1.4$ in dense gas), then the GK Big1 model ($N_{\rm
  SF}=1$). This is because regardless of the gas partioning recipe,
gas in these galaxies is dense enough that it is nearly all
molecular. Stellar driven winds cannot efficiently escape these deep
potential wells. Therefore there is a strong dependence on the gas
depletion time (star formation efficiency).

It is also clear from Fig.~\ref{fig:fstar_comp2} that modeling of
photo-ionization squelching will have an extremely degenerate effect
with that of gas partioning and stellar feedback on stellar
properties. However, observations of gas content should help break
these degeneracies.

\end{document}